# Berry Curvature, Semiclassical Electron Dynamics, and Topological Materials: Lecture Notes for Introduction to Solid State Physics

Daniel C. Ralph

Physics Department, Cornell University, Ithaca NY
and Kavli Institute at Cornell for Nanoscale Science, Ithaca, NY

Among the most exciting recent developments in solid state physics are the discoveries that the Berry curvature associated with electronic bands can have important physical consequences for the motion of electrons, and can even lead to new topological phases of electronic matter that had not previously been recognized. These developments are not yet well-represented in the most popular textbooks about solid state physics. These lecture notes aim to introduce these concepts at the introductory level, using the simplest possible mathematics. They were developed for a unit within the Cornell University course Phys 7635, a one-semester graduate-level survey at the level of the textbook by Ashcroft & Mermin.

Given that these are lecture notes, I have tried to make them self-contained, keeping the number of external references to a minimum. In the actual course, the lectures were not given consecutively, but other aspects of band dynamics and the limitations of this semiclassical model were interspersed. Related homework problems are attached.

Please send corrections or suggestion for improvement to me at dcr14@cornell.edu.

## I: Semiclassical Dynamics of Band Electrons

*Learning Goal: Understand the physical basis for the semiclassical equations of motion for wavepackets composed of electron Bloch states.*

Electrons in crystals are subject to a potential that is periodic with respect to the crystal's Bravais lattice. Whenever this is the case, we have seen that the energy eigenstates can be written in the Bloch form

$$\psi^n_{\boldsymbol{k},\sigma}(\boldsymbol{r}) = e^{i\boldsymbol{k}\cdot\boldsymbol{r}} u^n_{\boldsymbol{k},\sigma}(\boldsymbol{r}) \tag{1.1}$$

where the function $u^n_{\boldsymbol{k},\sigma}(\boldsymbol{r})$ is periodic with respect to the Bravais lattice, $\sigma$ is an index denoting the spin state, and $n$ is an index distinguishing different bands. As a function of changing the wavevector $\boldsymbol{k}$ these states map out a band with energy $\varepsilon^n_\sigma(\boldsymbol{k})$. (When I consider only states within the same band and the same spin I will often drop the $n$ and $\sigma$ indices.)

The question I wish to consider today is: How does interaction with a periodic lattice affect the dynamics of electrons?

To begin answering this question one can imagine constructing a wavepacket located near some position $\boldsymbol{r}_0$ in the sample and asking how it will move. We will form the wavepacket as a superposition of Bloch states corresponding to a narrow range of $\boldsymbol{k}$ about some value $\boldsymbol{k}_0$ (selecting states from a single band and one spin state), with the weight function $w(\boldsymbol{k}-\boldsymbol{k}_0)$ picked so that at time $t = 0$ the expectation value of the real-space position is some value $\boldsymbol{r}_0$:

$$\Psi_{WP}(\boldsymbol{r},t=0) = \int w(\boldsymbol{k}-\boldsymbol{k}_0)\psi^n_{\boldsymbol{k},\sigma}(\boldsymbol{r})d^3\boldsymbol{k} = \int w(\boldsymbol{k}-\boldsymbol{k}_0)e^{i\boldsymbol{k}\cdot\boldsymbol{r}}u^n_{\boldsymbol{k},\sigma}(\boldsymbol{r})d^3\boldsymbol{k}. \tag{1.2}$$

I will have more to say later about exactly how one must pick the weighting function in order to specify the desired real-space position. Given that the wavepacket is composed of wavevectors from a region of reciprocal space much smaller than the Brillouin zone ($\propto(\text{small factor})\pi/a$ where $a$ is the size of a unit cell), by the properties of Fourier transforms the width of the wavepacket in real space ($\propto a/[2\pi(\text{small factor})]$) will be much larger than a unit cell. Our approach will be semiclassical, in the sense that we will consider the wavepacket to be specified by well-defined values of both the wavevector $\boldsymbol{k}_0$ and position $\boldsymbol{r}_0$, with no worries about the uncertainty principle. Within this semiclassical approximation we will also neglect effects of quantum interference that might occur for a wave traveling over more than one classical path or within a closed orbit. Because I will generally consider only wavepackets made from Bloch states in one band and the same spin, in the following I will sometimes drop the $n$ and $\sigma$ indices on the Bloch function.

To specify the semiclassical dynamics of the wavepacket, we need to describe how the expectation values of the wavepacket $\langle\boldsymbol{k}\rangle$ and $\langle\boldsymbol{r}\rangle$ will change as a function of time from their initial values $\boldsymbol{k}_0$ and $\boldsymbol{r}_0$ in response to an external force $\boldsymbol{F}_{\text{ext}}$ due to applied electric ($\boldsymbol{E}$) and/or magnetic (***B***) fields. (For simplicity, I will assume that this external force is spatially uniform, and I will not consider any possible evolution of the spin state.) It turns out that these dynamics can be summarized in three simple rules:

Rule (#1) $\dfrac{d\langle \boldsymbol{k}\rangle}{dt} = \dfrac{1}{\hbar}\boldsymbol{F}_{\text{ext}} = \dfrac{-e}{\hbar}\left[\boldsymbol{E} + \dfrac{1}{c}\dfrac{d\langle \boldsymbol{r}\rangle}{dt}\times\boldsymbol{B}\right]$

Rule (#2) $\dfrac{d\langle \boldsymbol{r}\rangle}{dt} = \dfrac{1}{\hbar}\nabla_{\boldsymbol{k}}\varepsilon_{\sigma}^{n}(\boldsymbol{k})\Big|_{\boldsymbol{k}=\boldsymbol{k}_0} - \dfrac{d\langle \boldsymbol{k}\rangle}{dt}\times\boldsymbol{\Omega}_{\sigma}^{n}(\boldsymbol{k}_0)$

Rule (#3) Wavepackets stay in one band for sufficiently weak applied electric and magnetic fields.

Compared to older textbooks, rules #1 and #3 are unchanged, and the first term on the right hand side of rule #2 is simply the familiar expression for the group velocity of a wavepacket. The second term on the right hand side of rule #2 is based on newer understanding and is missing from many textbooks. The quantity $\boldsymbol{\Omega}_{\sigma}^{n}(\boldsymbol{k}_0)$ (to be defined below) is known as the Berry curvature.

***Rule #1***:

Before beginning our focus on the Berry-curvature term, it is worthwhile to sketch briefly the origin of rules #1 and #3. First, consider how the situation would be different if, instead of using Bloch states consistent with the lattice potential, we tried to make our electron wavepacket from pure free-electron plane wave states $\propto e^{i\boldsymbol{p}\cdot\boldsymbol{x}/\hbar}$, *i.e.*, states with definite values of the momentum, ***p***. As we will see, the semiclassical picture of a wavepacket moving with a reasonably well-defined momentum would break down as this wavepacket tried to propagate through a crystal, but initially the expectation values of the momentum and position would evolve as

$$\frac{d\langle \boldsymbol{p}\rangle}{dt} = \boldsymbol{F}_{\text{total}} = \boldsymbol{F}_{\text{ext}} + \boldsymbol{F}_{\text{lattice}} \tag{1.3}$$

$$\frac{d\langle \boldsymbol{r}\rangle}{dt} = \nabla_{p}\,\varepsilon(\boldsymbol{p})\Big|_{\boldsymbol{p}=\boldsymbol{p}_0} = \frac{\langle \boldsymbol{p}\rangle}{m}. \tag{1.4}$$

For the first expression, to calculate changes in the momentum of the wavepacket, by Newton's Laws one must consider all of the forces acting on the wavepacket -- in addition to the force from the externally-applied fields, this total force also includes the interaction of the electron with the lattice. Changes in the position of the wavepacket (Eq. (1.4)) would be determined initially just by the group velocity. On short time scales (on the scale the time needed to travel a distance corresponding to an atomic spacing) the lattice interaction would cause Bragg scattering of the type we considered when evaluating nearly-free-electron-model band structures. The wavepacket would then contain contributions from waves with momenta differing from the initial value $\boldsymbol{p}_0$ by many different reciprocal lattice vectors (times $\hbar$), and our semiclassical picture of a wavepacket centered at a single value of momentum would quickly break down.

The beauty of the semiclassical treatment of band electrons using wave packets made from Bloch states is that none of the complications of the electron-lattice interaction enter into the equation of motion for the crystal momentum, $\langle \boldsymbol{k} \rangle$. Instead, the electron-lattice interactions are taken into account (exactly) when solving for the Bloch states. By a neat mathematical trick, the consequences of the electron-lattice forces on the motion of the wavepacket can be accounted for entirely in terms of the form of the band structure $\varepsilon_\sigma^n(\boldsymbol{k})$ and the Berry curvature $\boldsymbol{\Omega}_\sigma^n(\boldsymbol{k})$ and their contributions to our Rule (#2).

So, what is the justification for Rule (#1)? This follows from lattice translation symmetry and the associated conservation of crystal momentum in a periodic potential. It can be proved by a few lines of operator algebra (see Appendix 1), but I don't find that this proof provides much physical insight. As a non-rigorous but more intuitive alternative, imagine that it is possible to alternately turn the external force and the lattice potential on and off. Consider an electron in any basis state corresponding to a definite value of $\boldsymbol{k}$. While the external force is on and the lattice potential is off, the electron will gain both momentum and crystal momentum with $\Delta \boldsymbol{p} = \hbar \Delta \boldsymbol{k} = \boldsymbol{F}_{\text{ext}} \Delta t$. Now imagine that the external force is turned off and the lattice potential is turned on. How will the crystal momentum evolve? In the presence of a periodic potential, all that can occur is Bragg scattering, so that the crystal momentum cannot undergo any arbitrary change, it can only change by reciprocal lattice vectors. If all states are specified by values of $\boldsymbol{k}$ within the first Brillouin zone, this means that scattering can occur only within states corresponding to the same value of $\boldsymbol{k}$. In general, the starting state can evolve into a superposition of states in different bands, but they must all share the same value of $\boldsymbol{k}$. Therefore, only the externally-applied force enters into determining the dynamics of $\boldsymbol{k}$ for the basis state, and the electron-lattice potential has no effect on this quantity. This means that our Rule (#1) applies for every basis state contributing to our wavepacket. For a time-independent weighting function, the expectation value of $\langle \boldsymbol{k} \rangle$ for the wavepacket will shift in the same way as the wavevectors of the basis states, establishing Rule (#1) for the wavepacket.

***Rule #3***:

If the externally-applied force is sufficiently weak, there is a further simplification, that by the general considerations associated with adiabatic evolution (*e.g.*, see a very nice discussion in Griffiths), each basis state will evolve continuously to a state in the same band, up to an arbitrary phase, with negligible coupling to states in different bands. This adiabatic evolution is the basis of Rule (#3). The idea is that if a system initially occupies one Bloch basis state with wavevector $\boldsymbol{k}_1$, and then the wavevector is shifted to another value $\boldsymbol{k}_2$ by the action of an external electric or magnetic field, the original state will no longer be an energy eigenstate and will therefore begin to evolve with time. In general, the new state can always be expressed as a superposition of states in different bands, all with the wavevector $\boldsymbol{k}_2$ (which make up a complete set). However, if the shift of the wavevector is sufficiently slow, the system can have enough time to adjust so that all of the weight of the superposition will remain in the Bloch state that comes from the same band as the original state. In this adiabatic limit, the system will therefore simply move between Bloch states in the same band, with at most a change in the phase of the wavefunction as a function of time. We will have more to say about this phase in the next lecture. The situations in which adiabatic evolution is most likely to fail are avoided crossings, where two bands approach each other with just a small energy gap, $\varepsilon_{\text{gap}}$. The condition required for the adiabatic approximation to hold is that the time required to traverse the avoided-crossing region in reciprocal space should be much longer than the time scale for oscillation in the amplitudes of quantum states across the gap, $\hbar / \varepsilon_{\text{gap}}$. As discussed in Ashcroft and Mermin (pages 219-220), this leads to conditions for the maximum values of electric or magnetic field that can be applied before the adiabatic assumption breaks down: Practically, large laboratory-scale magnetic fields can drive this breakdown in some systems with narrow gaps, but laboratory-scale electric fields are generally not strong enough.

***Rule #2***:

Rule #2 is the aspect of semiclassical electron dynamics that older textbooks got wrong – in that they did not recognize the existence of the term involving the Berry curvature. Rule #2 can be understood by analyzing effective Lagrangians (e.g., ref. 1,2) or by working through some operator algebra (Appendix 1), but I think the physics can also be understood most transparently just by analyzing the real-space motion of an electron wavepacket. We will first need to think carefully about how to construct a wavepacket from a superposition of Bloch states $\psi_{\boldsymbol{k},\sigma}(\boldsymbol{r}) = e^{i\boldsymbol{k}\cdot\boldsymbol{r}} u_{\boldsymbol{k},\sigma}(\boldsymbol{r})$ so that the superposition has a desired initial real-space position $\langle \boldsymbol{r} \rangle = \boldsymbol{r}_0$. To do this, you want to arrange the superposition so that the Bloch states in a range of $\boldsymbol{k}$ near $\boldsymbol{k}_0$ contribute in-phase to give constructive interference in the vicinity of the position $\boldsymbol{r} = \boldsymbol{r}_0$, while at positions differing from $\boldsymbol{r}_0$ by more than the real-space width of the wavepacket the different contributions get out of phase to give a negligible sum. You might guess that you could do this using a normalized weighting factor of the form $w(\boldsymbol{k}-\boldsymbol{k}_0) = w_R(\boldsymbol{k}-\boldsymbol{k}_0) e^{-i\boldsymbol{k}\cdot\boldsymbol{r}_0}$ with $w_R(\boldsymbol{k}-\boldsymbol{k}_0)$ a real-valued function sharply-peaked around $\boldsymbol{k} = \boldsymbol{k}_0$ and zero elsewhere. For this weighting function, the superposition would become

$$\begin{aligned} \Psi_{WP,0}(\boldsymbol{r},t=0) \quad ?=? \quad & \int w_R(\boldsymbol{k}-\boldsymbol{k}_0) e^{-i\boldsymbol{k}\cdot\boldsymbol{r}_0} \psi_{\boldsymbol{k}}(\boldsymbol{r}) d^3\boldsymbol{k} \\ = & \int w_R(\boldsymbol{k}-\boldsymbol{k}_0) e^{i\boldsymbol{k}\cdot(\boldsymbol{r}-\boldsymbol{r}_0)} u_{\boldsymbol{k}}(\boldsymbol{r}) d^3\boldsymbol{k}. \end{aligned} \tag{1.5}$$

If the functions $u_{\boldsymbol{k}}(\boldsymbol{r})$ for different $\boldsymbol{k}$ were all real for $\boldsymbol{r}$ near $\boldsymbol{r}_0$ (or shared the same phase), this form would be correct, because near $\boldsymbol{r} = \boldsymbol{r}_0$ the factor $e^{i\boldsymbol{k}\cdot(\boldsymbol{r}-\boldsymbol{r}_0)}$ would guarantee constructive interference for all values of $\boldsymbol{k}$, and for $\boldsymbol{r}$ sufficiently far from $\boldsymbol{r}_0$ the contributions for different values of $\boldsymbol{k}$ would give out-of-phase superpositions so that $\Psi_{WP,0}(\boldsymbol{r} \neq \boldsymbol{r}_0, t=0) \approx 0$. However, if $u_{\boldsymbol{k}}(\boldsymbol{r})$ does not have the same phase for all values of $\boldsymbol{k}$ for $\boldsymbol{r}$ near $\boldsymbol{r}_0$, this first guess is not right.

Let us allow for the possibility that the phase of $u_{\boldsymbol{k}}(\boldsymbol{r})$ depends on $\boldsymbol{k}$ for $\boldsymbol{r}$ near $\boldsymbol{r}_0$. This can happen simply as a result of using different phase conventions for defining the Bloch states, but we will also see in the next lecture that in some cases having such phase variations will be unavoidable. In this case, to make a wavepacket localized near the desired position $\boldsymbol{r}_0$ one must add a correction factor to the weighting function to counteract the changes in phase of the basis functions as a function of $\boldsymbol{k}$ and restore constructive interference in the vicinity of $\boldsymbol{r} = \boldsymbol{r}_0$. As a first step to give intuition, we can ask what is the correction factor that will give perfect constructive interference exactly at $\boldsymbol{r} = \boldsymbol{r}_0$ for all of the Bloch waves in the superposition so that maximum amount of constructive interference will be at this position. (We will see below that the value of $\langle \boldsymbol{r} \rangle$ need not be exactly equal to the position of maximum constructive interference if there are any asymmetries in the wavefunction about this point or if the magnitude of $u_{\boldsymbol{k}}(\boldsymbol{r})$ varies, but typically these differences will be small – less than a unit cell, and so much less than the width of the wavepacket.) Let $\phi(\boldsymbol{k})$ be the phase of $u_{\boldsymbol{k}}(\boldsymbol{r}_0)$ and consider a first-order Taylor expansion of this phase near $\boldsymbol{k}_0$: $\phi(\boldsymbol{k}) \approx \phi(\boldsymbol{k}_0) - \tilde{A}(\boldsymbol{k}_0)\cdot(\boldsymbol{k}-\boldsymbol{k}_0)$. We can then cancel this variation in phase by including a correction factor in the weighting with phase proportional to $+\tilde{\boldsymbol{A}}(\boldsymbol{k}_0)\cdot\delta\boldsymbol{k}$, that is, the weighting factor becomes $w(\boldsymbol{k}-\boldsymbol{k}_0) = w_R(\boldsymbol{k}-\boldsymbol{k}_0)e^{-i\boldsymbol{k}\cdot\boldsymbol{r}_0}e^{+i\left[(\boldsymbol{k}-\boldsymbol{k}_0)\cdot\tilde{A}(\boldsymbol{k}_0)\right]}$, accurate to linear order for values of $\boldsymbol{k}$ near $\boldsymbol{k}_0$ (those that contribute to the superposition). With this identification, our new guess for the superposition that will position the wavefunction at $\boldsymbol{r}_0$ is

$$\Psi_{WP}(\boldsymbol{r},t=0) \;\; ?=? \;\; \int w_R(\boldsymbol{k}-\boldsymbol{k}_0)e^{i\boldsymbol{k}\cdot(\boldsymbol{r}-\boldsymbol{r}_0)}\left[e^{i(\boldsymbol{k}-\boldsymbol{k}_0)\cdot\tilde{A}(\boldsymbol{k}_0)}u_{\boldsymbol{k}}(\boldsymbol{r})\right]d^3\boldsymbol{k}. \qquad (1.6)$$

The quantity in square brackets now has a fixed phase as a function of changing $\boldsymbol{k}$ when evaluated at $\boldsymbol{r}_0$ so that when $\boldsymbol{r} = \boldsymbol{r}_0$ all of the terms in the integrand for different values of $\boldsymbol{k}$ add up constructively. The key point is that the weighting factor must include this type of phase correction factor to generate constructive interference near a desired position $\boldsymbol{r}_0$.

It happens that the weighting function in Eq. (1.6) is very close to being correct, but not quite exactly right, because it is picked to put the point of maximum constructive interference at $\boldsymbol{r} = \boldsymbol{r}_0$, and this is not exactly equivalent to the condition that $\langle \boldsymbol{r} \rangle = \boldsymbol{r}_0$. To determine the phase correction factor that gives $\langle \boldsymbol{r} \rangle = \boldsymbol{r}_0$ we can assume a weighting function with a similar phase correction factor of the general form $w(\boldsymbol{k}-\boldsymbol{k}_0) = w_R(\boldsymbol{k}-\boldsymbol{k}_0)e^{-i\boldsymbol{k}\cdot\boldsymbol{r}_0}e^{+i(\boldsymbol{k}-\boldsymbol{k}_0)\cdot\boldsymbol{X}}$ (where $\boldsymbol{X}$ is an unknown coefficient to be determined) and calculate the expectation value of position directly. I do this in Appendix 2, and the answer is $\langle \boldsymbol{r} \rangle = \boldsymbol{r}_0 + \boldsymbol{A}(\boldsymbol{k}_0) - \boldsymbol{X}$, where

$$A(\boldsymbol{k}_0) = i \int_{\text{unit cell}} d^3\boldsymbol{r}\left(u_{\boldsymbol{k}}^*(\boldsymbol{r})\nabla_{\boldsymbol{k}}u_{\boldsymbol{k}}(\boldsymbol{r})\right)\Big|_{\boldsymbol{k}=\boldsymbol{k}_0} \qquad (1.7)$$

is a real-valued quantity known as the Berry connection. $\boldsymbol{A}(\boldsymbol{k})$ is very similar to our previous correction coefficient $\tilde{A}(\boldsymbol{k})$ except that while $\tilde{A}(\boldsymbol{k})$ corrects the phase at just one position in the real-space unit cell (at $\boldsymbol{r}_0$), $\boldsymbol{A}(\boldsymbol{k})$ corrects for an overall average variation in phase over the full real-space unit cell. (In fact, if one separates $u_{\boldsymbol{k}}(\boldsymbol{r})$ into a phase and real-valued magnitude, $u_{\boldsymbol{k}}(\boldsymbol{r}) = e^{i\phi(\boldsymbol{k},\boldsymbol{r})} u_{\boldsymbol{k}}^R(\boldsymbol{r})$, then one has simply $\boldsymbol{A}(\boldsymbol{k}_0) = -\int_{\text{unit cell}} \nabla_{\boldsymbol{k}} \phi(\boldsymbol{k},\boldsymbol{r})\Big|_{\boldsymbol{k}=\boldsymbol{k}_0} \left(u_{\boldsymbol{k}_0}^R(\boldsymbol{r})\right)^2 d^3\boldsymbol{r}$.) Given that the real-space width of the wavepacket is much larger than a unit cell, it makes sense that differences in phase conventions throughout the unit cell matter in determining $\langle \boldsymbol{r} \rangle$ rather than just the phase convention at one point within the unit cell $\boldsymbol{r}_0$. We can conclude that in order for value of $\langle \boldsymbol{r} \rangle$ for the wavepacket to be equal to $\boldsymbol{r}_0$ the weighting function must include the phase correction coefficient $\boldsymbol{X} = \boldsymbol{A}(\boldsymbol{k}_0)$, *i.e.*, we must use the weighting function

$$w(\boldsymbol{k}-\boldsymbol{k}_0) = w_R(\boldsymbol{k}-\boldsymbol{k}_0) e^{-i\boldsymbol{k}\cdot\boldsymbol{r}_0} e^{+i\left[(\boldsymbol{k}-\boldsymbol{k}_0)\cdot\boldsymbol{A}(\boldsymbol{k}_0)\right]} \tag{1.8}$$

so that the final correct superposition of Bloch waves within the wavepacket at $t = 0$ is

$$\Psi_{WP}(\boldsymbol{r},t=0) = \int w_R(\boldsymbol{k}-\boldsymbol{k}_0) e^{-i\boldsymbol{k}\cdot\boldsymbol{r}_0} e^{i(\boldsymbol{k}-\boldsymbol{k}_0)\cdot\boldsymbol{A}(\boldsymbol{k}_0)} e^{i\boldsymbol{k}\cdot\boldsymbol{r}} u_{\boldsymbol{k}}(\boldsymbol{r}) d^3\boldsymbol{k}. \tag{1.9}$$

We saw in our discussion of Rule (#1) that when an electric or magnetic field is applied, the wavevector $\boldsymbol{k}$ of each Bloch basis state will shift adiabatically $\boldsymbol{k} \to \boldsymbol{k} + \delta\boldsymbol{k}$. To compute the motion of the wavepacket as a function of time, it will be necessary to take into account how this shift will affect the phases of the basis states and hence their quality of constructive interference. The Berry connection enters this part of the analysis, as well. If the Berry connection $\boldsymbol{A}(\boldsymbol{k})$ is nonzero, this means that the overall phase convention used for defining the Bloch states varies as a function of $\boldsymbol{k}$, with shifts in the phase of $u_{\boldsymbol{k}}(\boldsymbol{r})$ of the form $\delta\phi(\boldsymbol{k}) \approx -\boldsymbol{A}(\boldsymbol{k}) \cdot \delta\boldsymbol{k}$. If we could take $\boldsymbol{A}(\boldsymbol{k}) = 0$, then to describe an adiabatic shift in wavevector $\boldsymbol{k} \to \boldsymbol{k} + \delta\boldsymbol{k}$ due to an external field we could simply relabel the Bloch states such that $\left[e^{i\boldsymbol{k}\cdot\boldsymbol{r}} u_{\boldsymbol{k}}(\boldsymbol{r})\right]_{\boldsymbol{k}\to\boldsymbol{k}+\delta\boldsymbol{k}} = e^{i(\boldsymbol{k}+\delta\boldsymbol{k})\cdot\boldsymbol{r}} u_{\boldsymbol{k}+\delta\boldsymbol{k}}(\boldsymbol{r})$. However, this is not the case if the basis states are defined using different phase conventions. Instead, to express $\left[e^{i\boldsymbol{k}\cdot\boldsymbol{r}} u_{\boldsymbol{k}}(\boldsymbol{r})\right]_{\boldsymbol{k}\to\boldsymbol{k}+\delta\boldsymbol{k}}$ in terms of $e^{i(\boldsymbol{k}+\delta\boldsymbol{k})\cdot\boldsymbol{r}} u_{\boldsymbol{k}+\delta\boldsymbol{k}}(\boldsymbol{r})$ we must also multiply by an additional phase correction factor to account for difference in phase between $u_{\boldsymbol{k}+\delta\boldsymbol{k}}(\boldsymbol{r})$ and $u_{\boldsymbol{k}}(\boldsymbol{r})$. If the overall phase for the basis states varies as $\delta\phi(\boldsymbol{k}) \approx -A(\boldsymbol{k}) \cdot \delta\boldsymbol{k}$, then the correct consequence of shifting the wave vector $\boldsymbol{k}$ is that $\left[e^{i\boldsymbol{k}\cdot\boldsymbol{r}} u_{\boldsymbol{k}}(\boldsymbol{r})\right]_{\boldsymbol{k}\to\boldsymbol{k}+\delta\boldsymbol{k}} = e^{+iA(\boldsymbol{k})\cdot\delta\boldsymbol{k}} \left(e^{i(\boldsymbol{k}+\delta\boldsymbol{k})\cdot\boldsymbol{r}} u_{\boldsymbol{k}+\delta\boldsymbol{k}}(\boldsymbol{r})\right)$. (We revisit this relationship next time and derive it systematically as part of a more general analysis of the relationship between adiabatic changes of quantum states and quantum phases.)

At this point we now have all the ingredients needed to understand the equation of motion for the real-space position of this wavepacket, by analyzing the dynamics over a short time interval $\delta t$. The weighting function that defines which basis states contribute to the superposition will stay fixed; the motion arises from how the basis states evolve in time. There are two contributions to these dynamics. The first is just the ordinary time evolution of each Bloch basis state, $u_{\boldsymbol{k}}(\boldsymbol{r},\delta t) = e^{-i\varepsilon(\boldsymbol{k})\delta t/\hbar} u_{\boldsymbol{k}}(\boldsymbol{r})$. This will give a shift in real-space position proportional to $\nabla_{\boldsymbol{k}} \varepsilon(\boldsymbol{k}_0)$, the familiar group velocity. The second contribution to the wavepacket dynamics is that during the time $\delta t$ an external force can shift the wavevector of each of the basis states $\boldsymbol{k} \to \boldsymbol{k} + \delta\boldsymbol{k}$, as we have just discussed. After this shift, if the basis states do not all have the same phase convention, the phase correction factor $e^{+i[(\boldsymbol{k}-\boldsymbol{k}_0)\cdot\boldsymbol{A}(\boldsymbol{k}_0)]}$ in our weighting function (Eq. 1.8) will no longer correctly cancel the phase differences between the shifted basis states in the immediate vicinity of the original position $\boldsymbol{r}_0$. This has the consequence that the real-space position where constructive interference occurs will undergo an additional shift by a term containing derivatives of the phase correction coefficient $\boldsymbol{A}(\boldsymbol{k})$ with respect to $\boldsymbol{k}$, or more precisely by the curl $\nabla\times\boldsymbol{A}(\boldsymbol{k})=\boldsymbol{\Omega}(\boldsymbol{k})$. (See details below.) This is the Berry curvature occurring in our Rule (#2).

These results can be understood quantitatively without too much effort. Starting with Eq. (1.9) and applying both the $\delta t$ and $\delta\boldsymbol{k}$ contributions to the dynamics, the form of wavepacket at time $\delta t$ becomes

$$\begin{aligned}\Psi_{WP}(\boldsymbol{r},\delta t) &= \int w_R(\boldsymbol{k}-\boldsymbol{k}_0)e^{-i\boldsymbol{k}\cdot\boldsymbol{r}_0}e^{i(\boldsymbol{k}-\boldsymbol{k}_0)\cdot\boldsymbol{A}(\boldsymbol{k}_0)}\left[e^{i\boldsymbol{k}\cdot\boldsymbol{r}}u_{\boldsymbol{k}}(\boldsymbol{r})\right]_{\boldsymbol{k}\to\boldsymbol{k}+\delta\boldsymbol{k}}e^{-i\varepsilon(\boldsymbol{k})\delta t/\hbar}d^3\boldsymbol{k} \\ &= \int w_R(\boldsymbol{k}-\boldsymbol{k}_0)e^{-i\boldsymbol{k}\cdot\boldsymbol{r}_0}e^{i(\boldsymbol{k}-\boldsymbol{k}_0)\cdot\boldsymbol{A}(\boldsymbol{k}_0)}e^{+i\boldsymbol{A}(\boldsymbol{k})\cdot\delta\boldsymbol{k}}\left(e^{i(\boldsymbol{k}+\delta\boldsymbol{k})\cdot\boldsymbol{r}}u_{\boldsymbol{k}+\delta\boldsymbol{k}}(\boldsymbol{r})\right)e^{-i\varepsilon(\boldsymbol{k})\delta t/\hbar}d^3\boldsymbol{k}.\end{aligned} \tag{1.10}$$

This expression contains all of the physics. It can be made more transparent by using some judicious Taylor expansions.

To simplify, let's first convert the integration variable to $\boldsymbol{k}'=\boldsymbol{k}+\delta\boldsymbol{k}$ and introduce the definition $\boldsymbol{k}_1=\boldsymbol{k}_0+\delta\boldsymbol{k}$ for the new peak position of the wavepacket in reciprocal space.

$$\begin{aligned}\Psi_{WP}(\boldsymbol{r},\delta t) &= \int w_R(\boldsymbol{k}'-\boldsymbol{k}_1)e^{-i(\boldsymbol{k}'-\delta\boldsymbol{k})\cdot\boldsymbol{r}_0}e^{i(\boldsymbol{k}'-\boldsymbol{k}_1)\cdot\boldsymbol{A}(\boldsymbol{k}_0)}e^{+i\boldsymbol{A}(\boldsymbol{k}'-\delta\boldsymbol{k})\cdot\delta\boldsymbol{k}}\left(e^{i\boldsymbol{k}'\cdot\boldsymbol{r}}u_{\boldsymbol{k}'}(\boldsymbol{r})\right)e^{-i\varepsilon(\boldsymbol{k}'-\delta\boldsymbol{k})\delta t/\hbar}d^3\boldsymbol{k}' \\ &= e^{i\delta\boldsymbol{k}\cdot\boldsymbol{r}_0}\int w_R(\boldsymbol{k}'-\boldsymbol{k}_1)e^{i\boldsymbol{k}'\cdot(\boldsymbol{r}-\boldsymbol{r}_0)}e^{i(\boldsymbol{k}'-\boldsymbol{k}_1)\cdot\boldsymbol{A}(\boldsymbol{k}_0)}e^{+i\boldsymbol{A}(\boldsymbol{k}'-\delta\boldsymbol{k})\cdot\delta\boldsymbol{k}}u_{\boldsymbol{k}'}(\boldsymbol{r})e^{-i\varepsilon(\boldsymbol{k}'-\delta\boldsymbol{k})\delta t/\hbar}d^3\boldsymbol{k}'\end{aligned} \tag{1.11}$$

Next, perform Taylor expansions relative to $\boldsymbol{k}_1$ (the new center of the wavepacket in $\boldsymbol{k}$-space) for the arguments of all the $\boldsymbol{k}'$-dependent exponential factors in the integrand.

$$\begin{aligned}&\Psi_{WP}(\boldsymbol{r},\delta t) \\ &= e^{i\delta\boldsymbol{k}\cdot\boldsymbol{r}_0}\int w_R(\boldsymbol{k}'-\boldsymbol{k}_1)e^{i\boldsymbol{k}'\cdot(\boldsymbol{r}-\boldsymbol{r}_0)}e^{i(\boldsymbol{k}'-\boldsymbol{k}_1)\cdot\boldsymbol{A}(\boldsymbol{k}_0)}e^{+i\boldsymbol{A}(\boldsymbol{k}'-\delta\boldsymbol{k})\cdot\delta\boldsymbol{k}}e^{-i\varepsilon(\boldsymbol{k}'-\delta\boldsymbol{k})\delta t/\hbar}u_{\boldsymbol{k}'}(\boldsymbol{r})d^3\boldsymbol{k}' \\ &= e^{i\Phi_c}\int w_R(\boldsymbol{k}'-\boldsymbol{k}_1)e^{i(\boldsymbol{k}'-\boldsymbol{k}_1)\cdot(\boldsymbol{r}-\boldsymbol{r}_0+\boldsymbol{Y})}u_{\boldsymbol{k}'}(\boldsymbol{r})d^3\boldsymbol{k}' \\ &= e^{i\Phi_c}e^{-i\boldsymbol{k}_1\cdot(\boldsymbol{r}-\boldsymbol{r}_0)}\int w_R(\boldsymbol{k}'-\boldsymbol{k}_1)e^{i\boldsymbol{k}'\cdot(\boldsymbol{r}-\boldsymbol{r}_0)}e^{i(\boldsymbol{k}'-\boldsymbol{k}_1)\cdot\boldsymbol{Y}}u_{\boldsymbol{k}'}(\boldsymbol{r})d^3\boldsymbol{k}'\end{aligned} \tag{1.12}$$

Here the constant $\boldsymbol{Y}$ functions as a new phase correction coefficient that contains the coefficients of all the linear terms in the Taylor expansions

$$\begin{aligned}\boldsymbol{Y} &= \boldsymbol{A}(\boldsymbol{k}_0)+\sum_\mu\nabla_{\boldsymbol{k}}[A(\boldsymbol{k}_1)]_\mu(\delta k)_\mu-\nabla_{\boldsymbol{k}}\varepsilon(\boldsymbol{k}_1)\delta t/\hbar \\ &= \left\{\boldsymbol{A}(\boldsymbol{k}_1)-\sum_\mu[\partial_{k_\mu}\boldsymbol{A}(\boldsymbol{k}_1)](\delta k)_\mu\right\}+\sum_\mu\nabla_{\boldsymbol{k}}[A(\boldsymbol{k}_1)]_\mu(\delta k)_\mu-\nabla_{\boldsymbol{k}}\varepsilon(\boldsymbol{k}_1)\delta t/\hbar \\ &= \boldsymbol{A}(\boldsymbol{k}_1)-\sum_\mu\Big[[\partial_{k_\mu}\boldsymbol{A}(\boldsymbol{k}_1)]-\nabla_{\boldsymbol{k}}[A(\boldsymbol{k}_1)]_\mu\Big](\delta k)_\mu-\nabla_{\boldsymbol{k}}\varepsilon(\boldsymbol{k}_1)\delta t/\hbar \\ &= \boldsymbol{A}(\boldsymbol{k}_1)+\delta\boldsymbol{k}\times\left[\nabla\times\boldsymbol{A}(\boldsymbol{k}_1)\right]-\nabla_{\boldsymbol{k}}\varepsilon(\boldsymbol{k}_1)\delta t/\hbar.\end{aligned} \tag{1.13}$$

Note how the Berry curvature $\nabla \times \boldsymbol{A}(\boldsymbol{k}) = \Omega(\boldsymbol{k})$ appears as part of this new phase correction coefficient $\boldsymbol{Y}$. The constant terms in the Taylor expansions result in an overall phase factor $e^{i\Phi_c} e^{-i\boldsymbol{k}_1 \cdot (\boldsymbol{r}-\boldsymbol{r}_0)}$ in Eq. (1.12) that has no effect on $\langle \boldsymbol{r} \rangle$ (and can be ignored).

Equation (1.12) now has now been reduced to exactly the same mathematical form we consider in Appendix 2, with the substitutions $\boldsymbol{k}_0 \to \boldsymbol{k}_1$ and $\boldsymbol{X} \to \boldsymbol{Y}$. The new expectation value of position is therefore by the calculation in Appendix 2 equal to $\langle \boldsymbol{r} \rangle = \boldsymbol{r}_0 - \boldsymbol{Y} + \boldsymbol{A}(\boldsymbol{k}_1)$, or

$$\langle \boldsymbol{r} \rangle = \boldsymbol{r}_0 - \delta \boldsymbol{k} \times \left[ \nabla_{\boldsymbol{k}} \times \boldsymbol{A}(\boldsymbol{k}_1) \right] + \nabla_{\boldsymbol{k}} \varepsilon(\boldsymbol{k}_1) \delta t / \hbar . \tag{1.14}$$

Dividing by $\delta t$, we have

$$\frac{d\langle \boldsymbol{r} \rangle}{dt} = -\frac{d\boldsymbol{k}}{dt} \times \left[ \nabla_{\boldsymbol{k}} \times \boldsymbol{A}(\boldsymbol{k}_1) \right] + \frac{1}{\hbar} \nabla_{\boldsymbol{k}} \varepsilon(\boldsymbol{k}_1). \tag{1.15}$$

With the identification that the Berry curvature is $\boldsymbol{\Omega}(\boldsymbol{k}) = \nabla_{\boldsymbol{k}} \times \boldsymbol{A}(\boldsymbol{k})$, this is our Rule (#2). (Note also that the center of the wavepacket in this expression is $\boldsymbol{k}_1$ is rather than $\boldsymbol{k}_0$, and $d\boldsymbol{k}/dt$ is the same for every basis state and is therefore also equal to $d\langle \boldsymbol{k} \rangle / dt$.)

Given that the physics at work here is essentially just an analysis of constructive interference within a wavepacket, why was the Berry curvature contribution to Rule (#2) neglected by most researchers (and textbook writers) for many decades? My guess is that there was an assumption that, given the arbitrariness of defining the phase of quantum-mechanical states, that it should always be possible to choose a phase convention such that for $\boldsymbol{r}$ in the vicinity of $\boldsymbol{r}_0$ $u_{\boldsymbol{k}}(\boldsymbol{r})$ has the same phase for all values of $\boldsymbol{k}$ contributing to the wavepacket, or equivalently that $\boldsymbol{A}(\boldsymbol{k})$ in our discussion above could be taken to be zero. This turns out not to be true – there are electronic band structures for which it is impossible to define Bloch-solution basis states so that they all have the same phase. In the next lecture we will consider in detail why this is, and what conditions are necessary in order for the Berry curvature to be nonzero.

***Anomalous velocity***:

Before we do that, however, we will say a few words about how the Berry curvature affects electron dynamics. Suppose a constant electric field $\boldsymbol{E}$ is applied to a sample, so that by our Rule (#1) $d\langle \boldsymbol{k} \rangle / dt = -e\boldsymbol{E} / \hbar$. Then, if the Berry curvature is non-zero,

$$\frac{d\langle \boldsymbol{r} \rangle}{dt} = \frac{1}{\hbar}\nabla_k \varepsilon_\sigma(\boldsymbol{k}_0) + \frac{e}{\hbar}\boldsymbol{E} \times \boldsymbol{\Omega}_\sigma(\boldsymbol{k}_0). \tag{1.16}$$

The conventional group velocity term (the first term on the right hand side) by itself can lead to fascinating and non-intuitive behaviors, but as $\langle \boldsymbol{k} \rangle$ shifts with time this term will generally produce a real-space acceleration, typically with a dominant component parallel to the applied electric field. The Berry curvature term, however, by virtue of the cross product always produces a real-space velocity that is *perpendicular* to the applied electric field. Because of this odd behavior, this contribution is termed the "anomalous velocity". The Berry curvature in some senses produces effects in real space that are like a magnetic field – it will deflect an electron trajectory to give a component of velocity perpendicular to an applied electric field. This reflects a deep mathematical analogy that is carried through in some of the notation; the Berry connection is generally given the symbol $\boldsymbol{A}$ in analogy to the vector potential of E&M. However, this deflection due to the Berry curvature can happen even in the absence of any applied magnetic field (or even in the absence of any breaking of time-reversal symmetry). Furthermore, as noted explicitly in our Rules for semiclassical dynamics, the Berry curvature (which contributes to the motion of $d\langle \boldsymbol{r} \rangle / dt$ via Rule (#2)) enters the mathematical description of the dynamics differently than a real magnetic field (which contributes to $d\langle \boldsymbol{k} \rangle / dt$ via Rule (#1)).

***Berry curvature and other kinds of waves***:

Finally, note that there is nothing particularly special about electron waves in our derivation of the Berry curvature and Rule (#2). The same equation of motion should apply to a wavepacket made from any type of wave. Why, then, did we not mention anything about Berry curvature and anomalous velocity when we discussed phonon wavepackets and thermal conductivity? There are two reasons: in order for the anomalous velocity $-d\boldsymbol{k}/dt \times \boldsymbol{\Omega}(\boldsymbol{k})$ to be nonzero, both $d\boldsymbol{k}/dt$ and $\boldsymbol{\Omega}(\boldsymbol{k})$ must be nonzero, and both conditions will be rare occurrences for phonon wavepackets. For electrons, easy-to-apply external fields (***E***, ***B***) can couple to the wavevector (via our Rule (#1)) to give nonzero values of $d\boldsymbol{k}/dt$. For a phonon wavepacket, though, the wavevector is usually just constant because it doesn't couple in the same way to any convenient external fields. In principle, spatially-varying or time-dependent strains might give a nonzero value of $d\boldsymbol{k}/dt$ for a phonon wavepacket, but these would be unusual cases. In regard to the conditions to have a non-zero value of Berry curvature, based on the arguments in the next lecture it follows that for spinless waves like phonons a non-zero Berry curvature can exist only when certain symmetries are broken within a crystal structure. For electrons, non-zero Berry curvatures can be generated either by these broken symmetries or by spin-orbit coupling even in high-symmetry lattices.

When $\mathbf{\Omega}(\boldsymbol{k})$ is non-zero for any type of wave (including, *e.g.*, phonons, magnons, or light) there is a possibility for non-trivial effects even if $d\boldsymbol{k}/dt = 0$ so that the Berry curvature does not directly influence the equation of motion. If the integral of $\mathbf{\Omega}(\boldsymbol{k})$ over the Brillouin zone is non-zero, the corresponding band will have a topological character with important physical consequences like topologically-required low-energy edge states. I will have more to say about this in lecture III.

## II: More about Berry curvature: When is it non-zero and what does it do?

*Learning Goals:*

- *Understand the idea of Berry phase for general quantum-mechanical states*
- *Applications of these principles to electron bands*
- *Conditions necessary for non-zero Berry curvatures*
- *Some physical consequences of Berry curvatures*

We have seen that if one makes a wavepacket from Bloch waves $e^{i\boldsymbol{k}\cdot\boldsymbol{r}}u_{\boldsymbol{k},\sigma}(\boldsymbol{r})$ defined so that as a function of $\boldsymbol{k}$ the contributions $u_{\boldsymbol{k},\sigma}(\boldsymbol{r})$ have different phase conventions (*i.e.*, a nonzero Berry connection $\boldsymbol{A}(\boldsymbol{k})$) then there can be some surprising physical consequences for the motion of the wavepacket in real space. For example, the wavepacket can experience an "anomalous velocity" transverse to an applied electric field of the form $\propto \boldsymbol{E}\times\boldsymbol{\Omega}(\boldsymbol{k}_0)$, where $\boldsymbol{\Omega}(\boldsymbol{k}_0)=\nabla_{\boldsymbol{k}}\times\boldsymbol{A}(\boldsymbol{k})\big|_{\boldsymbol{k}=\boldsymbol{k}_0}$.

This likely seems strange to you. The phase of a quantum-mechanical wavefunction can be gauge-transformed arbitrarily using any continuous real-valued function $\chi(\boldsymbol{k})$ by redefining $u_{\boldsymbol{k},\sigma}(\boldsymbol{r})\rightarrow e^{-i\chi(\boldsymbol{k})}u_{\boldsymbol{k},\sigma}(\boldsymbol{r})$, without causing any physical consequence. Why isn't it always possible to pick some function $\chi(\boldsymbol{k})$ so that the combination $e^{-i\chi(\boldsymbol{k})}u_{\boldsymbol{k},\sigma}(\boldsymbol{r})$ has the same phase near any desired real-space position for all values of $\boldsymbol{k}$ and so therefore $\boldsymbol{A}(\boldsymbol{k})=0$ everywhere and there is no need to worry about Berry curvatures? A short answer is that while it is always possible to choose a phase convention so that the wavefunction phase is the same for all states along a 1D path in $\boldsymbol{k}$-space (with no closed loops), it is not always possible to make the wavefunction phase the same throughout a full 2D or 3D region of $\boldsymbol{k}$-space. Since one must generally integrate over a 2D or 3D region to make a wavepacket, this means that a nonzero Berry curvature can be unavoidable. An even shorter answer is that even though the phases of quantum states can depend on the choice of gauge, the Berry curvature itself is a gauge-independent quantity, so that no matter what definition one picks for $\chi(\boldsymbol{k})$ this cannot change the value of the Berry curvature. In the first part of this lecture we will work to understand this state of affairs.

***The connection between Berry phase and adiabatic evolution of quantum states***:

To do this, let us consider a slightly more general quantum-mechanics problem. Imagine a quantum system where the Hamiltonian is a function of a continuous external parameter $\boldsymbol{\lambda}$ (*e.g.*, think magnetic field -- $\boldsymbol{\lambda}$ may be a many-component vector). The energy eigenstates will be functions of $\boldsymbol{\lambda}$ (Fig. 2.1), and they might have a phase convention that changes for different values of $\boldsymbol{\lambda}$. Suppose that the external parameter has the initial value $\boldsymbol{\lambda}_1$ and the system starts in the state $\left|\varepsilon_0(\boldsymbol{\lambda}_1)\right\rangle$. Then $\boldsymbol{\lambda}$ is made to change slowly so that the basis state $\left|\varepsilon_0(\boldsymbol{\lambda})\right\rangle$ evolves. We will assume that the different energy levels never touch and the quantum state evolves adiabatically. The question we want to answer is: How does the quantum state change when accounting for changes in the phase convention for the basis state? We will analyze whether or not it is always possible to choose a gauge so that the phase of the wavefunction stays the same for any possible trajectory of $\boldsymbol{\lambda}$.

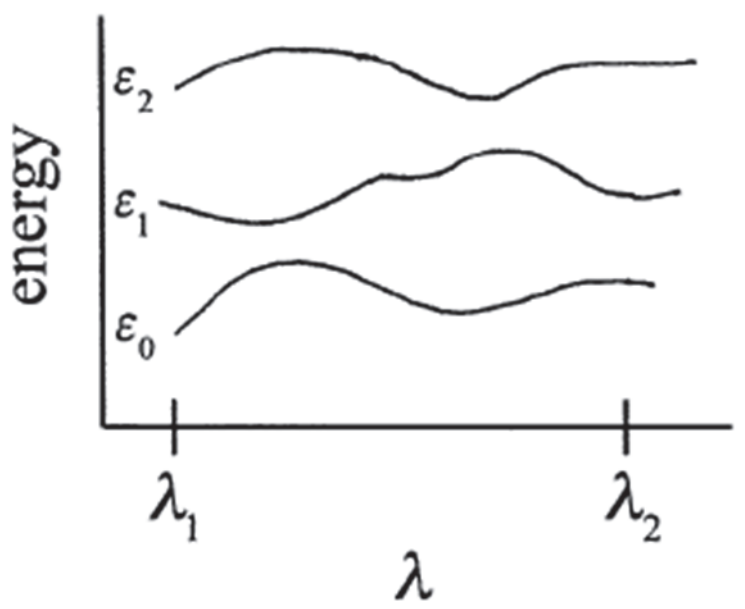

*Fig. 2.1. Adiabatic evolution of the states $|\varepsilon_i(\boldsymbol{\lambda})\rangle$ as a function of the external parameter $\boldsymbol{\lambda}$.*

This calculation can be performed explicitly as a function of time by simply using the Schrodinger equation:

$$i\hbar\frac{\partial}{\partial t}\left|\psi(t)\right\rangle = H\left(\boldsymbol{\lambda}(t)\right)\left|\psi(t)\right\rangle. \tag{2.1}$$

The wavefunction at any time can always be expanded in the energy eigenstates corresponding to the value of the parameter $\boldsymbol{\lambda}$ at that time:

$$\left|\psi(t)\right\rangle = \sum_n c_n(t)\left|\varepsilon_n(\boldsymbol{\lambda}(t))\right\rangle. \tag{2.2}$$

If we assume adiabatic evolution of a state that begins in the ground state, at later times the only state with non-zero amplitude in this expansion will be the ground state (see the nice discussion of adiabatic evolution in Griffiths), so in the adiabatic limit we can simplify to just a single term

$$\left|\psi(t)\right\rangle = c_0(t)\left|\varepsilon_0(\boldsymbol{\lambda}(t))\right\rangle. \tag{2.3}$$

Assuming the basis states are normalized, $c_0(t)$ will have magnitude 1, and only its phase will change. Let us apply the Schrodinger equation to this form of the wavefunction

$$\begin{aligned} &i\hbar\frac{\partial}{\partial t}\left[c_0(t)\left|\varepsilon_0(\boldsymbol{\lambda}(t))\right\rangle\right] = H\left(\boldsymbol{\lambda}(t)\right)\left[c_0(t)\left|\varepsilon_0(\boldsymbol{\lambda}(t))\right\rangle\right] \\ \Rightarrow \quad &i\hbar\left|\varepsilon_0(\boldsymbol{\lambda}(t))\right\rangle\frac{\partial}{\partial t}\left[c_0(t)\right] + i\hbar c_0(t)\frac{\partial}{\partial t}\left[\left|\varepsilon_0(\boldsymbol{\lambda}(t))\right\rangle\right] = \varepsilon_0(\boldsymbol{\lambda}(t))c_0(t)\left|\varepsilon_0(\boldsymbol{\lambda}(t))\right\rangle. \end{aligned} \tag{2.4}$$

Now project this equation onto the state $\left\langle\varepsilon_0(\boldsymbol{\lambda}(t))\right|$

$$i\hbar\frac{\partial}{\partial t}\left[c_0(t)\right] + i\hbar c_0(t)\left\langle\varepsilon_0(\boldsymbol{\lambda}(t))\right|\frac{\partial}{\partial t}\left|\varepsilon_0(\boldsymbol{\lambda}(t))\right\rangle = \varepsilon_0(\boldsymbol{\lambda}(t))c_0(t). \tag{2.5}$$

Let us define $A_t(t) \equiv i\left\langle\varepsilon_0(\boldsymbol{\lambda}(t))\right|\frac{\partial}{\partial t}\left|\varepsilon_0(\boldsymbol{\lambda}(t))\right\rangle$. This quantity is always real-valued, and accounts for any changes in phase convention for the basis state as a function of changes in time.
Equation (2.5) then becomes

$$\frac{\partial}{\partial t}\left[c_0(t)\right] = c_0(t)\left[\frac{-i\varepsilon_0(\boldsymbol{\lambda}(t))}{\hbar} + iA_t(t)\right] \tag{2.6}$$

with the solution

$$c_0(t) = c_0(0)\exp\left[\frac{-i}{\hbar}\int_0^t \varepsilon_0(\boldsymbol{\lambda}(t'))dt' + i\int_0^t A_t(t')\,dt'\right]. \tag{2.7}$$

The first term in the exponent on the right is the usual dynamical phase related to the energy of the quantum state. The second term, $\gamma(t) \equiv \int_0^t A_t(t')\,dt'$ is called the geometrical phase or the Berry phase, and all that it does is take into account any changes in the phase conventions of the eigenstates. Expressing the same idea using a notation similar to the one we used last time, we have locally that $\left[\left|\varepsilon_0(\boldsymbol{\lambda}(t))\right\rangle\right]_{t\to t+\delta t} = e^{iA_t(t)\delta t}\left|\varepsilon_0(\boldsymbol{\lambda}(t+\delta t))\right\rangle$ to account for the changing phase convention. The Berry phase is not gauge invariant, because it depends on any changes in in the phase conventions used to define the basis states.

Notice that because $\left|\varepsilon_0(\boldsymbol{\lambda}(t))\right\rangle$ depend on $t$ only through $\boldsymbol{\lambda}(t)$ one can also write

$$\begin{aligned}\gamma(t) &= \int_0^t A_t(t')\,dt' = i\int_0^t \left\langle \varepsilon_0(\boldsymbol{\lambda}(t'))\right|\frac{\partial}{\partial t'}\left|\varepsilon_0(\boldsymbol{\lambda}(t'))\right\rangle dt' \\ &= i\int_0^t \left\langle \varepsilon_0(\boldsymbol{\lambda})\right|\nabla_\lambda\left|\varepsilon_0(\boldsymbol{\lambda})\right\rangle\cdot\frac{d\boldsymbol{\lambda}}{dt'}\,dt' \\ &= i\int_{\text{path for }\boldsymbol{\lambda}} \left\langle \varepsilon_0(\boldsymbol{\lambda})\right|\nabla_\lambda\left|\varepsilon_0(\boldsymbol{\lambda})\right\rangle\cdot d\boldsymbol{\lambda}.\end{aligned} \tag{2.8}$$

That is, the geometric phase can be expressed as a line integral that depends only on the trajectory taken by the system in $\boldsymbol{\lambda}$-space, and not the speed of travel (as long as adiabatic evolution is maintained). The matrix element $A(\boldsymbol{\lambda}) \equiv i\left\langle \varepsilon_0(\boldsymbol{\lambda})\right|\nabla_\lambda\left|\varepsilon_0(\boldsymbol{\lambda})\right\rangle$is a generalization of the Berry connection defined in the previous lecture, and describes the dependence of the phase on the wavefunction as a function of changes in $\boldsymbol{\lambda}$.

***Redefining quantum phases using a gauge transformation***:

Now consider a gauge transformation that redefines the phase convention for the basis states in an arbitrary continuous manner: $\left|\varepsilon_0(\boldsymbol{\lambda})\right\rangle \to e^{-i\chi(\boldsymbol{\lambda})}\left|\varepsilon_0(\boldsymbol{\lambda})\right\rangle$. The new geometric phase accumulated over the path is

$$\begin{aligned}\gamma(t) &= i\int_{\text{path from }\boldsymbol{\lambda}_1\text{ to }\boldsymbol{\lambda}_2} \left\langle \varepsilon_0(\boldsymbol{\lambda})\right| e^{i\chi(\boldsymbol{\lambda})}\nabla_\lambda\left[e^{-i\chi(\boldsymbol{\lambda})}\left|\varepsilon_0(\boldsymbol{\lambda})\right\rangle\right]\cdot d\boldsymbol{\lambda} \\ &= \int_{\text{path from }\boldsymbol{\lambda}_1\text{ to }\boldsymbol{\lambda}_2} \nabla_\lambda\chi(\boldsymbol{\lambda})\cdot d\boldsymbol{\lambda} + \int_{\text{path from }\boldsymbol{\lambda}_1\text{ to }\boldsymbol{\lambda}_2} i\left\langle \varepsilon_0(\boldsymbol{\lambda})\right|\nabla_\lambda\left|\varepsilon_0(\boldsymbol{\lambda})\right\rangle\cdot d\boldsymbol{\lambda} \\ &= \chi(\boldsymbol{\lambda}_2) - \chi(\boldsymbol{\lambda}_1) + \int_{\text{path from }\boldsymbol{\lambda}_1\text{ to }\boldsymbol{\lambda}_2} i\left\langle \varepsilon_0(\boldsymbol{\lambda})\right|\nabla_\lambda\left|\varepsilon_0(\boldsymbol{\lambda})\right\rangle\cdot d\boldsymbol{\lambda}\end{aligned} \tag{2.9}$$

It follows that if we consider a trajectory through $\boldsymbol{\lambda}$-space between two different endpoints with no closed loops along the path, it is possible to define a function $\chi(\boldsymbol{\lambda})$ continuously along the path so that the change in geometrical phase is zero everywhere along the trajectory and the wavefunctions all have the same phase. [Just define $\chi(\boldsymbol{\lambda}) = \chi(\boldsymbol{\lambda}_1) - \int_{\text{path from }\boldsymbol{\lambda}_1\text{ to }\boldsymbol{\lambda}} i\left\langle \varepsilon_0(\boldsymbol{\lambda})\right|\nabla_\lambda\left|\varepsilon_0(\boldsymbol{\lambda})\right\rangle\cdot d\boldsymbol{\lambda}$ .] However, consider what happens if one follows a closed loop in parameter space so that $\boldsymbol{\lambda}_2 = \boldsymbol{\lambda}_1$. In that case, in order to satisfy the requirement that the wavefunction must be single valued, the phase factor $\chi(\boldsymbol{\lambda}_2)$ must return to its original value modulo $2\pi$. This means that the geometric phase evaluated for a path integral over a closed loop is gauge-invariant modulo $2\pi$. However, there are quantum systems for which the geometric phase evaluated over a close loop $i\oint\left\langle \varepsilon_0(\boldsymbol{\lambda})\right|\nabla_\lambda\left|\varepsilon_0(\boldsymbol{\lambda})\right\rangle\cdot d\boldsymbol{\lambda}$ is not equal to 0

modulo $2\pi$. (I will give a simple example in a few minutes.) When this is the case there is no continuous, single-valued function $\chi(\boldsymbol{\lambda})$ that can generate a gauge transformation that will make the phase accumulated over the loop equal zero, and it follows that there is no way to pick a phase convention such that all of the quantum states at different values of $\boldsymbol{\lambda}$ have the same phase. In other words: while a gauge transformation can be used to make phase of a quantum state constant over any individual 1-D trajectory with no closed loops, this is not always possible over closed loops. Since any 2D or 3D region can contain closed loops, there are quantum systems for which it is not possible to define a phase convention so that all of the quantum states have the same phase over a 2D or 3D region.

The overall message is that there are quantum systems that as a function of external parameters have a form of nontrivial built-in curvature such that the phase must rotate when the states are shifted along arbitrary trajectories of an external parameter. The mathematics is closely analogous to the rotation of classical vectors when they undergo parallel transport along curved 2D surfaces embedded within 3D space (*e.g.*, see Ref. 3).

***Berry curvature is gauge invariant***:

The mathematics at work can be understood even more simply by using Stokes Theorem. The geometric phase accumulated over a closed loop is

$$\gamma(t) = \oint_{\text{closed loop}} i\langle \varepsilon_0(\boldsymbol{\lambda}) | \nabla_\lambda | \varepsilon_0(\boldsymbol{\lambda}) \rangle \cdot d\boldsymbol{\lambda} = \int_{\text{area enclosed}} \nabla_\lambda \times i\langle \varepsilon_0(\boldsymbol{\lambda}) | \nabla_\lambda | \varepsilon_0(\boldsymbol{\lambda}) \rangle \cdot d\boldsymbol{S}_\lambda . \qquad (2.10)$$

The integrand in the second integral is a generalized Berry curvature $\nabla_\lambda \times \boldsymbol{A}(\boldsymbol{\lambda}) \equiv \boldsymbol{\Omega}(\boldsymbol{\lambda})$. This quantity is gauge invariant. To see this, suppose we redefine our phase conventions with a gauge transformation $|\varepsilon_0(\boldsymbol{\lambda})\rangle \to e^{-i\chi(\boldsymbol{\lambda})} |\varepsilon_0(\boldsymbol{\lambda})\rangle$. Then the value of the Berry curvature after the transformation is

$$\begin{aligned} \nabla_\lambda \times i\left[\left[\langle \varepsilon_0(\boldsymbol{\lambda}) | e^{i\chi(\boldsymbol{\lambda})}\right] \nabla_\lambda \left[e^{-i\chi(\boldsymbol{\lambda})} | \varepsilon_0(\boldsymbol{\lambda}) \rangle\right]\right] &= \nabla_\lambda \times \left[\nabla_\lambda \chi(\boldsymbol{\lambda}) + i\langle \varepsilon_0(\boldsymbol{\lambda}) | \nabla_\lambda | \varepsilon_0(\boldsymbol{\lambda}) \rangle\right] \\ &= \nabla_\lambda \times i\langle \varepsilon_0(\boldsymbol{\lambda}) | \nabla_\lambda | \varepsilon_0(\boldsymbol{\lambda}) \rangle, \end{aligned} \qquad (2.11)$$

independent of $\chi(\boldsymbol{\lambda})$ because the curl of a gradient of a differentiable function is identically zero. Therefore, if a quantum system has a nonzero Berry curvature, its value cannot be changed at all (much less shifted to zero) by a gauge transformation. This is appropriate and necessary, because Berry curvatures can contribute directly to measurable quantities (see our Rule (#2) for semiclassical band dynamics in the previous lecture), so they must not depend on the choice of gauge.

***A simple example -- a quantum system with non-zero Berry curvature***:

All of this probably seems very formal, but there is a simple example with non-zero Berry curvature that you can keep in mind. It will also be a very useful result for later calculations. A spin ½ particle in an external magnetic field $\boldsymbol{B}$ can be described by the general $2 \times 2$ matrix Hamiltonian

$$H = -\mu_B \boldsymbol{B} \cdot \boldsymbol{\sigma} \qquad (2.12)$$

where $\mu_B$ is the Bohr magneton and $\boldsymbol{\sigma}$ is the vector of Pauli matrices. Here the matrix is expressed in terms of the $s_z = \pm 1/2$ basis states, and the vector $\boldsymbol{B}$ corresponds to the external parameter $\boldsymbol{\lambda}$ in our discussion of the geometric phase. (This is in fact a completely general $2\times 2$ Hamiltonian, so it will also apply to many other systems if the parameter $\boldsymbol{B}$ is reinterpreted appropriately.) As the direction of $\boldsymbol{B}$ is rotated, the eigenstates of the Hamiltonian evolve, corresponding to states with spin components $\pm 1/2$ along the axis defined by the direction of $\boldsymbol{B}$, with energies $E = \pm \mu_B |\boldsymbol{B}|$. In problem #2 that is attached to these lecture notes, you will show that the Berry curvature, defined as

$$\nabla_{\boldsymbol{B}} \times i\langle \varepsilon(\boldsymbol{B}) | \nabla_{\boldsymbol{B}} | \varepsilon(\boldsymbol{B}) \rangle \text{ is } \pm \frac{1}{2}\frac{\hat{\boldsymbol{B}}}{B^2}, \tag{2.13}$$

which is manifestly non-zero. (The two signs correspond to the ground and excited spin states.) This result has a beautiful geometric interpretation, in that it corresponds to the field from a monopole of charge ±1/2 located at the origin. This makes it easy (using Stokes theorem) to calculate the geometric phase over any closed path for the evolution of the vector $\boldsymbol{B}$ simply by calculating the solid angle enclosed by the projection of the path onto the sphere -- the geometric phase equal to $\pm 1/2$ times the solid angle subtended by the path of $\boldsymbol{B}$ looking from the origin (see Fig. 2.2).

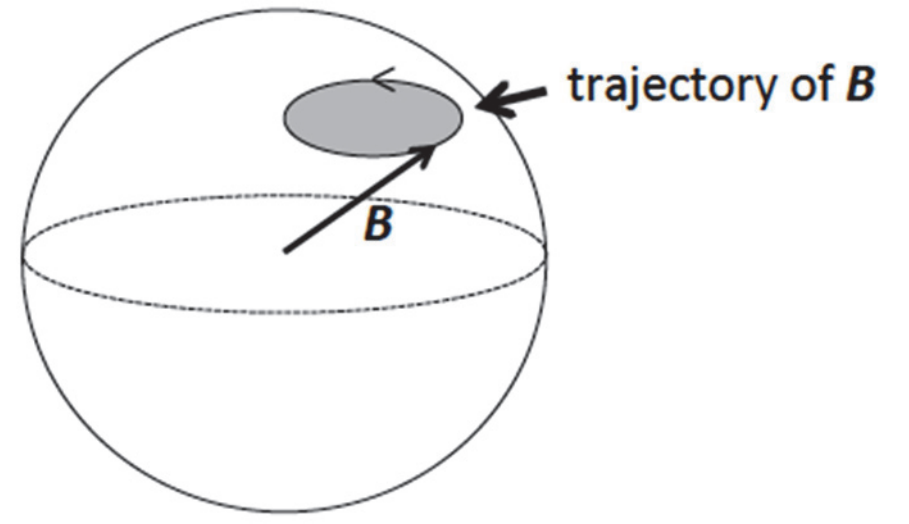

*Fig. 2.2. Relating the geometric phase for a closed path of **B** to the solid angle enclosed by the path.*

***Applications to electronic bandstructures***:

How does all of this relate to electronic bandstructures? The idea is that as one applies electric and magnetic fields, one tunes the wavevector $\boldsymbol{k}$, and this corresponds exactly to the type of adiabatic evolution problem we are considering with the vector $\boldsymbol{k}$ equivalent to our external parameter $\boldsymbol{\lambda}$. This can be phrased in terms of seeking solutions for the periodic-in-$\boldsymbol{k}$-space part $u^n_{\boldsymbol{k},\sigma}(\boldsymbol{r})$ of the Bloch wavefunction $\psi^n_{\boldsymbol{k},\sigma}(\boldsymbol{r}) = e^{i\boldsymbol{k}\cdot\boldsymbol{r}} u^n_{\boldsymbol{k},\sigma}(\boldsymbol{r})$. Assuming the Bloch form for the wavefunction, the Schrodinger equation is completely equivalent to

$$\left[\frac{\hbar^2}{2m}\left(\boldsymbol{k} + \frac{1}{i}\nabla_r\right)^2 + V(\boldsymbol{r})\right] u^n_{\boldsymbol{k},\sigma}(\boldsymbol{r}) = \varepsilon^n_{k,\sigma} u^n_{\boldsymbol{k},\sigma}(\boldsymbol{r}). \tag{2.14}$$

The expression in the square brackets on the left side can be viewed as an effective Hamiltonian $\hat{H}_{\boldsymbol{k}} \equiv e^{-i\boldsymbol{k}\cdot\boldsymbol{r}} \hat{H} e^{i\boldsymbol{k}\cdot\boldsymbol{r}}$ for the wavefunction $u^n_{\boldsymbol{k},\sigma}(\boldsymbol{r})$ in which the wavevector $\boldsymbol{k}$ is explicitly a controllable parameter. All of our previous results regarding geometric phases and Berry curvatures can be applied immediately. Consequently, when an electron follows a trajectory within $\boldsymbol{k}$-space there is a corresponding Berry connection (this is same quantity used for the phase correction coefficient in the previous lecture)

$$\boldsymbol{A}^n_\sigma(\boldsymbol{k}) = i\langle u^n_{\boldsymbol{k},\sigma} | \nabla_k | u^n_{\boldsymbol{k},\sigma} \rangle \tag{2.15}$$

so that

$$\left[u_{\boldsymbol{k}}(\boldsymbol{r})\right]_{\boldsymbol{k}\to\boldsymbol{k}+\delta\boldsymbol{k}} = e^{+i\boldsymbol{A}(\boldsymbol{k})\cdot\delta\boldsymbol{k}} u_{\boldsymbol{k}+\delta\boldsymbol{k}}(\boldsymbol{r}) \tag{2.16}$$

(as discussed last time) thereby generating a geometrical phase

$$\gamma = \int_{\text{path for } \boldsymbol{k}} \boldsymbol{A}_\sigma^n(\boldsymbol{k}) \cdot d\boldsymbol{k} \qquad (2.17)$$

and a Berry curvature that can be nonzero for some electron bands

$$\boldsymbol{\Omega}_\sigma^n(\boldsymbol{k}) = \nabla_{\boldsymbol{k}} \times \boldsymbol{A}_\sigma^n(\boldsymbol{k}). \qquad (2.18)$$

This Berry curvature can deflect electron trajectories in real space according to our Rule (#2) in the previous lecture and has other physical implications that I will mention later.

***When can an electron band have non-zero Berry curvature?***:

It turns out that the Berry curvature will be zero throughout many electron bands and will be nonzero in others. What are the conditions required for the Berry curvature to be nonzero for an electron bandstructure? This question is most easily addressed by analyzing the symmetry constraints present in our Rule (#2) of band dynamics:

$$\frac{d\langle \boldsymbol{r} \rangle}{dt} = \frac{1}{\hbar} \nabla_{\boldsymbol{k}} \varepsilon_\sigma^n(\boldsymbol{k})\Big|_{\boldsymbol{k}=\boldsymbol{k}_0} - \frac{d\langle \boldsymbol{k} \rangle}{dt} \times \boldsymbol{\Omega}_\sigma^n(\boldsymbol{k}_0) \qquad (2.19)$$

For the sake of argument, suppose $\boldsymbol{F}_{\text{ext}} = -e\boldsymbol{E}$ so that $d\langle \boldsymbol{k} \rangle / dt = -e\boldsymbol{E}/\hbar$ and our Rule (#2) becomes

$$\boldsymbol{v} = \frac{1}{\hbar} \nabla_{\boldsymbol{k}} \varepsilon_\sigma^n(\boldsymbol{k})\Big|_{\boldsymbol{k}=\boldsymbol{k}_0} + \frac{e}{\hbar} \boldsymbol{E} \times \boldsymbol{\Omega}_\sigma^n(\boldsymbol{k}_0) \qquad (2.20)$$

where $\boldsymbol{v}$ is the real-space velocity.

First, suppose that the quantum system under consideration satisfies inversion symmetry. Under inversion, the scalar $\varepsilon_\sigma^n$ should be unchanged. The vectors $\boldsymbol{v}$, $\boldsymbol{k}$, and $\boldsymbol{E}$ will change sign, but because spin is a pseudovector (like the orbital angular momentum $\boldsymbol{r} \times \boldsymbol{p}$), it will remain unchanged. It follows that the energy must obey $\varepsilon_\sigma^n(-\boldsymbol{k}) = \varepsilon_\sigma^n(\boldsymbol{k})$. Because the left hand side of Eq. (2.20), $\boldsymbol{v}$, changes sign, both terms on the right hand side must change sign, as well. Given $\varepsilon_\sigma^n(-\boldsymbol{k}) = \varepsilon_\sigma^n(\boldsymbol{k})$, the term $\nabla_{\boldsymbol{k}} \varepsilon_\sigma^n(\boldsymbol{k})\big|_{\boldsymbol{k}=\boldsymbol{k}_0}$ does indeed change sign, as it should. For the cross-product term, $\boldsymbol{E}$ changes sign, so in order that the overall cross product should change sign, we must have that the Berry curvature after the transformation should have the same sign as before. Consequently, if inversion symmetry holds, we must have $\boldsymbol{\Omega}_\sigma^n(-\boldsymbol{k}_0) = \boldsymbol{\Omega}_\sigma^n(\boldsymbol{k}_0)$.

Next, consider the analogous analysis for a quantum system that obeys time reversal symmetry. Under time reversal, the energy $\varepsilon$ should be unchanged and the vectors $\boldsymbol{v}$, $\boldsymbol{k}$, and spin will change sign, but the electric field $\boldsymbol{E}$ will not. In this case, in order for the cross product $\boldsymbol{E} \times \boldsymbol{\Omega}$ to change sign, we must have that the Berry curvature after the transformation should have the opposite sign from before. Therefore, if time-reversal symmetry holds, we must have $\boldsymbol{\Omega}_{-\sigma}^n(-\boldsymbol{k}_0) = -\boldsymbol{\Omega}_\sigma^n(\boldsymbol{k}_0)$ (meaning simply that the Kramers'-degenerate states related by time-reversal symmetry must always have opposite Berry curvatures).

If both inversion and time-reversal are good symmetries for our quantum system, we can combine these two results to conclude that $\boldsymbol{\Omega}_\sigma^n(\boldsymbol{k}_0) = \boldsymbol{\Omega}_\sigma^n(-\boldsymbol{k}_0) = -\boldsymbol{\Omega}_{-\sigma}^n(\boldsymbol{k}_0)$. That is, upon reversing just the spin, in this high-symmetry case the Berry curvature must change sign. This has different implications depending on whether spin-orbit coupling is weak or strong.

In the limit of zero spin-orbit coupling, the spatial wavefunction (the part that depends on $\boldsymbol{k}$) will be identical for both spin states, which means that the Berry curvature $\boldsymbol{\Omega}^n(\boldsymbol{k}) = \nabla_{\boldsymbol{k}} \times i\left\langle u_{\boldsymbol{k}}^n \middle| \nabla_{\boldsymbol{k}} \middle| u_{\boldsymbol{k}}^n \right\rangle$ will not depend on the spin state. The only way this can be consistent with the requirement that $\boldsymbol{\Omega}_\sigma^n(\boldsymbol{k}_0) = -\boldsymbol{\Omega}_{-\sigma}^n(\boldsymbol{k}_0)$ for a system in which inversion and time-reversal are good symmetries, is that $\boldsymbol{\Omega}_\sigma^n(\boldsymbol{k}) = 0$ for all $\boldsymbol{k}$.

However, if spin-orbit coupling is not weak, then the states $\left| u_{\boldsymbol{k},\sigma}^n \right\rangle$ and $\left| u_{\boldsymbol{k},-\sigma}^n \right\rangle$ will have different orbital wavefunctions, and there is no requirement that the Berry curvature be zero. In fact, for heavy metals that preserve both inversion and time-reversal symmetry, Berry curvatures can be large with important physical consequences (see below). The symmetry requirement $\boldsymbol{\Omega}_\sigma^n(\boldsymbol{k}_0) = -\boldsymbol{\Omega}_{-\sigma}^n(\boldsymbol{k}_0)$ remains in this case so, for example, states with opposite spins must have equal and opposite values of anomalous velocity.

The conclusion that can be drawn from this analysis is that the Berry curvature $\boldsymbol{\Omega}_\sigma^n(\boldsymbol{k})$ is allowed to be nonzero if:

- Inversion symmetry is broken (or/and)
- Time-reversal symmetry is broken (or/and)
- Spin-orbit coupling is strong.

***Examples: Physical effects of anomalous velocities from Berry curvature***:

These three scenarios can all be manifested in real materials and produce interesting physical effects. For example, the single-layer 2D semiconductor $MoS_2$ breaks inversion symmetry. It possesses regions of reciprocal space (*i.e.*, "valleys") with non-zero Berry curvature. Carriers in regions with a given sign of Berry curvature can be excited selectively using circularly-polarized light. If an in-plane electric field is then applied, the anomalous velocity caused by the Berry curvature deflects the carrier trajectories perpendicular to the electric field and can give rise to a transverse Hall-type voltage even in the absence of any applied magnetic field (Fig. 2.3(a)). This is known as the valley Hall effect.

Magnetic materials break time-reversal symmetry. Consequently, the bands in magnetic metals like Fe, Co, Ni, and their alloys can have non-zero Berry curvatures. This will also produce anomalous velocities transverse to an applied electric field, resulting in what is known as the anomalous Hall effect – a voltage perpendicular to both the magnetic moment and applied electric field that can be much larger than the ordinary Hall effect produced by an applied magnetic field (Fig 2.3(b)).

Heavy metals like Pt and Ta preserve both inversion and time-reversal symmetry, but they have strong spin-orbit coupling. They have non-zero Berry curvatures, subject to the requirement that $\mathbf{\Omega}^n_\sigma(\boldsymbol{k}_0) = -\mathbf{\Omega}^n_{-\sigma}(\boldsymbol{k}_0)$. When an electric field is applied to these materials, this causes electrons with opposite spins to be deflected in opposite directions (Fig. 2.3(c)). Because (given the time-reversal symmetry) the density of carriers must be the same for states with opposite spins, equal numbers of electrons are deflected to both sides and there is no net Hall voltage generated. However, the flows of spin angular momentum for the opposite spin states add, rather than subtract, so the Berry curvature can result in a large spin current transverse to the applied electric field. This is known as the spin Hall effect. In samples in which a thin-film magnetic device is adjacent to a thin film of a material with strong spin-orbit coupling, the spin current produced by the spin-orbit material can be absorbed by the magnet and exert a torque strong enough to reverse its magnetization or drive precessional dynamics. This effect is currently under investigation as perhaps the most efficient reliable mechanism to write information within emerging magnetic memory technologies.

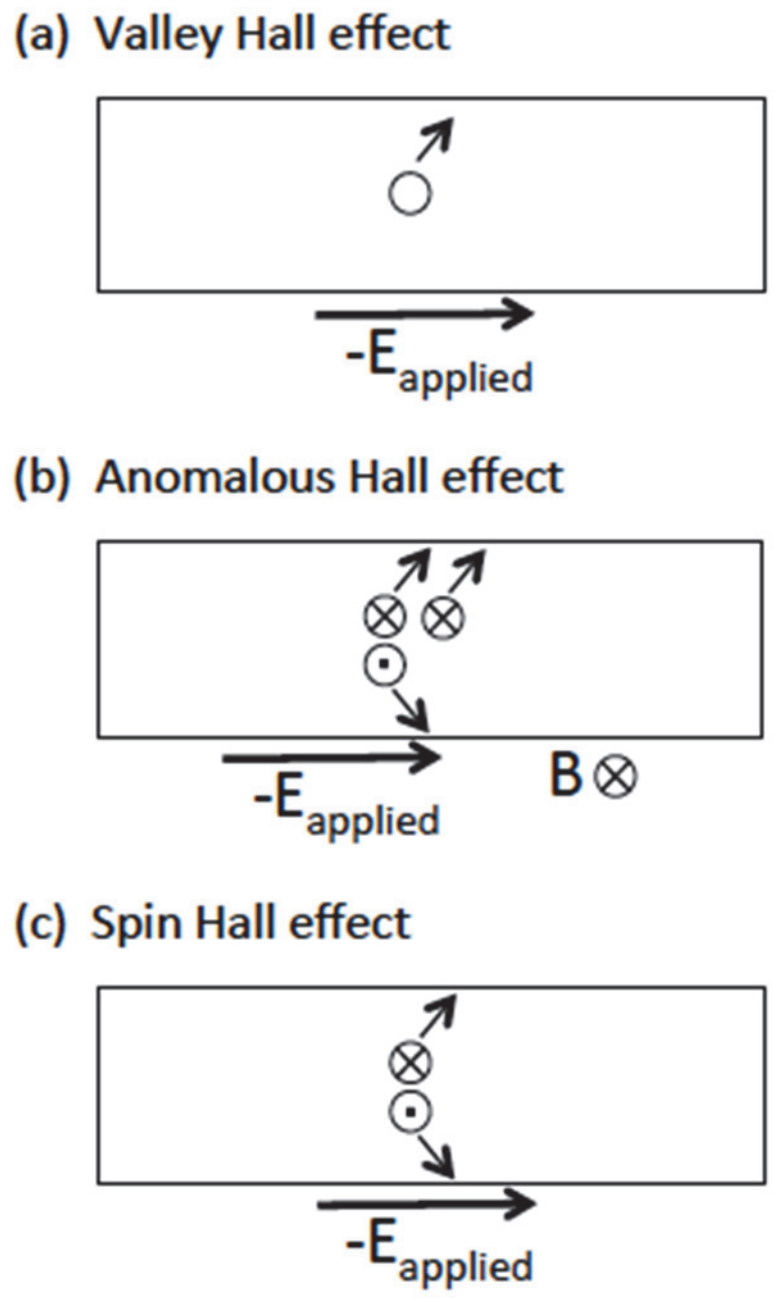

*Fig. 2.3. Effects arising from the Berry-curvature anomalous velocity. (a) Transverse charge current due to valley-polarized carriers in a material that breaks inversion symmetry. (b) Transverse charge and spin current in magentic materials that break time-reversal symmetry. (c) Transverse spin current in materials with strong spin-orbit coupling. (Crosses and dots represent out-of-plane spin directions.)*

One other complication that can sometimes occur is that there may be points in ***k***-space where two bands are degenerate. At such points the picture of adiabatic evolution breaks down and the Berry curvature may become singular – mathematically (via Stokes theorem) this means that these points can behave as monopole sources of flux. This type of degeneracy point plays a central role in understanding the generation of topological insulators, as we will discuss in a later lecture. There are also classes of 3D materials – Weyl or Dirac semi-metals -- where special degeneracy points exist somewhere within the Brillouin zone. If they occur close to the Fermi level, the singular Berry curvatures present in these bandstructures can produce huge anomalous velocities, along with a variety of other unusual phenomena. I will have a little bit more to say about these materials after we discuss topological insulators.

***Remarks about homework problems – practice in calculating Berry curvatures***:

I will leave it to the homework problems to give you some practice in calculating Berry curvatures, including for gapped graphene which is an important example of a Dirac-type Hamiltonian in a region of ***k***-space where two bands nearly touch. Here I will just summarize some of the key results you will find.

First, the Berry curvature will be zero everywhere for any tight-binding band composed from a single atomic orbital. The minimal model that gives non-zero Berry curvatures must couple at least two different bands (or in a tight-binding language, must have at least two atomic orbitals per unit cell). The Hamiltonian for this minimal model can be written completely generally in the form of the $2\times 2$ matrix

$$H = \boldsymbol{d}(\boldsymbol{k})\cdot\boldsymbol{\sigma}, \tag{2.21}$$

so solving for the Berry curvature for a spin in an applied magnetic field has immediate relevance for a wide variety of other problems, including calculating the Berry curvature relevant to gapped graphene.

While the Berry curvature can be calculated directly using the definition we have introduced, $\boldsymbol{\Omega}_\sigma^n(\boldsymbol{k}) = \nabla_{\boldsymbol{k}} \times i\left\langle u_{\boldsymbol{k},\sigma}^n \middle| \nabla_{\boldsymbol{k}} \middle| u_{\boldsymbol{k},\sigma}^n \right\rangle$, this is not always the most convenient approach, particularly for numerical calculations, because it requires derivatives of the eigenstates. With some operator algebra (see Appendix 3) it is possible to show that an equivalent expression for the Berry curvature is

$$\boldsymbol{\Omega}_\sigma^n(\boldsymbol{k}) = i\sum_{m\neq n} \frac{\left\langle u_{\boldsymbol{k}}^n \middle| (\nabla_{\boldsymbol{k}} H) \middle| u_{\boldsymbol{k}}^m \right\rangle \times \left\langle u_{\boldsymbol{k}}^m \middle| (\nabla_{\boldsymbol{k}} H) \middle| u_{\boldsymbol{k}}^n \right\rangle}{(\varepsilon_{\boldsymbol{k}}^m - \varepsilon_{\boldsymbol{k}}^n)^2}, \tag{2.22}$$

where $H$ is the effective Hamiltonian for determining $u_{\boldsymbol{k}}^n(\boldsymbol{r})$ and the band indices $m$ and $n$ are meant to count over spin as well as orbital bands in the case of strong spin-orbit coupling. In Eq. (2.22), the derivatives are performed on the Hamiltonian, which can generally be done analytically, rather than on the eigenstates. This expression also makes clearer some of the underlying physics in ways that our original definition did not. First, Berry curvature requires coupling between bands; it is zero in single-band problems. Second, the Berry curvature tends to be largest near points of reciprocal space where the gaps between bands are smallest, i.e., near avoided crossings where $\varepsilon_{\boldsymbol{k},\sigma}^m - \varepsilon_{\boldsymbol{k},\sigma}^n$ is a minimum. You will see that this is the case when solving for the band structure of gapped graphene. Third, the sum of the Berry curvatures at a given value of $\boldsymbol{k}$ over all the electron bands in a problem must be zero. This can be seen by summing Eq. (2.22) over $n$, and noting that there is a cancellation in the double sum between pairs of indices ($n$,$m$) and ($m$,$n$) due to a change in sign of the cross product.

***Other physical consequences of Berry physics***:

Finally, I want to make clear that our focus so far on how the Berry curvature causes the anomalous velocity in semiclassical band dynamics really only begins to scratch the surface on the physical implications of Berry physics. Berry connections and Berry curvatures are also intimately related to a variety of other physical phenomena, including electrical polarization, magnetoelectric polarizability, optical selection rules, electron interference conditions, multiferroic materials, the quantum Hall effect, and (as we will see in the next lecture) topological materials. In materials with Berry curvature, wavepackets will also typically have non-zero values of orbital magnetic moment associated with a real-space circulation of charge current about the wavepacket's center of mass. This magnetic moment will shift band energies when there is an applied magnetic field and will also induce an additional term in the semiclassical wavepacket dynamics if one goes beyond the assumptions we have employed so far to consider a spatially-varying applied magnetic field. Berry-generated magnetic moments are an exception to the rule of thumb that the electron-lattice interaction tends to suppress orbital magnetic moments by disrupting rotational symmetry – with Berry physics the electron-lattice interaction can instead induce orbital magnetic moments that are a function of the wavepacket's wavevector $\boldsymbol{k}$.

A good place to begin further reading about these topics is a recent book by David Vanderbilt, *Berry Phases in Electronic Structure Theory: Electric Polarization, Orbital Magnetization, and Topological Insulators*, Cambridge University Press, 2019 [4].

## III: Topological Invariants, Topological Insulators, and Edge States

(These notes are adapted in part from lecture notes by Ion Garate (Univ. of Sherbrooke, unpublished) and by J. K. Asbóth, L. Oroszlány, and A. Pályi (Ref. (5).)

*Learning Goals:*

- *What is a topological invariant?*
- *Understand the Su-Schrieffer-Heeger Model as a simple example that can realize a topological insulator*
- *See how level crossing within a Dirac-type Hamiltonian can result in topological transitions*
- *See how at the interface between materials with unequal topological invariants the band gap must go to zero because an interface state must exist.*

You are no doubt familiar with the idea that to a topologist a donut is equivalent to a coffee cup, in that they are both closed surfaces threaded by a single hole. This is a reference to the Gauss-Bonnet theorem from geometry, that if you integrate the Gaussian curvature $\boldsymbol{K}(\boldsymbol{r})$ over a closed 2D surface, the result takes only discrete values:

$$\int_{\text{closed surface}} \boldsymbol{K}(\boldsymbol{r}) \cdot d\boldsymbol{S} = 4\pi(1-g) \tag{3.1}$$

where $g$ is an integer corresponding to the number of holes threading the surface.

Similar mathematics can come into play when analyzing the properties of electron bands. Within the repeated zone scheme, states on the edge of the first Brillouin zone that are connected by reciprocal lattice vectors are equivalent. One can therefore think of having a form of periodic boundary conditions at the boundary of the first Brillouin zone, so that electron bands within the first Brillouin zone are defined on closed manifolds (specifically, d-dimensional tori). There are quantities associated with the Berry curvature of electron bands that are analogous to the Gaussian curvature of a 2D surface, in that when they are integrated over the closed manifold associated with first Brillouin zone they give discrete values. The different discrete values are called topological invariants. These ideas are defined rigorously only for insulators because integrating occupied states over the entire Brillouin zone requires that there not be any partially-filled bands, and there should also not be any touching of different bands at the same value of $\boldsymbol{k}$ in order that the Berry connection is always well-defined. Therefore, materials in which the electron bands have a non-zero topological invariant are called topological insulators. Topological insulators represent forms of matter distinct from non-topological insulators. They have a variety of neat properties, with one of the most interesting being that if two insulators have different values for a topological invariant then at an interface between these materials the band gap must go to zero to allow an edge state localized near the interface.

### *Topological invariants in 1D, 2D, and 3D*:

There is more than one type of topological invariant. Invariants take slightly different forms in different spatial dimensions (although they are related within the language of differential geometry). Even within a given spatial dimension there can be more than one type. The 2D case is most like the Gauss-Bonnet theorem. In this case the Berry curvature really is a "curvature" analogous to a Gaussian curvature, so that integrating the Berry curvature over an entire 2D electron band can only give certain discrete values:

$$\int_{BZ} \frac{d^2\boldsymbol{k}}{(2\pi)^2} \Omega_\sigma^n(\boldsymbol{k}) = \frac{c_{n,\sigma}}{2\pi} , \tag{3.2}$$

where $c_{n,\sigma}$ is an integer known as the (first) Chern number. This Chern number has an immediate physical consequence in that it is proportional to the Hall conductivity. This is because the Hall conductivity of an insulator is calculated by integrating the transverse anomalous velocity ($\propto \Omega_\sigma^n(\boldsymbol{k})$) over all of the states in the filled bands, which immediately gives the result:

$$\sigma_H \equiv \frac{|j_\perp|}{E} = \frac{e^2}{\hbar} \sum_{\text{filled bands } n} \int_{BZ} \frac{d^2\boldsymbol{k}}{(2\pi)^2} \Omega_\sigma^n(\boldsymbol{k}) = \frac{e^2}{2\pi\hbar} \sum_{\text{filled bands } n} c_{n,\sigma} . \tag{3.3}$$

This means that in a 2D insulator with a non-zero Chern number the Hall effect is automatically quantized. The quantum Hall effect can be understood within this context (if one takes the applied magnetic field into account from the beginning in solving for the electron bands), and a quantized non-zero value of the Hall conductivity can also occur in 2D materials that break time reversal symmetry even in the absence of an applied magnetic field, in which case the phenomenon has been called the quantum anomalous Hall effect. It is important to note that when insulators have a band with a nonzero Chern number it is not the case that the filled band is inert, carrying no current. A filled band cannot transport any current parallel to an electric field, but there can still be a nonzero Hall conductivity associated with charge currents and electric fields that are perpendicular (*e.g.*, the band can support a flowing charge current that generates a transverse voltage, but no longitudinal voltage).

The total Chern number (summed over all filled bands) of a 2D material must be zero if time reversal symmetry applies, because in this case $\boldsymbol{\Omega}_{-\sigma}(-\boldsymbol{k}_0) = -\boldsymbol{\Omega}_\sigma(\boldsymbol{k}_0)$ (recall from Lecture 2), and integration over Kramers'-degenerate bands will add up to zero, even if the Chern numbers for individual bands might be non-zero. However, there is also a different topological invariant first defined by Kane and Mele that can be non-zero even when time-reversal symmetry is preserved, dubbed a $Z_2$ invariant. In cartoon terms, this can correspond to a spin-up band having a Chern number of +1 and a spin-down band having a Chern number of -1, but it can be generalized rigorously to models in which $s_z$ is not conserved. This topological invariant is the key to defining non-magnetic topological insulators, a phenomenon known as the quantum spin Hall effect. Topological crystalline insulators are related, but with a discrete crystal symmetry (reflection or rotation) playing an essential role in protecting the topological state.

In 1D, it is possible to define the phases of wavefunctions such that the Berry connection $A_\sigma^n(k) = i\left\langle u_{k,\sigma}^n \middle| \frac{d}{dk} \middle| u_{k,\sigma}^n \right\rangle$ is periodic as a function of *k* over the Brillouin zone, in which case the integral of the Berry connection over the 1D Brillouin zone

$$\gamma = \int_{BZ} A_\sigma^n(k)\, dk \tag{3.4}$$

can be understood as a geometric phase integrated over a closed loop, which must therefore be gauge invariant. Physically, this quantity is related to the electrical polarization within a unit cell. We will see by example in this lecture that this can be a topological invariant.

3D topological invariants are integrals over a 3D Brillouin zone of a generalization of Berry curvature, that can be expressed rigorously in the language of vector bundles within differential geometry. There exist topological invariants in 3D analogous to both Chern numbers and $Z_2$ invariants in 2D.

***Su-Schrieffer-Heeger Model: Example of a 1D topological insulator***:

To get a flavor of some of the physics that can result with topological materials, I will consider the simplest example, a 1D problem known as the Su-Schrieffer-Heeger (SSH) model, originally designed to understand the electronic properties of the 1D polymer polyacetylene. It is also the simplest model for the formation of a charge density wave. I will highlight properties that generalize to higher-dimensional topological materials as I work through this example.

The SSH model consists of a 1D tight-binding chain with one electron per atomic site, but in which the positions of the atoms have some freedom to shift in 1D while keeping the average spacing between atoms fixed. The overall Hamiltonian is

$$H = \sum_n t_{n+1,n} \left[ \, |n+1\rangle\langle n| \; + \; |n\rangle\langle n+1| \, \right] \; + \; \frac{1}{2}\sum_n K(u_{n+1} - u_n)^2 \; , \tag{3.5}$$

where $u_n$ is the displacement of the nth atom from its position in an equally-spaced chain and the tight-binding overlap integral is assumed to depend linearly on the spacing between atoms:

$$t_{n+1,n} = t_0 - \alpha(u_{n+1} - u_n) . \tag{3.6}$$

The first sum in Eq. (3.5) is a standard 1D tight-binding problem to solve for the electronic states. The second sum is an elastic energy cost associated with having the atoms shift away from their equally-spaced positions. I will assume an even number of atomic sites. The Hamiltonian is independent of spin, so each electronic state will be two-fold degenerate and we will not bother with a spin index.

I will have you work through this problem in the homework to show that the chain is unstable to dimerization. The system can lower its electronic energy by allowing the atoms to shift their positions so that there are alternately larger and smaller spacings between atoms. This opens up a gap at the edge of the Brillouin zone of the dimerized chain (at $k = \pm\pi / a$ if $a/2$ is the spacing between atoms in the equally-spaced chain), and if the tight-binding band contains one electron per atom (or two electrons per dimerized unit cell), the shifting of energy levels near the gap lowers the total electronic energy (Fig. 3.1(a)). This comes at a cost of increased elastic energy, but you will see that the minimum-energy state has a non-zero amount of dimerization. For a finite chain there are two different ways this dimerization can occur – the leftmost two atoms in the chain can have either the larger or smaller value of the spacing (Fig. 3.1(b)). It turns out that these two different forms of dimerization are topologically distinct, corresponding to different values of a topological invariant. For an infinite chain or a chain with periodic boundary conditions, this is not very interesting, because the two states are related simply by a translation by $a/2$. However in a finite chain with two ends these two states are qualitatively different because of topologically-induced edge states. Our main goals for this lecture are to calculate the topological invariants for the two states and to understand the properties of the edge states.

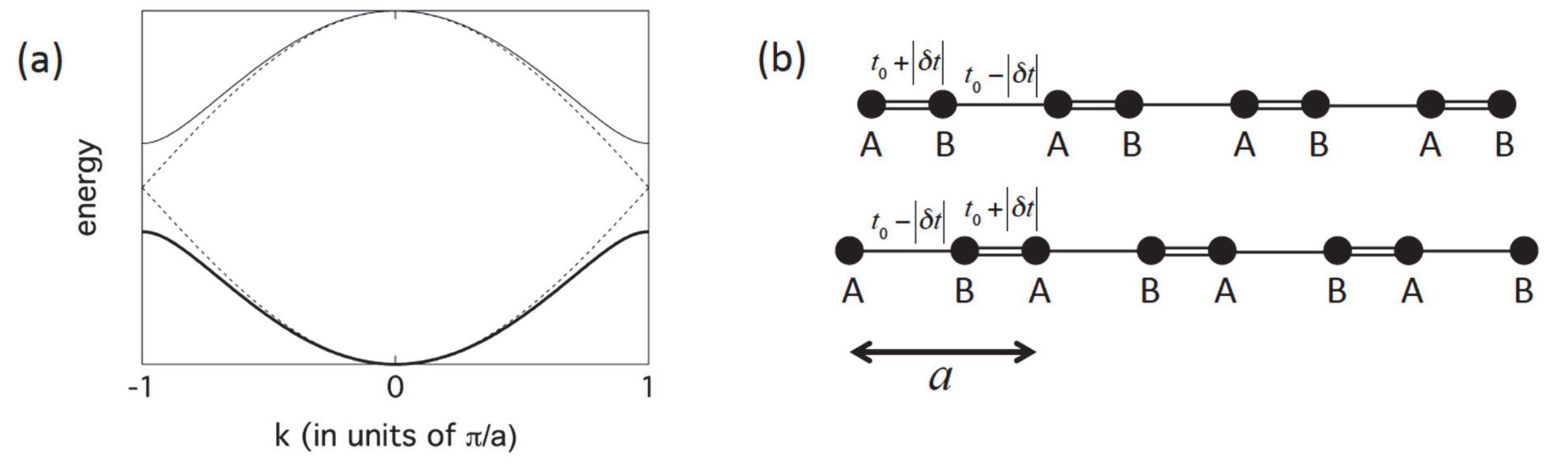

*Fig. 3.1. (a) Tight-binding bands for the dimerized chain with periodic boundary conditions (solid lines) compared to the undimerized chain (dotted lines). The energy of the ground-state band (darker solid line) is lowered by the dimerization. (b) Two different possibilities for dimerization of a finite 1D chain.*

To begin, let us focus on bulk properties by assuming that the 1D chain has periodic boundary conditions with a period of $2N$ atoms. We will also assume that the individual atomic states are sufficiently localized to be orthogonal to a good approximation, and that there is a fixed amount of dimerization in the chain with the overlap integrals alternating between the two values $t_0 + \delta t$ and $t_0 - \delta t$ (we will always assume $t_0 > 0$). Therefore $\delta t > 0$ gives one flavor of dimerization and < 0 for the other (Fig. 3.1(b)). The tight binding model will require two atomic states per unit cell (I will call them A and B), so I can assume a tight-binding Bloch state of the form

$$\psi_k(r) = \frac{1}{\sqrt{N}} \sum_R e^{ikR} [a(k)\phi_A(r-R) + b(k)\phi_B(r-(R+\Delta r))] \tag{3.7}$$

where $\Delta r$ is the spacing between atoms within the unit cell and $R$ denotes the Bravais lattice vectors of the different unit cells. Eq. (3.7) makes a specific choice for the phase convention within the unit cell that is nice because it will allow us to define an effective Hamiltonian with which to solve for Berry curvature that will be periodic in $k$ over the Brillouin zone. This will be convenient for the calculation.*

To solve for the electronic structure, we can follow exactly the same procedure we used to determine the tight-binding electronic states of graphene: apply the Hamiltonian to the Bloch wave function and then project onto each of the two atomic basis states. This yields a system of two linear equations, giving an effective Hamiltonian matrix that depends on $k$.

---

* One could use alternative intra-cell phase conventions in the tight-binding wavefunction, e.g., $\psi_k(r) = \frac{1}{\sqrt{N}} \sum_R e^{ikR} [a(k)\phi_A(r-R) + b(k) e^{ik\Delta r} \phi_B(r-(R+\Delta r))]$, but then $H_{\text{eff}}(\boldsymbol{k})$ in the basis of the coefficients $a$ and $b$ would no longer have the periodicity of the Brillouin zone, which would be inconvenient for our purposes.

$$\begin{pmatrix} 0 & (t_0+\delta t)+(t_0-\delta t)e^{-ika} \\ (t_0+\delta t)+(t_0-\delta t)e^{ika} & 0 \end{pmatrix}\begin{pmatrix} a \\ b \end{pmatrix} = E(k)\begin{pmatrix} a \\ b \end{pmatrix}. \quad (3.8)$$

The term proportional to $(t+\delta t)$ in each off-diagonal matrix element corresponds to the overlap integral between the A and B atoms in the same unit cell, while the terms proportional to $(t-\delta t)$ correspond to overlap integrals between A and B atoms in different (*i.e.*, adjacent) unit cells. Reducing the problem to this $2\times 2$ matrix Hamiltonian is great, because in our previous work we already found the solutions for the energies and Berry curvatures for a general $2\times 2$ Hamiltonian, and our problem now has precisely this form. The effective Hamiltonian in Eq. (3.8) can be written using Pauli matrices in the generic form

$$H_{\text{eff}}(\boldsymbol{k}) = \boldsymbol{d}(\boldsymbol{k})\cdot\boldsymbol{\sigma} \quad (3.9)$$

with

$$\begin{aligned} d_x &= (t_0+\delta t)+(t_0-\delta t)\cos(ka) \\ d_y &= (t_0-\delta t)\sin(ka) \\ d_z &= 0. \end{aligned} \quad (3.10)$$

The energy eigenvalues of the two electron bands are therefore just

$$E(k) = \pm\left|\boldsymbol{d}(k)\right| = \pm\sqrt{2(t_0^2+(\delta t)^2)+2(t_0^2-(\delta t)^2)\cos(ka)}\ . \quad (3.11)$$

Note for later use that the energy gap between the two bands only goes to zero if and only if both $\delta t = 0$ and *k* is at the edge of the Brillouin zone ($=\pm\pi/a$).

The topological invariant we are interested in calculating is the geometric phase determined by integrating the Berry connection for the lower-energy band over the first Brillouin zone.

$$\gamma = \int_{BZ} i\left\langle u_{k,-}(r)\right|\frac{d}{dk}\left|u_{k,-}(r)\right\rangle dk = \int_{-\pi/a}^{\pi/a} i\left\langle u_{k,-}(r)\right|\frac{d}{dk}\left|u_{k,-}(r)\right\rangle dk. \quad (3.12)$$

We could calculate this directly by solving the $2\times 2$ Hamiltonian to determine the coefficients $a_-(k)$ and $b_-(k)$ for the lower-energy band, using them to construct $\left|u_{k,-}(r)\right\rangle$, and working through algebra. However, there is a more elegant approach by which we can recast the integral in terms of quantities we already know. Given that the vector ***d***(*k*) in the effective Hamiltonian matrix is a smooth function of *k*, we can rewrite the integral for the geometric phase in terms of a line integral in ***d***-space:

$$\gamma = \int_{-\pi/a}^{\pi/a} i\left\langle u_{k,-}(r)\right|\nabla_{\boldsymbol{d}}\left|u_{k,-}(r)\right\rangle\cdot\frac{d\vec{\boldsymbol{d}}}{dk}dk = \int_{\text{path for } \boldsymbol{d}(k)} i\left\langle u_{k,-}(r)\right|\nabla_{\boldsymbol{d}}\left|u_{k,-}(r)\right\rangle\cdot d\vec{\boldsymbol{d}}. \quad (3.13)$$

The path for ***d***(*k*) is a closed loop, since by Eq. (3.10) $\boldsymbol{d}(k=\pi/a)=\boldsymbol{d}(k=-\pi/a)=\hat{x}2\delta t$. Therefore, by applying Stokes theorem, we can convert the path integral to a surface integral of a Berry curvature over a surface in ***d***-space bounded by the contour of ***d***(*k*).

$$\gamma = \int_{\text{area enclosed by } \boldsymbol{d}(k)} \nabla_{\boldsymbol{d}}\times i\left\langle u_{k,-}(r)\right|\nabla_{\boldsymbol{d}}\left|u_{k,-}(r)\right\rangle\cdot d\boldsymbol{S}_{\boldsymbol{d}} = \int_{\text{area enclosed by } \boldsymbol{d}(k)} \Omega_-(\boldsymbol{d})\cdot d\boldsymbol{S}_{\boldsymbol{d}}. \quad (3.14)$$

We have not yet calculated this Berry curvature $\Omega_-(\boldsymbol{d})$, but this is straightforward. If we define the Bloch bands composed of just the A sites and just the B sites

$$\psi_k^A(r) = \frac{1}{\sqrt{N}} \sum_R e^{ikR} \phi_A(r-R) = e^{ikr} u_k^A(r) \text{ and } \psi_k^B(r) = \frac{1}{\sqrt{N}} \sum_R e^{ikR} \phi_B(r-(R+\Delta r)) = e^{ikr} u_k^B(r), \tag{3.15}$$

then we can write $u_{k,-}(r)$ as a superposition of $u_k^A(r)$ and $u_k^B(r)$

$$u_{k,-}(r) = a_- u_k^A(r) + b_- u_k^B(r), \tag{3.16}$$

with the same coefficients $a_-(k)$ and $b_-(k)$ (or, equivalently, $a_-(\boldsymbol{d})$ and $b_-(\boldsymbol{d})$) that solve Eq. (3.8). What's more, since $u_k^A(r)$ and $u_k^B(r)$ are each associated with tight-binding bands made from a single atomic orbital, individually they have no Berry curvature. We are assuming that the A and B atomic orbitals are effectively orthogonal, which means that $\left\langle u_k^B(r) \middle| \nabla_{\boldsymbol{d}} \middle| u_k^A(r) \right\rangle = 0$ and $\left\langle u_k^A(r) \middle| \nabla_{\boldsymbol{d}} \middle| u_k^B(r) \right\rangle = 0$. It follows that the Berry curvature we need to calculate has the simple form

$$\Omega_-(\boldsymbol{d}) = \nabla_{\boldsymbol{d}} \times i \left\langle a_- u_k^A(r) + b_- u_k^A(r) \middle| \nabla_{\boldsymbol{d}} \middle| a_- u_k^A(r) + b_- u_k^A(r) \right\rangle = \nabla_{\boldsymbol{d}} \times i \left[ a_-^* \nabla_{\boldsymbol{d}} a_- + b_-^* \nabla_{\boldsymbol{d}} b_- \right], \tag{3.17}$$

which is merely the Berry curvature associated with the simple effective $2 \times 2$ Hamiltonian in Eq. (3.9), which from the homework we already know has the value

$$\Omega_-(\boldsymbol{d}) = \frac{\hat{\boldsymbol{d}}}{2d^2}. \tag{3.18}$$

You will recall that this has a simple geometrical interpretation – a spherically-symmetric flux from a monopole of charge ½ located at the origin. Therefore the geometric phase to be calculated by the last integral in Eq. (3.14) is just the integrated flux from the monopole of charge ½ through the area enclosed by $\boldsymbol{d}(k)$. This is ½ times the solid angle that the trajectory $\boldsymbol{d}(k)$ subtends about the origin. This is something that is very easy to calculate, as follows.

From Eq. (3.10), As $k$ evolves through the Brillouin zone, the contour $\boldsymbol{d}(k)$ is a circle passing progressively through the points $\boldsymbol{d}(k=-\pi/a) = (2\delta t)\hat{x}$, $\boldsymbol{d}(k=-\pi/2a) = (t_0+\delta t)\hat{x} - (t_0-\delta t)\hat{y}$, $\boldsymbol{d}(k=0) = (2t_0)\hat{x}$, $\boldsymbol{d}(k=\pi/2a) = (t_0+\delta t)\hat{x} + (t_0-\delta t)\hat{y}$, and back to $\boldsymbol{d}(k=\pi/a) = (2\delta t)\hat{x}$. The contour for the case of $\delta t > 0$ (recall $t_0 > 0$) corresponds to Fig. (3.2(a)). The circle lies entirely in the *x*-*y* plane with the origin on the outside of the circle. Because the circle is oriented edge-on to the origin, none of the flux from the monopole source at the origin can penetrate through the contour. Therefore, by Eq. (3.14) the geometric phase is simply 0 for $\delta t > 0$. This is a non-topological insulator.

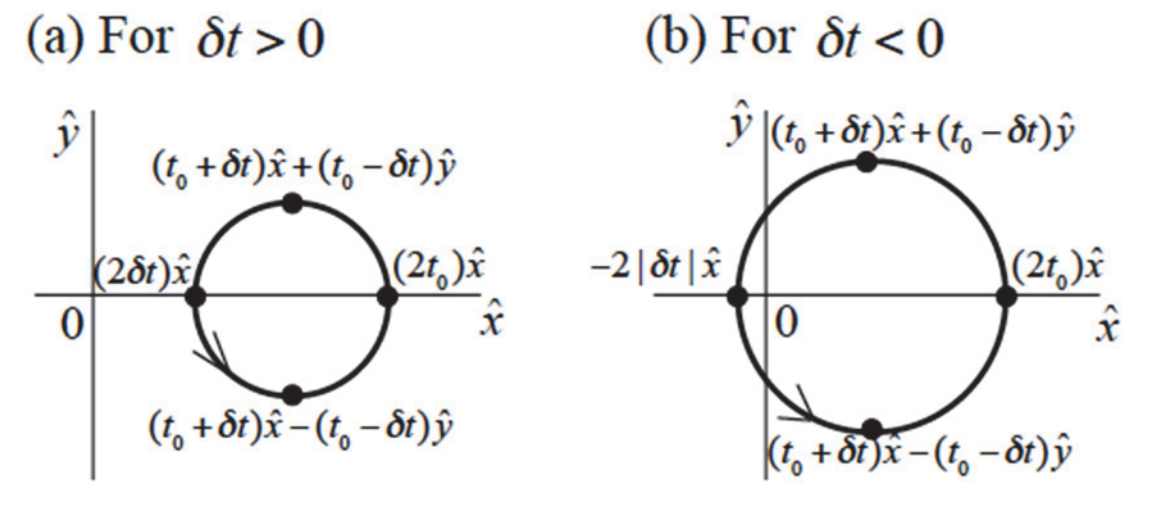

*Fig. 3.2 Contours for calculating geometric phases in the SSH model.*

On the other hand, for the case when $\delta t < 0$ the circle $\boldsymbol{d}(k)$ is shown in Fig. (3.2(b)). Now $\boldsymbol{d}(k=-\pi/a)$ shifts to negative values of x so that the origin is within the circle. The flux going through the contour will intersect the entire half of a sphere at $z > 0$, so the solid angle enclosed is $2\pi$, and the geometric phase from Eq. (3.14) is $\pi$. This is not equal to 0 modulo $2\pi$, so this phase represents a topological invariant that is different from the $\delta t > 0$ case. This case of $\delta t < 0$ corresponds to a 1D topological insulator.

This result is topological in the sense that one can imagine changing parameters in the Hamiltonian continuously to charge the trajectory ***d***(*k*) (e.g., Fig. 3.3). As long as $d_z$ is always zero and the contour does not cross through the point $(d_x, d_y) = (0,0)$, the geometric phase cannot not change its value – contours that do not enclose the origin will have a geometric phase of zero indicating a non-topological state, while contours that do enclose the origin will have a phase of $\pi$. The only way to cross between these values is to have a contour intersect the origin, but when ***d*** = 0 by Eq. (3.11) the energy gap goes to zero. This means we no longer have a well-defined insulating state. The conclusion to be drawn is quite general – insulating bands cannot change their topological character except by processes where a bandgap at least momentarily closes and reopens at some value (or values) of ***k***, so that there is no way that an insulator can continuously change from one topological state to another while always remaining an insulator.

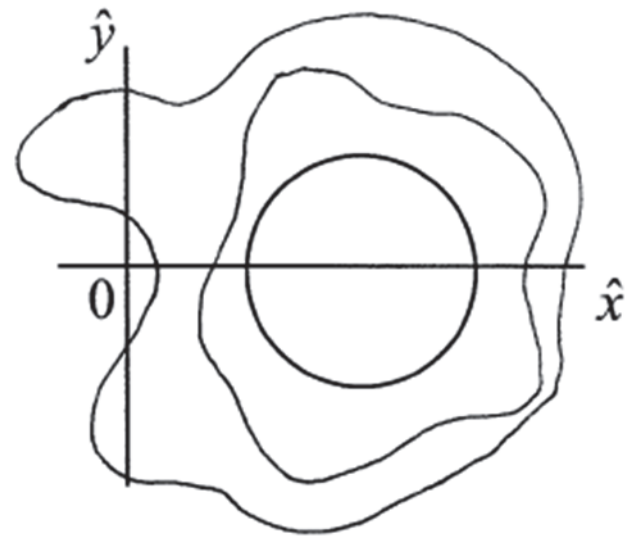

*Fig. 3.3. If the Hamiltonian is changed the trajectory **d(k)** may be deformed, but if $d_z$ remains zero and the trajectory never crosses the origin, the geometric phase is unchanged.*

In 2D or 3D, the mathematics can be expressed in an entirely analogous way. Using higher-dimensional versions of Stokes theorem, topological invariants can be expressed as a flux through a higher-dimensional surface. The points where ***d*** = 0 continue to act as monopole sources of flux, so the value of the topological invariant is determined by whether or not the surface captures the flux from these monopoles. If the surface sweeps through a ***d*** = 0 point, the bandgap will momentarily go to zero and the topological invariant can change its value.

***The relationship between topological transitions and energy-level crossings***:

For the SSH model, another view of what happens as one tunes through the zero-bandgap condition from the non-topological state ($\delta t > 0$) to the topological state ($\delta t < 0$) is shown in Fig. 3.4. When $\delta t > 0$, the Berry connection within each band can be nonzero, with different values a function of *k*, but there are sign variations so that the total geometric phase integrated over each band must be equal to zero (modulo $2\pi$). For example, for the choice of gauge used in the homework, this cancellation arises in the ground-state band because the phase near the bottom of the band is less than zero, while the phase near the top of the ground-state band (near $k = \pi / a$) is greater than zero. Phases in the excited-state band can be chosen to be equal and opposite to those in the ground-state band (see homework). As the bandgap narrows and approaches zero, the phase near the top of the ground-state band becomes increasingly concentrated near $k = \pi / a$, with an integrated phase of $\pi / 2$ occurring just in the narrow avoided-crossing region. The excited-state band develops a corresponding integrated phase of $-\pi / 2$ near $k = \pi / a$. As the two bands touch and cross to re-open the band gap, the integrated phases from the two bands in the avoided-crossing region are interchanged, so that now the top of the ground state band has an integrated phase of $-\pi/2$ rather than $\pi/2$. Consequently, both the top and bottom parts of the ground-state band contribute negative phases, and the total geometric phase of the ground state is no longer 0 (modulo $2\pi$). This can all be worked out explicitly using results from the homework. If we Taylor-expand $H_{\text{eff}}(k)$ near the minimum-bandgap point $k=\pi/a$ and also in the limit $\delta t \ll t_0$ we have

$$H_{\text{eff}}(q)=\begin{pmatrix} 0 & 2\delta t+it_0aq \\ 2\delta t-it_0aq & 0 \end{pmatrix}=2\delta t\sigma_x - t_0aq\sigma_y, \tag{3.19}$$

where I have defined the shifted wavevector $q=k-\pi/a$ and identified $d_x=2\delta t$ and $d_y=-t_0aq$. Expressing $d_x$, $d_y$ in spherical polar coordinates, we have $\theta=\pi/2$ and $\tan\phi=d_y/d_x=-t_0aq/(2\delta t)$. From the homework, we know that for any coordinate $j$ the Berry connection in the ground state can be written $A_{j,-}=-\cos^2(\theta/2)\partial_j\phi$. Thus, after some algebra,

$$A_{q,-}=-\frac{1}{2}\frac{d\phi}{dq}=\frac{1}{2}\frac{\left(2\delta t/t_0a\right)^2}{q^2+\left(2\delta t/t_0a\right)^2}, \tag{3.20}$$

and integrating over the minimum-bandgap region from $-\Delta q$ to $\Delta q$, the geometric phase from this region is

$$\int_{-\Delta q}^{\Delta q} A_{q,-}\,dq=\tan^{-1}\left(\frac{at_0\Delta q}{2\delta t}\right). \tag{3.21}$$

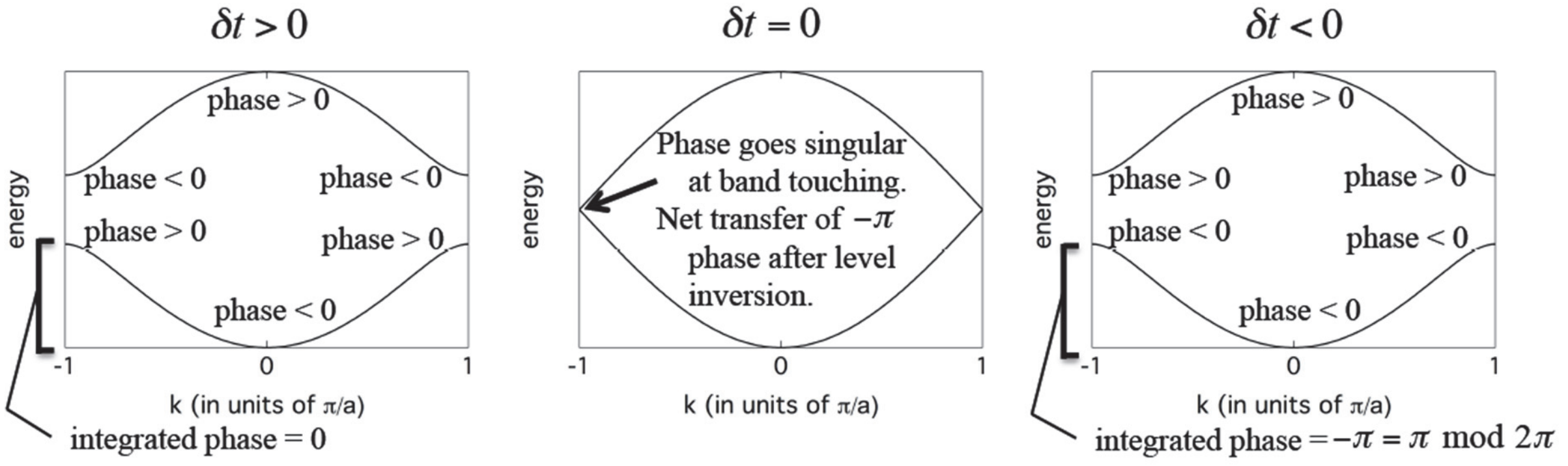

*Fig. 3.4. Contributions to the geometric phase from different parts of the bands in the SSH model, and how these phases changes due to the level crossing when the bandgap goes through zero. The signs correspond to the gauge convention from the homework.*

Very near the zero-bandgap condiction, when $\delta t\approx 0$, this is equal to $(\pi/2)\text{sign}(\delta t)$. Therefore, immediately before the crossing, when $\delta t>0$, the integrated phase from this avoided crossing region in the ground state is $+\pi/2$ (and in the excited state it is $-\pi/2$), while immediately after the crossing, when $\delta t<0$, there is a transfer so that in the ground state the integrated phase is $-\pi/2$ (and the excited state $+\pi/2$), as noted above. For states with values of *k* well away from the avoided crossing region, such small changes in $\delta t$ have negligible effect on the eigenstates and therefore also negligible effect on the Berry connection (and so the integrated phase in the ground state away from the band-gap minimum must stay = $-\pi/2$). Therefore the effect of the sign change in $\delta t$ is a net change in geometric phase for the ground state from 0 to $-\pi$ (and a change in the geometric phase of the excited state from from 0 to $\pi$). The signs here do not matter, the values are the same modulo $2\pi$, and could be switched by a gauge transformation. The result is therefore fully consistent with our previous calculation based on the contours in Fig. 3.2.

Similar level crossings can also generate topological transitions in 2D and 3D. In 2D, this can be illustrated using the results of the gapped-graphene homework problem. We saw that for gapped graphene, the effective Hamiltonian near the minimums in the bandgap could be written

$$H_{\text{eff}}(\boldsymbol{k}) = \begin{pmatrix} m & U(k_x - \pm ik_y) \\ U(k_x \pm ik_y) & -m \end{pmatrix} = \boldsymbol{d}(\boldsymbol{k}) \cdot \sigma, \tag{3.22}$$

corresponding to $d_z = m$, $d_x = Uk_x$, $d_y = \pm Uk_y$, and with the two signs corresponding to the K and K' points in reciprocal space. (The wavevector is measured relative to these points.) Note the similarity of Eq. (3.22) to the effective Hamiltonian for the SSH model – near the minimum bandgap one component of the vector $\boldsymbol{d}$ is a small constant ($\delta t$ or *m*) not depending on $\boldsymbol{k}$ and the other components are linearly proportional to components of $\boldsymbol{k}$ measured relative to the bandgap minimum – this is the generic form for a Dirac Hamiltonian with a small mass term. In the gapped-graphene homework problem, for the small-gap case that $|m|/(Uk_0) << 1$, we saw that integrating the Berry curvature for the ground state in the 2D region of $\boldsymbol{k}$-space around the bandgap minimum gives $\gamma_- = \pm\pi \text{ sign}(m)$, while the integrated Berry curvature in the excited state is equal and opposite. If one tunes the sign of *m* through zero (meaning that the there is an energy inversion between the starting atomic levels at the A and B sites of the graphene lattice) the integrated Berry curvatures about the K and K' points (which are equal and opposite) both flip sign, and so there is no net change in the total integrated Berry curvature within the ground state. However, if it can be arranged to flip the sign of the band-gap parameter *m* just for one of the K or K' points and not the other (which requires a perturbation that breaks time-reversal symmetry rather than the broken inversion symmetry of ordinary gapped graphene), this would cause a net change in the integrated Berry curvature of $2\pi$, changing the topological invariant Chern number

$$c_n = \frac{1}{2\pi} \int_{BZ} d^2\boldsymbol{k}\,\Omega^n(\boldsymbol{k}) \tag{3.23}$$

from 0 to $\pm 2\pi$, creating a quantum anomalous Hall state.

Let's consider how all this relates to topological transitions more generally. For any dimension of $\boldsymbol{k}$ we have seen that the eigenstates in a 2-band model are governed by an effective Hamiltonian of the form (Eq. 3.9)

$$H_{\text{eff}}(\boldsymbol{k}) = \boldsymbol{d}(\boldsymbol{k}) \cdot \boldsymbol{\sigma} = d_x(\boldsymbol{k})\sigma_x + d_y(\boldsymbol{k})\sigma_y + d_z(\boldsymbol{k})\sigma_z \tag{3.24}$$

with $d_x$, $d_y$, and $d_z$ real-valued and with energy eigenvalues $E = \pm\sqrt{d_x^2(\boldsymbol{k}) + d_y^2(\boldsymbol{k}) + d_z^2(\boldsymbol{k})}$. In order to create a topological transition, one must somehow drive the bandgap between the two bands to zero at one or more points in $\boldsymbol{k}$-space, because if not the topological invariant is (as the name implies) invariant. Just trying to push the bands together at some arbitrary value of $\boldsymbol{k}$ (i.e., tuning $d_z \to 0$) will generally not work by itself, because off-diagonal matrix elements coupling the two bands will cause an avoided crossing, rather than a crossing. (That is, one needs all three of $d_x$, $d_y$, and $d_z$ to go to zero simultaneously to create a band touching.) In practice, this means that even with careful tuning of external parameters the bands can only be made to touch at certain values of $\boldsymbol{k}$ where symmetries of the system require that there is zero off-diagonal coupling between the bands. Let us call any one of these points $\boldsymbol{k} = \boldsymbol{k}_0$. Then near $\boldsymbol{k}_0$ the effective Hamiltonian can be approximated by a Taylor expansion.

$$\begin{aligned} H_{\text{eff}}(\boldsymbol{k}) &\approx \nabla_{\boldsymbol{k}} d_x(\boldsymbol{k}_0)\cdot(\boldsymbol{k}-\boldsymbol{k}_0)\sigma_x + \nabla_{\boldsymbol{k}} d_y(\boldsymbol{k}_0)\cdot(\boldsymbol{k}-\boldsymbol{k}_0)\sigma_y + \nabla_{\boldsymbol{k}} d_z(\boldsymbol{k}_0)\cdot(\boldsymbol{k}-\boldsymbol{k}_0)\sigma_z \\ &= \sum_{i=x,y,z} \boldsymbol{v}_i \cdot \delta\boldsymbol{k}\ \sigma_i = \sum_{i=x,y,z} \left|\boldsymbol{v}_i\right| \left(\delta\boldsymbol{k}\right)_{\boldsymbol{v}_i}\ \sigma_i \end{aligned} \tag{3.25}$$

where $\hbar\boldsymbol{v}_i = \nabla_{\boldsymbol{k}} d_i(\boldsymbol{k}_0)$ and $\delta\boldsymbol{k} = \boldsymbol{k} - \boldsymbol{k}_0$. The usual result, in the absence of any symmetry constraints, is once again simply a generic Dirac Hamiltonian with a linear dispersion curve near the bands-touching point, with effective velocities $\boldsymbol{v}_i$ serving as (possibly non-orthogonal) axes defining the displacements $\left(\delta\boldsymbol{k}\right)_{\boldsymbol{v}_i}$. If as a function of an external parameter $\lambda$ one generates a level crossing during which a bandgap goes to zero at $\boldsymbol{k} = \boldsymbol{k}_0$ and reopens, this can be expressed with the same mathematics we have seen for the SSH and gapped-graphene models – one component of $H_{\text{eff}}(\boldsymbol{k})$ has the form $m(\lambda)\sigma_i$ with $m(\lambda)$ going from a non-zero value through 0 and changing sign to give a level crossing at $\boldsymbol{k}_0$, while the other components keep their linear dependence on $\left(\delta\boldsymbol{k}\right)_{\boldsymbol{v}_i}$. When this happens, the equal-and-opposite integrated Berry curvatures in the two bands near the bandgap minimum are interchanged, making the total Berry curvature of the ground state nonzero by the same process as we have seen for the gapped-graphene example, and generating a topological phase change.

(Note for experts: While an effective Dirac Hamiltonian is the usual case in $\boldsymbol{k}$-space near where two bands touch, it is not required in all circumstances -- if there are zeros in the effective velocities $\boldsymbol{v}_i$ or they are not linearly independent then the locus of bands-touching might be lines or surfaces in $\boldsymbol{k}$-space rather than simple points, and the dispersions as a function of $\delta\boldsymbol{k}$ could have higher powers than linear. If you are interested in further reading, these cases can have interesting topological consequences, too.)

Given the connection between level crossings and topological states, a key to identifying examples of time-reversal symmetric 2D and 3D topological insulators (*i.e.*, quantum spin Hall insulators) has been to find materials with a suitable form of band inversion. Suppose that in a non-topological material one has two bands close to touching near the Fermi level. If spin-orbit coupling is initially weak, spin-up and spin-down bands will have the same Berry curvatures. If the bands can be made to shift by increasingly strong spin-orbit coupling (a time-reversal symmetric perturbation) so that the band gap goes to zero and then reopens, one can have a similar transfer of Berry curvature for the two spin states (in this case with opposite signs) resulting in a new topological state. Consequently, a productive strategy to search for new non-magnetic topological insulators has been to look for materials in which the order of bands is different than for widely-separated atoms because of strong spin-orbit coupling. In particular, it has productive to look for situations when the bands that cross have different parities so that effective Hamiltonians with matrix elements linear in ***k*** are allowed, analogous to our Dirac Hamiltonians.

***Topological Edge States:***

I now want to turn to the subject of edge states – the idea that there are guaranteed to be special states with energy in the bandgap that are localized in real space near the edge of a topological insulator (adjacent to vacuum) or more generally at the interface between any two insulators with different values of a topological invariant. We will again start with the SSH model as a simple concrete example. To consider edges, we will need to switch from the assumption of periodic boundary conditions that we used to calculate bulk properties to the assumption of a finite chain (still with $2N$ atoms) and no periodic boundary condition. The existence of a difference in electronic properties at the edges between the non-topological ($\delta t > 0$) and topological ($\delta t < 0$) states is most obvious for the "fully-dimerized" cases when $\delta t = \pm t_0$ (Fig. 3.5). When $\delta t = t_0$ (greater than zero), there is coupling with strength $2t_0$ between the A and B atoms in each unit cell, with no coupling between neighboring unit cells. The resulting energy eigenstates will form two flat bands with all energies equal to $E = \pm|2t_0|$ (each coupled pair of atoms gives just an energy splitting due to the off-diagonal matrix element of a $2 \times 2$ matrix). For the topological state with $\delta t = -t_0$, however, there will be coupling with strength $2t_0$ between the A and B atoms in different unit cells, with no coupling within the unit cells. In the middle of the chain, dimerization will produce lots of states all with the same energies $E = \pm|2t_0|$ as before, but now there will also be two atoms not coupled to any others at the edges of the chain – an uncoupled A atom on the left and an uncoupled B atom on the right. These will have just the same energy as the starting atomic states, which we had defined to be $E = 0$, right in the center of the bandgap.

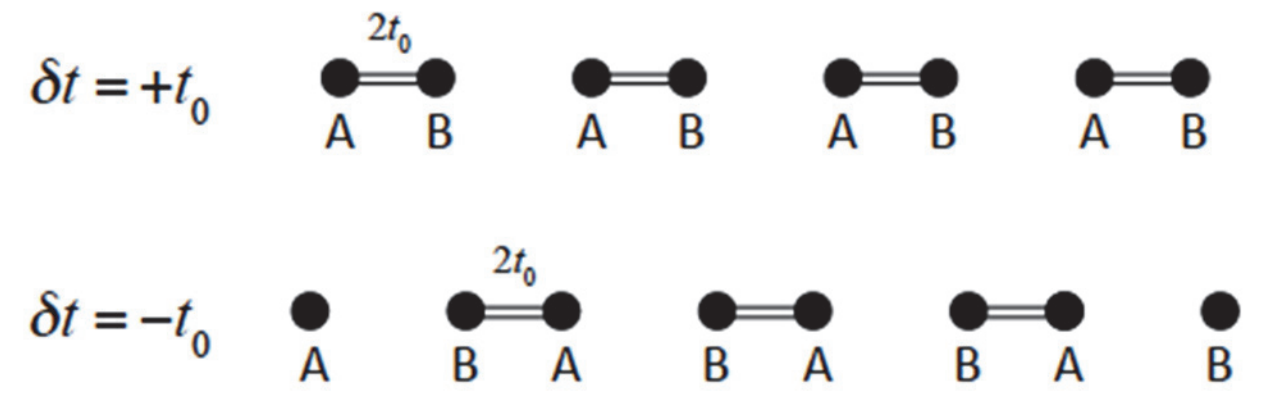

*Fig. 3.5. Fully-dimerized SSH chains for the non-topological and topological cases.*

This result so far is pretty trivial, but the remarkable thing is that these edge states with energy $E = 0$ are topologically protected. If one tunes $\delta t$ away from $-t_0$ to turn on coupling between all of the atoms, there remain edge states locked to the energy $E = 0$ as long as the sign of $\delta t$ is always negative. The edge states do not remain localized just on the last atom; they become evanescent waves decaying exponentially as they penetrate toward the bulk of the chain, but their energies do not shift. (This is true for a sufficiently long chain. If the evanescent waves from the two ends overlap significantly, they can couple to each other to produce a small splitting.) What's even more remarkable is that the parameters $t_0$ and $\delta t$ do not need to remain fixed as a function of position – they can vary with position and the edge state will still exist and remain pinned at $E = 0$ as long as $\delta t$ never becomes positive. If $\delta t$ does switch sign at some location, this represents a boundary between different topological phases, and there will be an $E=0$ edge state localized there, too. This state of affairs is sufficiently surprising that I suggest you play around with direct matrix diagonalization of the tight binding model on a finite chain to observe the edge states yourself. (This will be assigned as a homework problem.) For those wishing understand these phenomena at a deeper level, these properties are intimately connected to an underlying symmetry of the SSH model called a chiral symmetry.

Here, I will show analytically that there are edge states with energy $E=0$ and how to solve for their amplitude as a function of position. I will consider a situation where $\delta t$ changes slowly with position $x$ with a sign change at $x = 0$, such that at $x < 0$ $\delta t$ is positive and at $x > 0$ $\delta t$ is negative. Throughout the boundary region, $|\delta t| << t_0$. Since translational invariance is broken, the energy eigenstates will no longer be Bloch states with a single value of $k$, but the low-energy eigenstates (near $E=0$) can still be constructed as wavepackets in the form of superpositions of low-energy Bloch states. For the SSH model, those will be the states near the edge of the Brillouin zone $k=\pi/a$, for which the linearized effective Hamiltonian is Eq. (3.19). To make the physics a little more transparent, let us redefine variables $m(x)=2\delta t(x)$ and $\hbar v = at_0$ so that the effective Hamiltonian becomes

$$H_{\text{eff}}(q)=\begin{pmatrix} 0 & m(x)+i\hbar vq \\ m(x)-i\hbar vq & 0 \end{pmatrix}=m(x)\sigma_x-\hbar vq\sigma_y. \tag{3.26}$$

This is a Dirac-type Hamiltonian with an effective mass $m(x)$ and a velocity $v$, of the generic form discussed in Eq. (3.25). In this situation where the effective mass varies with position there is a standard trick (called the envelope function approximation, see ref. 5.) for solving for the spatial wavefunctions of the low-energy eigenstates The spatial wavefunction written in terms of the real-space amplitude on the A and B sites $\begin{pmatrix} \psi_A(x) \\ \psi_B(x) \end{pmatrix}$ can be determined by writing $H_{\text{eff}}(q)$ in terms of the operator $q\to\frac{1}{i}\frac{d}{dx}$ and solving the eigenvalue equation

$$\begin{pmatrix} 0 & m(x)+\hbar v\dfrac{d}{dx} \\ m(x)-\hbar v\dfrac{d}{dx} & 0 \end{pmatrix}\begin{pmatrix} \psi_A(x) \\ \psi_B(x) \end{pmatrix}=E\begin{pmatrix} \psi_A(x) \\ \psi_B(x) \end{pmatrix}. \tag{3.27}$$

This is analogous to the usual procedure of substituting $\boldsymbol{p}\to(\hbar/i)\nabla_{\boldsymbol{x}}$ in the conventional Schrodinger equation when considering the real-space wavefunction as a wavepacket formed as the superposition of momentum eigenstates.

Let's set $E = 0$ in Eq. (3.27), and see if indeed there is always a solution. A zero-energy solution requires

$$m(x)\psi_A(x)-\hbar v\frac{d\psi_A(x)}{dx}=0 \text{ and } m(x)\psi_B(x)+\hbar v\frac{d\psi_B(x)}{dx}=0. \tag{3.28}$$

or

$$\frac{d\psi_A(x)}{dx} = +\frac{m(x)}{\hbar v}\psi_A(x) \text{ and } \frac{d\psi_B(x)}{dx} = -\frac{m(x)}{\hbar v}\psi_B(x). \qquad (3.29)$$

Both equations can be solved by direct integration

$$\psi_A(x) = \psi_A(0)\exp\left[+\frac{1}{\hbar v}\int_0^x m(x')\,dx'\right] \text{ and } \psi_B(x) = \psi_B(0)\exp\left[-\frac{1}{\hbar v}\int_0^x m(x')\,dx'\right]. \qquad (3.30)$$

Now in the case we are considering ( $m(x) > 0$ for $x < 0$ and $m(x) < 0$ for $x > 0$), $\psi_B(x)$ grows in magnitude indefinitely as $x$ moves away from 0 in both directions, and so that part of the solution in not normalizable or physical. But that is OK. We will still have a solution if we set $\psi_B(x) = 0$ everywhere because in this case $\psi_A(x)$ does not need to be 0 – the solution for $\psi_A(x)$ decreases in magnitude as $x$ moves away from 0 in both directions (e.g., Fig. 3.6) and is normalizable. So we do indeed have a solution with $E = 0$ even if $m(x)$ (or $\delta t(x)$) varies with position – the wavefunction has zero weight on the B sites but decays evanescently to both sides on the A sites to be localized near the interface. (The solution must decay evanescently because the energy $E$ is in the bulk bandgap, so it could not yield a propagating wavefunction.) If the sign change of $m(x)$ had the opposite character, increasing through $x$ = 0, the weight of the wavefunction would instead be entirely on the B sites.

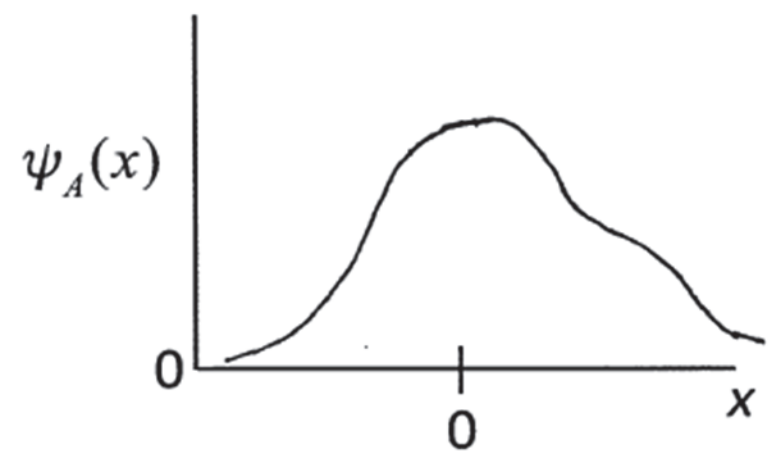

*Fig. 3.6. Generic interfacial wavefunction corresponding to a solution of Eq. (3.30), that decays to zero on both sides of the interface positioned at x = 0.*

This result can also be generalized to problems in higher dimensions. Whenever the effective Hamiltonian near the minimum bandgap can be linearized to the form of Dirac Hamiltonian with an effective mass term and the effective mass term changes sign to produce a level inversion and a change in a topological invariant, the envelope function approximation gives an eigenvalue equation analogous to Eq. (3.26). Integration in the spatial direction perpendicular to the interface will give a solution localized to the interface with energy $E = 0$ in the gap. In 2D and 3D, this state can be delocalized in the directions parallel to the interface (still with a linear Dirac dispersion in these directions) to give an edge mode in 2D or a surface state in 3D. In 3D materials, the linear Dirac dispersion of the surface state can be measured directly using angle-resolved photoemission spectroscopy (ARPES) (see Fig. 3.7).

These topological edge modes (in 2D) or surface states (in 3D) have many interesting experimental properties that are being actively explored. For 2D material with a nonzero Chern number, the edge modes are effectively 1-way streets that are always metallic because backscattering of the electron states can be forbidden. For 3D nonmagnetic topological insulators, the electrons in the surface state can have the unusual property that the direction of spin is locked to the direction of the in-plane

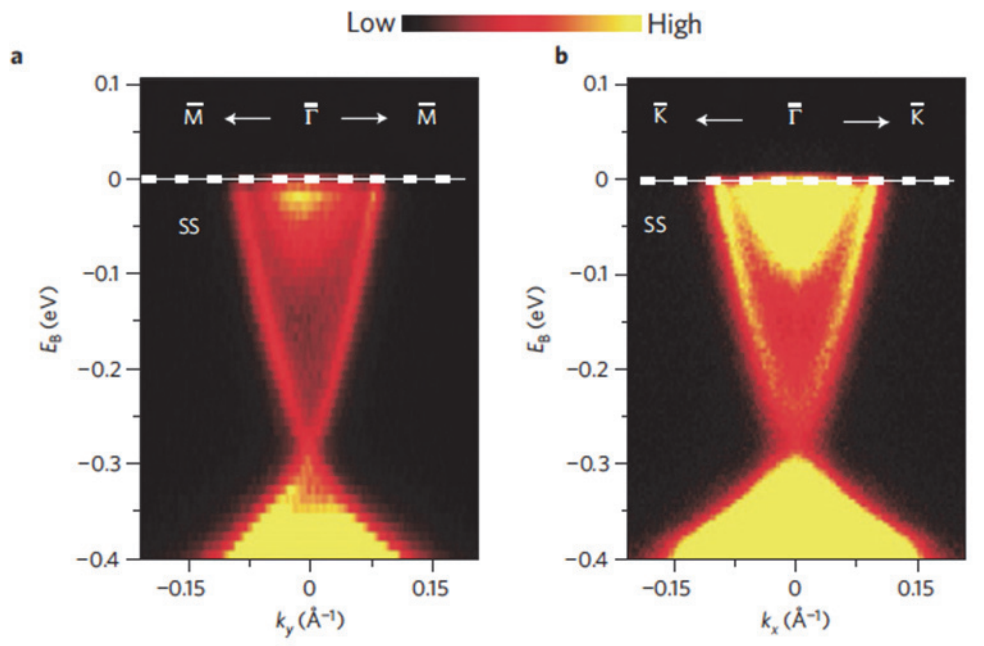

*Fig. 3.7. ARPES measurements of the surface electronic band dispersion on $Bi_2Se_3$(111), showing cuts in two directions through the Dirac point of the surface state. (from ref. [6])*

wavevector. This not only suppresses scattering processes but it also leads to interesting magneto-electric effects – because of this locking, an in-plane current (resulting form an imbalance in the occupation of momentum states) is necessarily accompanied by an interfacial spin accumulation (the resulting imbalance in the occupation of spin states).

***A brief comment about topological semi-metals***:

Finally, I'll mention another application of topology in solid state physics. The same mathematical framework that we have applied to topological insulators is also useful when thinking about Weyl and Dirac semimetals, but utilized in a quite different way. We introduced these classes of materials briefly when we discussed the experimental consequences of Berry curvature – they have 3D bands that touch at a finite number of points in ***k***-space, at each of which the Berry curvature can be singular, acting like a monopole source of flux in ***k***-space. In this case the relevant topological calculations are not integrations over the Brillouin zone – instead they are integrations over the Fermi surface, which is a 2D closed surface (or a collection of closed surfaces) within ***k***-space. If a Fermi surface encloses a band-touching point, the component of Berry curvature perpendicular to the Fermi surface integrated over the Fermi surface is a topological invariant, a different example of a Chern number, that can be nonzero if either time-reversal or inversion symmetry is broken. Just as for topological insulators, this nonzero topological invariant can have non-trivial consequences for surface states and magnetoelectric properties. A very nice review article about these interesting semimetals has recently been published by Armitage, Mele, and Vishwanath in *Rev. Mod. Phys*. [7].

**Acknowledgment**:
These notes are posted as a contribution to the broader impact of NSF grant DMR-1708499.

**Appendix 1: Derivation of the Rules of Band Dynamics Using Operator Algebra Techniques**

**I. Rule (#1) of band dynamics:** $\dfrac{d\langle \boldsymbol{k} \rangle}{dt} = \dfrac{1}{\hbar}\boldsymbol{F}_{\text{ext}}$.

(This discussion is adapted from Kittel, Appendix E. See also ref. 8.)

We will use that the crystal obeys lattice translation symmetry, so that the Hamiltonian $\hat{H}_0$ in the absence of any externally applied field commutes with the lattice translation operator, $\hat{T}$, defined so that $\hat{T}\psi(\boldsymbol{r}) = \psi(\boldsymbol{r}+\boldsymbol{a})$ where $\psi$ is an arbitrary wavefunction and $\boldsymbol{a}$ is a Bravais lattice vector. We will also use that over the short time $\delta t$ in which we want to compute the change in $\langle \boldsymbol{k} \rangle$ that the external force $\boldsymbol{F}$ can be approximated as a constant in both time and space.

If we operate with $\hat{T}$ on a Bloch wavefunction, we have by the property of Bloch wavefunctions that

$$\hat{T}\psi_{\boldsymbol{k}}(\boldsymbol{r}) = \psi_{\boldsymbol{k}}(\boldsymbol{r}+\boldsymbol{a}) = e^{i\boldsymbol{k}\cdot\boldsymbol{a}}\psi_{\boldsymbol{k}}(\boldsymbol{r}).$$

Therefore, the expectation value of the operator $\hat{T}$ in such a Bloch state is

$$\langle \hat{T} \rangle = \langle \psi_{\boldsymbol{k}} | \hat{T} | \psi_{\boldsymbol{k}} \rangle = e^{i\boldsymbol{k}\cdot\boldsymbol{a}}. \tag{A1.1}$$

If we take a time derivative of this expression, we have

$$\frac{d\langle \hat{T} \rangle}{dt} = e^{i\boldsymbol{k}\cdot\boldsymbol{a}} i\boldsymbol{a}\cdot\frac{d\boldsymbol{k}}{dt}, \tag{A1.2}$$

since the only factor that can depend on time is $\boldsymbol{k}$.

Now, from the general principles of quantum mechanics, we can also relate this same time dependence to an expectation value of the commutator of the Hamiltonian operator with the operator $\hat{T}$:

$$\frac{d\langle \hat{T} \rangle}{dt} = \frac{i}{\hbar}\langle [\hat{H},\hat{T}] \rangle, \tag{A1.3}$$

where here $\hat{H}$ is the full Hamiltonian including the external force: $\hat{H} = \hat{H}_0 - \boldsymbol{F}_{\text{ext}}\cdot\boldsymbol{r}$ where $\boldsymbol{F}_{\text{ext}}$ is an external force (assumed constant in our short time interval) and the bare crystal Hamiltonian commutes with $\hat{T}$: $[\hat{H}_0,\hat{T}] = 0$.

To evaluate the commutator, consider its action on any arbitrary function $\psi(\boldsymbol{r})$:

$$\begin{aligned}[\hat{H},\hat{T}]\psi(\boldsymbol{r}) &= (\hat{H}_0 - \boldsymbol{F}_{\text{ext}}\cdot\boldsymbol{r})\hat{T}\psi(\boldsymbol{r}) - \hat{T}(\hat{H}_0 - \boldsymbol{F}_{\text{ext}}\cdot\boldsymbol{r})\psi(\boldsymbol{r}) \\ &= \boldsymbol{F}_{\text{ext}}\cdot(-\boldsymbol{r}\hat{T}\psi(\boldsymbol{r}) + \hat{T}\boldsymbol{r}\psi(\boldsymbol{r})),\end{aligned}$$

where the last line follows since $\boldsymbol{F}_{\text{ext}}$ is a constant vector and $[\hat{H}_0,\hat{T}] = 0$. Then when we operate with $\hat{T}$,

$$
\begin{aligned}
[\hat{H},\hat{T}]\psi(\boldsymbol{r}) &= \boldsymbol{F}_{\text{ext}}\cdot(-\boldsymbol{r}\psi(\boldsymbol{r}+\boldsymbol{a})+(\boldsymbol{r}+\boldsymbol{a})\psi(\boldsymbol{r}+\boldsymbol{a})) \\
&= \boldsymbol{F}_{\text{ext}}\cdot\boldsymbol{a}\psi(\boldsymbol{r}+\boldsymbol{a}) \\
&= \boldsymbol{F}_{\text{ext}}\cdot\boldsymbol{a}\hat{T}\psi(\boldsymbol{r}).
\end{aligned}
$$

This implies that as an operator equation

$$[\hat{H},\hat{T}] = \boldsymbol{F}_{\text{ext}}\cdot\boldsymbol{a}\hat{T}.$$

Therefore, using Eqs. (A1.3) and (A1.1) we have

$$\frac{d\langle\hat{T}\rangle}{dt} = \frac{i}{\hbar}\langle[\hat{H},\hat{T}]\rangle = \frac{i}{\hbar}\langle\boldsymbol{F}_{\text{ext}}\cdot\boldsymbol{a}\hat{T}\rangle = \frac{i}{\hbar}\boldsymbol{F}_{\text{ext}}\cdot\boldsymbol{a}\langle\hat{T}\rangle = \frac{i}{\hbar}\boldsymbol{F}_{\text{ext}}\cdot\boldsymbol{a}e^{i\boldsymbol{k}\cdot\boldsymbol{a}}.$$

But from Eq. (A1.2), we also have that this same quantity is equal to

$$\frac{d\langle\hat{T}\rangle}{dt} = e^{i\boldsymbol{k}\cdot\boldsymbol{a}}i\boldsymbol{a}\cdot\frac{d\boldsymbol{k}}{dt}.$$

Since these expressions hold for arbitrary values of the Bravais lattice vector $\boldsymbol{a}$, we must have that

$$\frac{d\boldsymbol{k}}{dt} = \boldsymbol{F}_{\text{ext}}/\hbar.$$

This holds for every Bloch state which is included in the superposition to make the wavepacket. The force $\boldsymbol{F}$ is just a constant. Therefore, when evaluating the change in the expectation value $\langle\boldsymbol{k}\rangle$ for the full wavepacket the same expression holds:

$$\frac{d\langle\boldsymbol{k}\rangle}{dt} = \boldsymbol{F}_{\text{ext}}/\hbar.$$

This is the Rule (#1) that we wanted to demonstrate.

**II. Rule (#2) of band dynamics:** $\frac{d\langle\boldsymbol{r}\rangle}{dt} = \frac{1}{\hbar}\nabla_{\boldsymbol{k}}\varepsilon^n_\sigma(\boldsymbol{k})\Big|_{\boldsymbol{k}=\boldsymbol{k}_0} - \frac{\boldsymbol{F}_{\text{ext}}}{\hbar}\times\boldsymbol{\Omega}^n_\sigma(\boldsymbol{k}_0)$

In the following I will write Bloch states in the form $|\psi_k\rangle = e^{i\boldsymbol{k}\cdot\boldsymbol{r}}|n,\boldsymbol{k}\rangle$, where $|n,\boldsymbol{k}\rangle$ is the periodic part of the Bloch wavefunction and $n$ is a band index. I will use the normalization convention that $\langle n,\boldsymbol{k}|n,\boldsymbol{k}\rangle = 1$ when integrated over one unit cell.

We will analyze the problem within the Heisenberg representation of the position operator $\hat{\boldsymbol{r}}$ using first-order perturbation theory to analyze how the external force $\boldsymbol{F}_{\text{ext}}$ alters the eigenstate. With or without an external force, the operator that describes the time rate of change of the position of a wavepacket is the velocity operator

$$\hat{\boldsymbol{v}} = \frac{d\hat{\boldsymbol{r}}}{dt} = \frac{i}{\hbar}\left[\hat{H},\hat{\boldsymbol{r}}\right]. \tag{A1.4}$$

We want to evaluate $\langle\hat{\boldsymbol{v}}\rangle \equiv \langle\psi_{\boldsymbol{k}}|\hat{\boldsymbol{v}}|\psi_{\boldsymbol{k}}\rangle / \langle\psi_{\boldsymbol{k}}|\psi_{\boldsymbol{k}}\rangle = \langle n,\boldsymbol{k}|e^{-i\boldsymbol{k}\cdot\hat{\boldsymbol{r}}}\hat{\boldsymbol{v}}e^{i\boldsymbol{k}\cdot\hat{\boldsymbol{r}}}|n,\boldsymbol{k}\rangle = \langle n,\boldsymbol{k}|\hat{\boldsymbol{v}}_{\boldsymbol{k}}|n,\boldsymbol{k}\rangle$,

where we have defined $\hat{\boldsymbol{v}}_{\boldsymbol{k}} = e^{-i\boldsymbol{k}\cdot\hat{\boldsymbol{r}}}\hat{\boldsymbol{v}}e^{i\boldsymbol{k}\cdot\hat{\boldsymbol{r}}}$.

It is also convenient to define $\hat{H}_{\boldsymbol{k}} = e^{-i\boldsymbol{k}\cdot\hat{\boldsymbol{r}}}\hat{H}e^{i\boldsymbol{k}\cdot\hat{\boldsymbol{r}}}$ as we have in lecture 2, describing the effective Hamiltonian (within which the wavevector $\boldsymbol{k}$ is a parameter) for which $\left|n,\boldsymbol{k}\right\rangle$ is an eigenstate. With this definition, we can also write

$$\hat{\boldsymbol{v}}_{\boldsymbol{k}} = \frac{1}{\hbar}\nabla_{\boldsymbol{k}}\hat{H}_{\boldsymbol{k}} \tag{A1.5}$$

because

$$\frac{1}{\hbar}\nabla_{\boldsymbol{k}}\hat{H}_{\boldsymbol{k}} = \frac{1}{\hbar}\nabla_{\boldsymbol{k}}\left(e^{-i\boldsymbol{k}\cdot\hat{\boldsymbol{r}}}\hat{H}e^{i\boldsymbol{k}\cdot\hat{\boldsymbol{r}}}\right) = \frac{1}{\hbar}\left(-ie^{-i\boldsymbol{k}\cdot\hat{\boldsymbol{r}}}\hat{\boldsymbol{r}}\hat{H}e^{i\boldsymbol{k}\cdot\hat{\boldsymbol{r}}} + ie^{-i\boldsymbol{k}\cdot\hat{\boldsymbol{r}}}\hat{H}\hat{\boldsymbol{r}}e^{i\boldsymbol{k}\cdot\hat{\boldsymbol{r}}}\right) = e^{-i\boldsymbol{k}\cdot\hat{\boldsymbol{r}}}\frac{i}{\hbar}\left(\hat{H}\hat{\boldsymbol{r}} - \hat{\boldsymbol{r}}\hat{H}\right)e^{i\boldsymbol{k}\cdot\hat{\boldsymbol{r}}} = \hat{\boldsymbol{v}}_{\boldsymbol{k}}.$$

We also have that $\hat{\boldsymbol{v}}_{\boldsymbol{k}} = \frac{i}{\hbar}\left[\hat{H}_{\boldsymbol{k}},\hat{\boldsymbol{r}}\right]$ since $\hat{\boldsymbol{r}}$ and $e^{i\boldsymbol{k}\cdot\hat{\boldsymbol{r}}}$ commute.

I will present two arguments. The first is the operator analysis that is most commonly presented, and it gives the correct answer, but it is not entirely rigorous. In the second argument, I will sketch how a more-rigorous calculation goes.

For the first argument, let us apply standard first-order perturbation theory to evaluate the expectation value of the velocity

$$\left\langle\hat{\boldsymbol{v}}\right\rangle = \left\langle n,\boldsymbol{k}\right|\hat{\boldsymbol{v}}_{\boldsymbol{k}}\left|n,\boldsymbol{k}\right\rangle = \frac{1}{\hbar}\left\langle n,\boldsymbol{k}\right|\nabla_{\boldsymbol{k}}\hat{H}_{\boldsymbol{k}}\left|n,\boldsymbol{k}\right\rangle$$

under the circumstances that the eigenstates are perturbed by a constant external force $\boldsymbol{F}_{\text{ext}}$; *i.e.*, we will perform perturbation theory with the perturbation $\Delta U = -\boldsymbol{F}_{\text{ext}}\cdot\hat{\boldsymbol{r}}$.

The first-order change in the eigenstates is $\left|n,\boldsymbol{k}\right\rangle \to \left|n,\boldsymbol{k}\right\rangle + \delta\left|n,\boldsymbol{k}\right\rangle$ with

$$\delta\left|n,\boldsymbol{k}\right\rangle = \sum_{m\neq n}\left|m,\boldsymbol{k}\right\rangle\frac{\left\langle m,\boldsymbol{k}\right|-\boldsymbol{F}_{\text{ext}}\cdot\hat{\boldsymbol{r}}\left|n,\boldsymbol{k}\right\rangle}{E_{n,\boldsymbol{k}} - E_{m,\boldsymbol{k}}} = -\boldsymbol{F}_{\text{ext}}\cdot\sum_{m\neq n}\left|m,\boldsymbol{k}\right\rangle\frac{\left\langle m,\boldsymbol{k}\right|\hat{\boldsymbol{r}}\left|n,\boldsymbol{k}\right\rangle}{E_{n,\boldsymbol{k}} - E_{m,\boldsymbol{k}}}.$$

There is a convenient trick for evaluating the matrix element of the position operator. Since $\hat{\boldsymbol{v}}_k = \frac{i}{\hbar}\left[\hat{H}_{\boldsymbol{k}},\hat{\boldsymbol{r}}\right]$, we have

$$\left\langle m,\boldsymbol{k}\right|\hat{\boldsymbol{v}}_k\left|n,\boldsymbol{k}\right\rangle = \frac{i}{\hbar}\left\langle m,\boldsymbol{k}\right|\left(\hat{H}_{\boldsymbol{k}}\hat{\boldsymbol{r}} - \hat{\boldsymbol{r}}\hat{H}_{\boldsymbol{k}}\right)\left|n,\boldsymbol{k}\right\rangle = \frac{i}{\hbar}\left(E_{m,\boldsymbol{k}} - E_{n,\boldsymbol{k}}\right)\left\langle m,\boldsymbol{k}\right|\hat{\boldsymbol{r}}\left|n,\boldsymbol{k}\right\rangle,$$

and therefore $\left\langle m,\boldsymbol{k}\right|\hat{\boldsymbol{r}}\left|n,\boldsymbol{k}\right\rangle = \frac{\hbar}{i}\frac{\left\langle m,\boldsymbol{k}\right|\hat{\boldsymbol{v}}_k\left|n,\boldsymbol{k}\right\rangle}{(E_{m.\boldsymbol{k}} - E_{n,\boldsymbol{k}})}$.

Consequently, for the perturbed eigenstate we have

$$\delta\left|n,\boldsymbol{k}\right\rangle = -\boldsymbol{F}_{\text{ext}}\cdot\frac{\hbar}{-i}\sum_{m\neq n}\left|m,\boldsymbol{k}\right\rangle\frac{\left\langle m,\boldsymbol{k}\right|\boldsymbol{v}_k\left|n,\boldsymbol{k}\right\rangle}{(E_{n,\boldsymbol{k}}-E_{m,\boldsymbol{k}})^2} = -\boldsymbol{F}_{\text{ext}}\cdot i\sum_{m\neq n}\left|m,\boldsymbol{k}\right\rangle\frac{\left\langle m,\boldsymbol{k}\right|\nabla_k H_k\left|n,\boldsymbol{k}\right\rangle}{(E_{n,\boldsymbol{k}}-E_{m,\boldsymbol{k}})^2}.$$

In the last step we have used Eq. (A1.5). Evaluating the expectation value of the velocity operator in the perturbed eigenstates, to first order we have (here $\left|n,\boldsymbol{k}\right\rangle$ and $\left|m,\boldsymbol{k}\right\rangle$ are unperturbed eigenstates)

$$\begin{aligned}
\left\langle\hat{\boldsymbol{v}}\right\rangle &= \frac{1}{\hbar}\left(\left\langle n,\boldsymbol{k}\right| + \delta\left\langle n,\boldsymbol{k}\right|\right)\nabla_{\boldsymbol{k}}\hat{H}_{\boldsymbol{k}}\left(\left|n,\boldsymbol{k}\right\rangle + \delta\left|n,\boldsymbol{k}\right\rangle\right)\\
&= \frac{1}{\hbar}\left\langle n,\boldsymbol{k}\right|\nabla_{\boldsymbol{k}}\hat{H}_{\boldsymbol{k}}\left|n,\boldsymbol{k}\right\rangle + \frac{1}{\hbar}\left\langle n,\boldsymbol{k}\right|\nabla_{\boldsymbol{k}}\hat{H}_{\boldsymbol{k}}\left(\delta\left|n,\boldsymbol{k}\right\rangle\right) + \frac{1}{\hbar}\left(\delta\left\langle n,\boldsymbol{k}\right|\right)\nabla_{\boldsymbol{k}}\hat{H}_{\boldsymbol{k}}\left|n,\boldsymbol{k}\right\rangle\\
&= \frac{1}{\hbar}\left\langle n,\boldsymbol{k}\right|\nabla_{\boldsymbol{k}}\hat{H}_{\boldsymbol{k}}\left|n,\boldsymbol{k}\right\rangle + \frac{1}{\hbar}i\sum_{m\neq n}\frac{\left\langle n,\boldsymbol{k}\right|\nabla_{\boldsymbol{k}}\hat{H}_{\boldsymbol{k}}\left|m,\boldsymbol{k}\right\rangle\left[-\boldsymbol{F}_{\text{ext}}\cdot\left\langle m,\boldsymbol{k}\right|\nabla_{\boldsymbol{k}}\hat{H}_{\boldsymbol{k}}\left|n,\boldsymbol{k}\right\rangle\right]}{(E_{n,\boldsymbol{k}}-E_{m,\boldsymbol{k}})^2}\\
&\quad -\frac{1}{\hbar}i\sum_{m\neq n}\frac{\left\langle m,\boldsymbol{k}\right|\nabla_{\boldsymbol{k}}\hat{H}_{\boldsymbol{k}}\left|n,\boldsymbol{k}\right\rangle\left[-\boldsymbol{F}_{\text{ext}}\cdot\left\langle n,\boldsymbol{k}\right|\nabla_{\boldsymbol{k}}\hat{H}_{\boldsymbol{k}}\left|m,\boldsymbol{k}\right\rangle\right]}{(E_{n,\boldsymbol{k}}-E_{m,\boldsymbol{k}})^2}\\
&= \frac{1}{\hbar}\left\langle n,\boldsymbol{k}\right|\nabla_{\boldsymbol{k}}\hat{H}_{\boldsymbol{k}}\left|n,\boldsymbol{k}\right\rangle - \frac{1}{\hbar}\boldsymbol{F}_{\text{ext}}\times\left[i\sum_{m\neq n}\frac{\left\langle n,\boldsymbol{k}\right|\nabla_{\boldsymbol{k}}\hat{H}_{\boldsymbol{k}}\left|m,\boldsymbol{k}\right\rangle\times\left\langle m,\boldsymbol{k}\right|\nabla_{\boldsymbol{k}}\hat{H}_{\boldsymbol{k}}\left|n,\boldsymbol{k}\right\rangle}{(E_{n,\boldsymbol{k}}-E_{m,\boldsymbol{k}})^2}\right].
\end{aligned} \tag{A1.6}$$

In the last step we have used the vector identity $\boldsymbol{a}\times(\boldsymbol{b}\times\boldsymbol{c}) = \boldsymbol{b}(\boldsymbol{a}\cdot\boldsymbol{c}) - \boldsymbol{c}(\boldsymbol{a}\cdot\boldsymbol{b})$.

But the quantity is square brackets is just the Berry curvature:

$$\boldsymbol{\Omega}^n_\sigma(\boldsymbol{k}) = i\sum_{m\neq n}\frac{\left\langle n,\boldsymbol{k},\boldsymbol{\sigma}\right|\left(\nabla_{\boldsymbol{k}}\hat{H}_{\boldsymbol{k}}\right)\left|m,\boldsymbol{k},\boldsymbol{\sigma}\right\rangle\times\left\langle m,\boldsymbol{k},\boldsymbol{\sigma}\right|\left(\nabla_{\boldsymbol{k}}\hat{H}_{\boldsymbol{k}}\right)\left|n,\boldsymbol{k},\boldsymbol{\sigma}\right\rangle}{(E_{m,\boldsymbol{k}}-E_{n,\boldsymbol{k}})^2},$$

where we have re-introduced the spin label, $\boldsymbol{\sigma}$. (See Eq. (A3.4) in Appendix 3 for this expression of the Berry curvature.)

The first term in the last line of Eq. (A1.6) can be identified as $\frac{1}{\hbar}\nabla_{\boldsymbol{k}}\varepsilon^n_\sigma(\boldsymbol{k})$ since

$$\begin{aligned}
\frac{1}{\hbar}\nabla_{\boldsymbol{k}}\varepsilon^n_\sigma(\boldsymbol{k}) &= \frac{1}{\hbar}\nabla_{\boldsymbol{k}}\left(\left\langle n,\boldsymbol{k},\boldsymbol{\sigma}\right|\hat{H}_{\boldsymbol{k}}\left|n,\boldsymbol{k},\boldsymbol{\sigma}\right\rangle\right)\\
&= \frac{1}{\hbar}\left(\nabla_{\boldsymbol{k}}\left\langle n,\boldsymbol{k},\boldsymbol{\sigma}\right|\right)\hat{H}_{\boldsymbol{k}}\left|n,\boldsymbol{k},\boldsymbol{\sigma}\right\rangle + \frac{1}{\hbar}\left\langle n,\boldsymbol{k},\boldsymbol{\sigma}\right|\left(\nabla_{\boldsymbol{k}}\hat{H}_{\boldsymbol{k}}\right)\left|n,\boldsymbol{k},\boldsymbol{\sigma}\right\rangle + \frac{1}{\hbar}\left\langle n,\boldsymbol{k},\boldsymbol{\sigma}\right|\hat{H}_{\boldsymbol{k}}\left(\nabla_{\boldsymbol{k}}\left|n,\boldsymbol{k},\boldsymbol{\sigma}\right\rangle\right)\\
&= \frac{1}{\hbar}E_{n,\boldsymbol{k}}\left(\nabla_{\boldsymbol{k}}\left\langle n,\boldsymbol{k},\boldsymbol{\sigma}\right|\right)\left|n,\boldsymbol{k},\boldsymbol{\sigma}\right\rangle + \frac{1}{\hbar}\left\langle n,\boldsymbol{k},\boldsymbol{\sigma}\right|\left(\nabla_{\boldsymbol{k}}\hat{H}_{\boldsymbol{k}}\right)\left|n,\boldsymbol{k},\boldsymbol{\sigma}\right\rangle + \frac{1}{\hbar}E_{n,\boldsymbol{k}}\left\langle n,\boldsymbol{k},\boldsymbol{\sigma}\right|\left(\nabla_{\boldsymbol{k}}\left|n,\boldsymbol{k},\boldsymbol{\sigma}\right\rangle\right)\\
&= \frac{1}{\hbar}\left\langle n,\boldsymbol{k},\boldsymbol{\sigma}\right|\left(\nabla_{\boldsymbol{k}}\hat{H}_{\boldsymbol{k}}\right)\left|n,\boldsymbol{k},\boldsymbol{\sigma}\right\rangle + \frac{1}{\hbar}E_{n,\boldsymbol{k}}\nabla_{\boldsymbol{k}}\left(\left\langle n,\boldsymbol{k},\boldsymbol{\sigma}\middle|n,\boldsymbol{k},\boldsymbol{\sigma}\right\rangle\right)
\end{aligned}$$

and $\nabla_{\boldsymbol{k}}\left(\left\langle n,\boldsymbol{k},\boldsymbol{\sigma}\middle|n,\boldsymbol{k},\boldsymbol{\sigma}\right\rangle\right) = 0$ because $\left\langle n,\boldsymbol{k},\boldsymbol{\sigma}\middle|n,\boldsymbol{k},\boldsymbol{\sigma}\right\rangle$ is identically 1.

So we have, finally

$$\langle \hat{\boldsymbol{v}} \rangle = \frac{d\langle \hat{\boldsymbol{r}} \rangle}{dt} = \frac{1}{\hbar}\nabla_{\boldsymbol{k}}\varepsilon_{\sigma}^{n}(\boldsymbol{k}) - \frac{\boldsymbol{F}_{\text{ext}}}{\hbar}\times \boldsymbol{\Omega}_{\sigma}^{n}(\boldsymbol{k}).$$

This is Rule (#2) that we wanted to demonstrate.

This is the correct answer, but it is not entirely rigorous, because strictly speaking $\hat{\boldsymbol{r}}$ is not a well-defined operator in the space of plane-wave-like Bloch states and also the perturbation $\Delta U = -\boldsymbol{F}_{\text{ext}} \cdot \hat{\boldsymbol{r}}$ is not small for an unbounded sample. For those who are interested, a more-rigorous second argument can begin by analyzing more generally how the periodic part of a Bloch state changes when ***dk/dt*** is nonzero due to any kind of external force. One can use a form of perturbation theory to include a first-order correction in ***dk/dt*** to the adiabatic approximation. (See ref. [4], pages 97-101.) The time-dependent wavefunction $|n,\boldsymbol{k},t\rangle$ can then be written in terms of the unperturbed wavefunctions $|n,\boldsymbol{k}\rangle$ in the form

$$|n,\boldsymbol{k},t\rangle = e^{i\int A_n(t)dt}e^{-iE(t)/\hbar}\left[|n,\boldsymbol{k}\rangle + \frac{\hbar}{i}\frac{d\boldsymbol{k}}{dt}\cdot\sum_{m\neq n}|m,\boldsymbol{k}\rangle\frac{\langle m,\boldsymbol{k}|\nabla_{\boldsymbol{k}}\hat{H}_{\boldsymbol{k}}|n,\boldsymbol{k}\rangle}{\left(E_{n,\boldsymbol{k}}-E_{m,\boldsymbol{k}}\right)^2}\right].$$

The expectation value of the velocity is (keeping terms up to first order in *dk/dt*)

$$\begin{aligned}\langle \hat{\boldsymbol{v}} \rangle &= \frac{1}{\hbar}\langle n,\boldsymbol{k},t|\nabla_{\boldsymbol{k}}\hat{H}_{\boldsymbol{k}}|n,\boldsymbol{k},t\rangle \\ &= \frac{1}{\hbar}\langle n,\boldsymbol{k}|\nabla_{\boldsymbol{k}}\hat{H}_{\boldsymbol{k}}|n,\boldsymbol{k}\rangle - i\sum_{m\neq n}\frac{\langle n,\boldsymbol{k}|\nabla_{\boldsymbol{k}}\hat{H}_{\boldsymbol{k}}|m,\boldsymbol{k}\rangle\left[\left(d\boldsymbol{k}/dt\right)\cdot\langle m,\boldsymbol{k}|\nabla_{\boldsymbol{k}}\hat{H}_{\boldsymbol{k}}|n,\boldsymbol{k}\rangle\right]}{(E_{n,\boldsymbol{k}}-E_{m,\boldsymbol{k}})^2} \\ &\quad + i\sum_{m\neq n}\frac{\left[\left(d\boldsymbol{k}/dt\right)\cdot\langle n,\boldsymbol{k}|\nabla_{\boldsymbol{k}}\hat{H}_{\boldsymbol{k}}|m,\boldsymbol{k}\rangle\right]\langle m,\boldsymbol{k}|\nabla_{\boldsymbol{k}}\hat{H}_{\boldsymbol{k}}|n,\boldsymbol{k}\rangle}{(E_{n,\boldsymbol{k}}-E_{m,\boldsymbol{k}})^2}\end{aligned}$$

$$= \frac{1}{\hbar}\langle n,\boldsymbol{k}|\nabla_{\boldsymbol{k}}\hat{H}_{\boldsymbol{k}}|n,\boldsymbol{k}\rangle - \frac{d\boldsymbol{k}}{dt}\times\left[i\sum_{m\neq n}\frac{\langle n,\boldsymbol{k}|\nabla_{\boldsymbol{k}}\hat{H}_{\boldsymbol{k}}|m,\boldsymbol{k}\rangle\times\langle m,\boldsymbol{k}|\nabla_{\boldsymbol{k}}\hat{H}_{\boldsymbol{k}}|n,\boldsymbol{k}\rangle}{(E_{n,\boldsymbol{k}}-E_{m,\boldsymbol{k}})^2}\right],$$

where again in the last step we have used the vector identity $\boldsymbol{a}\times\left(\boldsymbol{b}\times\boldsymbol{c}\right)=\boldsymbol{b}\left(\boldsymbol{a}\cdot\boldsymbol{c}\right)-\boldsymbol{c}\left(\boldsymbol{a}\cdot\boldsymbol{b}\right)$.

Therefore we reach the same result as before:

$$\langle \hat{\boldsymbol{v}} \rangle = \frac{1}{\hbar}\nabla_{k}\varepsilon_{\sigma}^{n}(\boldsymbol{k}) - \frac{d\boldsymbol{k}}{dt}\times \boldsymbol{\Omega}_{\sigma}^{n}(\boldsymbol{k}).$$

## Appendix 2: Calculating the expectation value of position for wavepacket made from a superposition of Bloch waves.

I will consider a wavepacket defined by a weighting function with the general form

$$w(\boldsymbol{k}-\boldsymbol{k}_0) = w_R(\boldsymbol{k}-\boldsymbol{k}_0)e^{-i\boldsymbol{k}\cdot\boldsymbol{r}_0}e^{+i(\boldsymbol{k}-\boldsymbol{k}_0)\cdot\boldsymbol{X}} \tag{A2.1}$$

where $w_R(\boldsymbol{k}-\boldsymbol{k}_0)$ is a normalized real-valued function that is sharply-peaked around $\boldsymbol{k}=\boldsymbol{k}_0$ and zero elsewhere. The coefficient $\boldsymbol{X}$ serves as a phase correction factor as discussed in the main lecture notes. We will use the normalization convention that the periodic part of the Bloch wavefunction is normalized within one unit cell:

$$\int_{\text{unit cell}} \left|u_{\boldsymbol{k}}(\boldsymbol{r})\right|^2 d^3\boldsymbol{r} = 1 \ . \tag{A2.2}$$

To keep track of all normalizations, it is easier to do the calculation assuming a finite crystal with periodic boundary conditions and $N$ unit cells in total. The full normalized wavepacket is then

$$\begin{aligned}\Psi_{WP,0} &= \frac{1}{\sqrt{N}} \sum_{\boldsymbol{k}} w_R(\boldsymbol{k}-\boldsymbol{k}_0)e^{i(\boldsymbol{k}-\boldsymbol{k}_0)\cdot\boldsymbol{X}}e^{i\boldsymbol{k}\cdot(\boldsymbol{r}-\boldsymbol{r}_0)}u_{\boldsymbol{k}}(\boldsymbol{r}) \\ &= \frac{1}{\sqrt{N}} e^{i\boldsymbol{k}_0\cdot(\boldsymbol{r}-\boldsymbol{r}_0)}\sum_{\boldsymbol{k}} w_R(\boldsymbol{k}-\boldsymbol{k}_0)e^{i(\boldsymbol{k}-\boldsymbol{k}_0)\cdot\left[\boldsymbol{X}+\boldsymbol{r}-\boldsymbol{r}_0\right]}u_{\boldsymbol{k}}(\boldsymbol{r}).\end{aligned} \tag{A2.3}$$

The final results will apply also for the continuum limit

For mathematical convenience, I will actually calculate the expectation value of $\boldsymbol{r}-\boldsymbol{r}_0$:

$$\begin{aligned}\left\langle \boldsymbol{r}-\boldsymbol{r}_0\right\rangle &= \int d^3\boldsymbol{r}\Psi^*_{WP,0}(\boldsymbol{r})\left(\boldsymbol{r}-\boldsymbol{r}_0\right)\Psi_{WP,0}(\boldsymbol{r}) \\ &= \frac{1}{N}\int d^3\boldsymbol{r} \sum_{\boldsymbol{k},\boldsymbol{k}'} w_R(\boldsymbol{k}'-\boldsymbol{k}_0)w_R(\boldsymbol{k}-\boldsymbol{k}_0)u^*_{\boldsymbol{k}'}(\boldsymbol{r})u_{\boldsymbol{k}}(\boldsymbol{r})e^{-i(\boldsymbol{k}'-\boldsymbol{k}_0)\cdot\left[\boldsymbol{X}+\boldsymbol{r}-\boldsymbol{r}_0\right]}e^{i(\boldsymbol{k}-\boldsymbol{k}_0)\cdot\left[\boldsymbol{X}+\boldsymbol{r}-\boldsymbol{r}_0\right]}\left(\boldsymbol{r}-\boldsymbol{r}_0\right) \\ &= \frac{1}{i}\frac{1}{N}\int d^3\boldsymbol{r} \sum_{\boldsymbol{k},\boldsymbol{k}'} w_R(\boldsymbol{k}'-\boldsymbol{k}_0)w_R(\boldsymbol{k}-\boldsymbol{k}_0)u^*_{\boldsymbol{k}'}(\boldsymbol{r})u_{\boldsymbol{k}}(\boldsymbol{r})e^{-i(\boldsymbol{k}'-\boldsymbol{k}_0)\cdot\left[\boldsymbol{X}+\boldsymbol{r}-\boldsymbol{r}_0\right]}e^{i(\boldsymbol{k}-\boldsymbol{k}_0)\cdot\left[\boldsymbol{X}\right]}\left(\nabla_{\boldsymbol{k}}e^{i(\boldsymbol{k}-\boldsymbol{k}_0)\cdot(\boldsymbol{r}-\boldsymbol{r}_0)}\right).\end{aligned} \tag{A2.4}$$

At this point I will integrate by parts.

$$\begin{aligned}&= -\frac{1}{i}\frac{1}{N}\int d^3\boldsymbol{r} \sum_{\boldsymbol{k},\boldsymbol{k}'} w_R(\boldsymbol{k}'-\boldsymbol{k}_0)u^*_{\boldsymbol{k}'}(\boldsymbol{r})e^{-i(\boldsymbol{k}'-\boldsymbol{k}_0)\cdot\left[\boldsymbol{X}+\boldsymbol{r}-\boldsymbol{r}_0\right]}e^{i(\boldsymbol{k}-\boldsymbol{k}_0)\cdot(\boldsymbol{r}-\boldsymbol{r}_0)}\left(\nabla_{\boldsymbol{k}}\left[w_R(\boldsymbol{k}-\boldsymbol{k}_0)u_{\boldsymbol{k}}(\boldsymbol{r})e^{i(\boldsymbol{k}-\boldsymbol{k}_0)\cdot\left[\boldsymbol{X}\right]}\right]\right) \\ &= -\frac{1}{i}\frac{1}{N}\int d^3\boldsymbol{r} \sum_{\boldsymbol{k},\boldsymbol{k}'} w_R(\boldsymbol{k}'-\boldsymbol{k}_0)u^*_{\boldsymbol{k}'}(\boldsymbol{r})e^{i(\boldsymbol{k}-\boldsymbol{k}')\cdot(\boldsymbol{r}-\boldsymbol{r}_0)}e^{-i(\boldsymbol{k}'-\boldsymbol{k}_0)\cdot\left[\boldsymbol{X}\right]}\left(\nabla_{\boldsymbol{k}}\left[w_R(\boldsymbol{k}-\boldsymbol{k}_0)u_{\boldsymbol{k}}(\boldsymbol{r})e^{i(\boldsymbol{k}-\boldsymbol{k}_0)\cdot\left[\boldsymbol{X}\right]}\right]\right).\end{aligned} \tag{A2.5}$$

The integral over $\boldsymbol{r}$ is similar to the one we have considered in the context of x-ray scattering. It is zero unless $\boldsymbol{k}'=\boldsymbol{k}$ because otherwise the contributions from different unit cells add up out of phase to give perfect destructive interference. We can thus divide the integral over $\boldsymbol{r}$ up into a lattice sum times structure factor (that is, an integral over one unit cell), with the lattice sum = $N\delta_{\boldsymbol{k},\boldsymbol{k}'}$.

$$\left\langle \boldsymbol{r}-\boldsymbol{r}_0 \right\rangle = -\frac{1}{i} \int_{\text{unit cell}} d^3\boldsymbol{r} \sum_{\boldsymbol{k},\boldsymbol{k}'} \delta_{\boldsymbol{k},\boldsymbol{k}'} w_R(\boldsymbol{k}'-\boldsymbol{k}_0) u^*_{\boldsymbol{k}'}(\boldsymbol{r}) e^{i(\boldsymbol{k}-\boldsymbol{k}')\cdot(\boldsymbol{r}-\boldsymbol{r}_0)} e^{-i(\boldsymbol{k}'-\boldsymbol{k}_0)\cdot[X]} \left( \nabla_{\boldsymbol{k}} \left[ w_R(\boldsymbol{k}-\boldsymbol{k}_0) u_{\boldsymbol{k}}(\boldsymbol{r}) e^{i(\boldsymbol{k}-\boldsymbol{k}_0)\cdot[X]} \right] \right)$$

$$= -\frac{1}{i} \int_{\text{unit cell}} d^3\boldsymbol{r} \sum_{\boldsymbol{k}} w_R(\boldsymbol{k}-\boldsymbol{k}_0) u^*_{\boldsymbol{k}}(\boldsymbol{r}) e^{-i(\boldsymbol{k}-\boldsymbol{k}_0)\cdot[X]} \left( \nabla_{\boldsymbol{k}} \left[ w_R(\boldsymbol{k}-\boldsymbol{k}_0) u_{\boldsymbol{k}}(\boldsymbol{r}) e^{i(\boldsymbol{k}-\boldsymbol{k}_0)\cdot[X]} \right] \right). \tag{A2.6}$$

Here we take the gradient of the three terms in the square brackets and evaluate them at $\boldsymbol{k} = \boldsymbol{k}_0$, because the sharply-peaked weighting function $w_R(\boldsymbol{k}-\boldsymbol{k}_0)$ makes the integrand zero everywhere except near $\boldsymbol{k} = \boldsymbol{k}_0$. There is no contribution from $\nabla_{\boldsymbol{k}} w_R(\boldsymbol{k}-\boldsymbol{k}_0)$ because $w_R(\boldsymbol{k}-\boldsymbol{k}_0)$ has a maximum (with zero gradient) at $\boldsymbol{k} = \boldsymbol{k}_0$. So there are just two contributions from the gradient:

$$\left\langle \boldsymbol{r}-\boldsymbol{r}_0 \right\rangle = -\frac{1}{i} \int_{\text{unit cell}} d^3\boldsymbol{r} \sum_{\boldsymbol{k}} \left| w_R(\boldsymbol{k}-\boldsymbol{k}_0) \right|^2 u^*_{\boldsymbol{k}}(\boldsymbol{r}) e^{-i(\boldsymbol{k}-\boldsymbol{k}_0)\cdot[X]} \left( e^{i(\boldsymbol{k}-\boldsymbol{k}_0)\cdot[X]} \nabla_{\boldsymbol{k}} u_{\boldsymbol{k}}(\boldsymbol{r}) + u_{\boldsymbol{k}}(\boldsymbol{r}) e^{i(\boldsymbol{k}-\boldsymbol{k}')\cdot[X]} (i\boldsymbol{X}) \right)$$

$$= -\frac{1}{i} \left[ \left( \sum_{\boldsymbol{k}} \left| w_R(\boldsymbol{k}-\boldsymbol{k}_0) \right|^2 \int_{\text{unit cell}} d^3\boldsymbol{r} \left( u^*_{\boldsymbol{k}}(\boldsymbol{r}) \nabla_{\boldsymbol{k}} u_{\boldsymbol{k}}(\boldsymbol{r}) \right) \Big|_{\boldsymbol{k}=\boldsymbol{k}_0} \right) + \left( i\boldsymbol{X} \sum_{\boldsymbol{k}} \left| w_R(\boldsymbol{k}-\boldsymbol{k}_0) \right|^2 \int_{\text{unit cell}} \left| u_{\boldsymbol{k}}(\boldsymbol{r}) \right|^2 d^3\boldsymbol{r} \right) \right]. \tag{A2.7}$$

Since both $\left| w_R(\boldsymbol{k}-\boldsymbol{k}_0) \right|^2$ and $\left| u_{\boldsymbol{k}}(\boldsymbol{r}) \right|^2$ are normalized,

$$\begin{aligned} \left\langle \boldsymbol{r}-\boldsymbol{r}_0 \right\rangle &= -\frac{1}{i} \left[ \int_{\text{unit cell}} d^3\boldsymbol{r} \left( u^*_{\boldsymbol{k}}(\boldsymbol{r}) \nabla_{\boldsymbol{k}} u_{\boldsymbol{k}}(\boldsymbol{r}) \right) \Big|_{\boldsymbol{k}=\boldsymbol{k}_0} + i\boldsymbol{X} \right] \\ &= i \int_{\text{unit cell}} d^3\boldsymbol{r} \left( u^*_{\boldsymbol{k}}(\boldsymbol{r}) \frac{\partial}{\partial \boldsymbol{k}} u_{\boldsymbol{k}}(\boldsymbol{r}) \right) \Bigg|_{\boldsymbol{k}=\boldsymbol{k}_0} - \boldsymbol{X} \\ &= \boldsymbol{A}(\boldsymbol{k}_0) - \boldsymbol{X}. \end{aligned} \tag{A2.8}$$

## Appendix 3: Alternative expressions for Berry curvature

In the following, I use the notation $\left|n\right\rangle$ or $\left|m\right\rangle$ to refer to energy eigenstates of the same Hamiltonian, with that Hamiltonian having the same value of some external parameter (*e.g.*, the parameter $\boldsymbol{\lambda}$ or $\boldsymbol{k}$ in the notes). The gradients $\nabla$ are taken with respect to this parameter $\boldsymbol{\lambda}$ or $\boldsymbol{k}$.

The general formula for the Berry curvature can also be transformed into additional sometimes-more-convenient forms:

$$\Omega_n \equiv \nabla \times i\left\langle n\middle|\nabla\middle|n\right\rangle \tag{A3.1}$$

$$= i\left(\nabla\left\langle n\right|\right)\times\left(\nabla\left|n\right\rangle\right) \tag{A3.2}$$

$$= -i\sum_{m\neq n}\left\langle n\middle|\nabla\middle|m\right\rangle\times\left\langle m\middle|\nabla\middle|n\right\rangle \tag{A3.3}$$

$$= i\sum_{m\neq n}\frac{\left\langle n\right|\left(\nabla H\right)\left|m\right\rangle\times\left\langle m\right|\left(\nabla H\right)\left|n\right\rangle}{(E_m-E_n)^2}. \tag{A3.4}$$

Auation (A3.1) is the definition of the Berry curvature. Equation (A3.2) can be shown easily by expanding both Eq. (A3.1) and Eq. (A3.2) and comparing terms. Equation (A3.3) follows from Eq. (A3.2) by inserting a complete set of states plus some algebra, as follows:
Inserting a complete set of energy eigenstates $\left|m\right\rangle$ into Eq. (A3.2) yields

$$\Omega_n = i\sum_m\left(\nabla\left\langle n\right|\right)\left[\left|m\right\rangle\left\langle m\right|\right]\times\left(\nabla\left|n\right\rangle\right) = i\sum_m\left(\nabla\left\langle n\right|\right)\left|m\right\rangle\times\left\langle m\middle|\nabla\middle|n\right\rangle. \tag{A3.5}$$

If the states $\left|n\right\rangle$ and $\left|m\right\rangle$ are different they are orthogonal, so that $\left\langle n\middle|m\right\rangle = 0$. If they are the same, then $\left\langle n\middle|m\right\rangle = 1$. In either case, taking the gradient of $\left\langle n\middle|m\right\rangle$ gives zero, so that

$$0 = \nabla\left\langle n\middle|m\right\rangle = \left(\nabla\left\langle n\right|\right)\left|m\right\rangle + \left\langle n\middle|\nabla\middle|m\right\rangle. \tag{A3.6}$$

Therefore

$$\left(\nabla\left\langle n\right|\right)\left|m\right\rangle = -\left\langle n\middle|\nabla\middle|m\right\rangle \tag{A3.7}$$

and we can substitute this into Eq. (A3.5) to yield

$$\Omega_n = -i\sum_m\left\langle n\middle|\nabla\middle|m\right\rangle\times\left\langle m\middle|\nabla\middle|n\right\rangle. \tag{A3.8}$$

We can restrict the sum to $m \neq n$ since the $m = n$ term gives a cross product that is identically zero. This proves Eq. (A3.3).

The final expression for the Berry curvature, Eq. (A3.4) results from using in Eq. (A3.3) the identity

$$\left\langle n\middle|\nabla\middle|m\right\rangle = \frac{\left\langle n\right|\left(\nabla H\right)\left|m\right\rangle}{E_m - E_n} \quad \text{for } n \neq m\ . \tag{A3.9}$$

This identity follows from first noting that for an energy eigenstate $\left|m\right\rangle$

$$\nabla\left\langle n\middle|H\middle|m\right\rangle = \nabla\left[E_m\left\langle n\,|\,m\right\rangle\right] = 0 \tag{A3.10}$$

(where in the last step we have used Eq. (A3.6)).

But then using Eq. (A3.10) we can also evaluate

$$\begin{aligned}0 &= \nabla\left\langle n\middle|H\middle|m\right\rangle = \left(\nabla\left\langle n\right|\right)H\left|m\right\rangle + \left\langle n\middle|\left(\nabla H\right)\middle|m\right\rangle + \left\langle n\right|H\left(\nabla\left|m\right\rangle\right) \\ &= E_m\left(\nabla\left\langle n\right|\right)\left|m\right\rangle + \left\langle n\middle|\left(\nabla H\right)\middle|m\right\rangle + E_n\left\langle n\right|\left(\nabla\left|m\right\rangle\right) \\ &= -E_m\left\langle n\middle|\nabla\middle|m\right\rangle + \left\langle n\middle|\left(\nabla H\right)\middle|m\right\rangle + E_n\left\langle n\middle|\nabla\middle|m\right\rangle,\end{aligned} \tag{A3.11}$$

where in the final step we have used Eq. (A3.7). The final equality in Eq. (A3.11) is equivalent to Eq. (A3.9). Substituting Eq. (A3.9) twice into Eq. (A3.3) gives

$$\Omega_n = -i\sum_{m\neq n}\frac{\left\langle n\middle|\left(\nabla H\right)\middle|m\right\rangle}{E_m - E_n}\times\frac{\left\langle m\middle|\left(\nabla H\right)\middle|n\right\rangle}{E_n - E_m} = i\sum_{m\neq n}\frac{\left\langle n\middle|\left(\nabla H\right)\middle|m\right\rangle\times\left\langle m\middle|\left(\nabla H\right)\middle|n\right\rangle}{\left(E_m - E_n\right)^2}. \tag{A3.12}$$

This is Eq. (A3.4).

Equation (A3.4) can be particularly useful because in many situations taking gradients of eigenstates (as in Eq. A3.1) is cumbersome, but taking gradients of the Hamiltonian itself can be much easier. We will see an example of this in a problem set. Also, for numerical calculations Eq. (A3.4) is generally much more useful than the definition in Eq. (A3.1) because numerical solutions for the eigenstates may not naturally give a smoothly-varying phase for the eigenstates, making numerical evaluation of the Berry connection awkward and difficult.

**Homework problems**

**Berry Curvature and Anomalous Velocity**

**1. Regarding minimal models for non-zero Berry curvature.** (Easy, but gives some useful insights.)

**(a)** If one makes a Bloch state in a tight-binding model using only a single atomic basis-state orbital $\phi(\boldsymbol{r})$ within each unit cell, one has $\psi_{\boldsymbol{k}}(\boldsymbol{r}) = \sum_{\boldsymbol{R}} e^{i\boldsymbol{k}\cdot\boldsymbol{R}}\phi(\boldsymbol{r}-\boldsymbol{R}) = e^{i\boldsymbol{k}\cdot\boldsymbol{r}}\sum_{\boldsymbol{R}} e^{-i\boldsymbol{k}\cdot(\boldsymbol{r}-\boldsymbol{R})}\phi(\boldsymbol{r}-\boldsymbol{R})$.

Therefore the periodic part of the solution can be chosen to be $u_{\boldsymbol{k}}(\boldsymbol{r}) = \sum_{\boldsymbol{R}} e^{-i\boldsymbol{k}\cdot(\boldsymbol{r}-\boldsymbol{R})}\phi(\boldsymbol{r}-\boldsymbol{R})$ and in the unit cell near $\boldsymbol{R}$=0 we have just $u_{\boldsymbol{k}}(\boldsymbol{r}) = e^{-i\boldsymbol{k}\cdot\boldsymbol{r}}\phi(\boldsymbol{r})$ assuming well-localized orbitals. By evaluating the Berry connection $\boldsymbol{A}(\boldsymbol{k}) = i\left\langle u_{\boldsymbol{k}}(\boldsymbol{r})\middle|\nabla_{\boldsymbol{k}}u_{\boldsymbol{k}}(\boldsymbol{r})\right\rangle$ (integrating over the first unit cell) and the Berry curvature, show that the Berry curvature is always zero for a tight-binding band made from a single atomic basis state in each unit cell.

**(b)** If $g_{\boldsymbol{k}}(\boldsymbol{r})$ is a normalized quantum state (i.e., $\int\left|g_{\boldsymbol{k}}(\boldsymbol{r})\right|^2 d^3\boldsymbol{r} = 1$) with zero Berry curvature, show that the most general wavefunction obtained by a gauge transformation, $e^{i\chi(\boldsymbol{k})}g_{\boldsymbol{k}}(\boldsymbol{r})$, also has zero Berry curvature for any choice of the real function $\chi(\boldsymbol{k})$.

**(c)** If $f_{\boldsymbol{k}}(\boldsymbol{r})$ and $g_{\boldsymbol{k}}(\boldsymbol{r})$ are orthogonal normalized states with zero Berry curvature and also $\left\langle g_{\boldsymbol{k}}(\boldsymbol{r})\middle|\nabla_{\boldsymbol{k}}\middle|f_{\boldsymbol{k}}(\boldsymbol{r})\right\rangle = \left\langle f_{\boldsymbol{k}}(\boldsymbol{r})\middle|\nabla_{\boldsymbol{k}}\middle|g_{\boldsymbol{k}}(\boldsymbol{r})\right\rangle = 0$, show that the normalized superposition $u_{\boldsymbol{k}}(\boldsymbol{r}) = a(\boldsymbol{k})f_{\boldsymbol{k}}(\boldsymbol{r}) + b(\boldsymbol{k})g_{\boldsymbol{k}}(\boldsymbol{r})$ can have a non-zero Berry curvature. State its value in terms of $a(\boldsymbol{k})$ and $b(\boldsymbol{k})$, and show that it is real-valued.

> [From the results here, we can conclude that the minimal model to obtain a non-zero Berry curvature is one that includes mixing between at least two basis states. Problem #2 of this problem set will analyze this model. Analogously, the minimal model to obtain an electron band with non-zero Berry curvature requires mixing between at least two single-orbital bands, because within the unit cell at the origin the periodic part of the tight-binding wavefunction made from two basis states per unit cell has precisely the form $u_{\boldsymbol{k}}(\boldsymbol{r}) = a(\boldsymbol{k})f_{\boldsymbol{k}}(\boldsymbol{r}) + b(\boldsymbol{k})g_{\boldsymbol{k}}(\boldsymbol{r})$ considered in part (c). Problem #3 will analyze a specific realization.]

**2. Berry phase and Berry curvature for a general two-level system**.

The simplest model with a non-zero Berry curvature is a two-level system. (See problem 1). Therefore, consider a two-level system described by the general Hamiltonian

$$H = \varepsilon \mathbf{1}_2 + \boldsymbol{d}\cdot\boldsymbol{\sigma} = \begin{pmatrix} \varepsilon + d_z & d_x - id_y \\ d_x + id_y & \varepsilon - d_z \end{pmatrix},$$

where $\mathbf{1}_2$ is the $2\times 2$ identity matrix, $\boldsymbol{d} = (d_x, d_y, d_z)$ is a real vector with length $d$, and $\boldsymbol{\sigma}$ is a vector of Pauli matrices. If one takes $\boldsymbol{d} = -\mu_B \boldsymbol{B}$, where $\boldsymbol{B}$ is the magnetic field, this is the Hamiltonian of a spin-1/2 particle in a magnetic field. If one takes $\boldsymbol{d} = v_0 \boldsymbol{k}$, where $\boldsymbol{k}$ is an electron wavevector, this corresponds to the Hamiltonian of a Weyl fermion.

**(a)** Using the spherical representation $\boldsymbol{d} = d[\sin(\theta)\cos(\phi),\ \sin(\theta)\sin(\phi),\ \cos(\theta)]$, show that the Hamiltonian has eigenvalues $E_\pm(\boldsymbol{d}) = \varepsilon \pm d$ with normalized eigenstates that can be written

$$\left|+(\boldsymbol{d})\right\rangle = \begin{pmatrix} \cos\frac{\theta}{2} \\ e^{i\phi}\sin\frac{\theta}{2} \end{pmatrix}, \quad \left|-(\boldsymbol{d})\right\rangle = \begin{pmatrix} -\sin\frac{\theta}{2} \\ e^{i\phi}\cos\frac{\theta}{2} \end{pmatrix}. \tag{1}$$

**(b)** Show that for any coordinate $j$ that the $j$-th component of the Berry connection for the two states are given by

$$A_{j,+} \equiv i\left\langle +(\boldsymbol{d})\middle|\partial_j\middle| +(\boldsymbol{d})\right\rangle = -\sin^2\left(\frac{\theta}{2}\right)\partial_j\phi$$

$$A_{j,-} \equiv i\left\langle -(\boldsymbol{d})\middle|\partial_j\middle| -(\boldsymbol{d})\right\rangle = -\cos^2\left(\frac{\theta}{2}\right)\partial_j\phi\ .$$

**(c)** Suppose that one rotates the vector $\boldsymbol{d}$ along a closed loop at a fixed angle $\theta_0$, from $\phi = 0$ to $\phi = 2\pi$. Evaluate the Berry phase accumulated for the two states

$$\gamma_\pm = \oint \boldsymbol{A}_\pm \cdot d\boldsymbol{R}$$

in two ways: by directly evaluating the line integral, and by using Stokes theorem to convert the line integral to a surface integral

$$\gamma_\pm = \int_0^{2\pi} d\phi \int_0^{\theta_0} d\theta \left(\partial_\theta A_{\phi,\pm} - \partial_\phi A_{\theta,\pm}\right).$$

Are your answers consistent? Note that the Berry phases are only defined modulo $2\pi$, and we also have $\gamma_+ = -\gamma_-$ modulo $2\pi$.

**(d)** The integrand within the surface integral in (c) is not exactly the Berry curvature $\Omega_\pm \equiv \nabla_{\boldsymbol{d}} \times \boldsymbol{A}_\pm$ because the curl in the Berry curvature vector involves derivatives over linear displacements rather than angles. Using the expressions for the gradient and curl in spherical polar coordinates $\{d,\theta,\phi\}$ (*e.g.*, see the back cover of Jackson) to operate on the eigenstates (Eq. 1) show that the Berry curvature for the two-level system takes the simple form

$$\nabla_{\boldsymbol{d}} \times \boldsymbol{A}_{\pm} = -\pm\frac{1}{2}\frac{\vec{\boldsymbol{d}}}{d^3} = -\pm\frac{1}{2}\frac{\hat{\boldsymbol{d}}}{d^2}. \qquad (2)$$

This quantity is equivalent to the field produced by a monopole of charge $-\pm 1/2$ located at the origin. From this, relate your answer for the Berry phase in (c) to the flux through the area on the sphere bounded by the path of the rotating vector ***d***.

**(e)** Repeat the calculation of $A_{j,\pm}$, $\gamma_{\pm}$, and $\nabla_{\boldsymbol{d}} \times \boldsymbol{A}_{\pm}$ in parts (b-d) for a different gauge, obtained by transforming $|\pm(\boldsymbol{d})\rangle \rightarrow e^{i\chi(\theta,\phi)}|\pm(\boldsymbol{d})\rangle$, where $\chi$ is a doubly-differentiable function. Show that although the Berry connection changes, the geometric phase over a closed loop remains unchanged (modulo $2\pi$) and the Berry curvature is unchanged.

**(f)** Using the alternative expression for the Berry curvature mentioned in the Lecture Notes,

$$\Omega_n = i\sum_{m\neq n}\frac{\langle n|(\nabla H)|m\rangle \times \langle m|(\nabla H)|n\rangle}{(E_m - E_n)^2}, \qquad (3)$$

provide an alternative derivation of the Berry curvature $\Omega_{\pm} \equiv \nabla_{\boldsymbol{d}} \times \boldsymbol{A}_{\pm}$ for this problem (Eq. (2)). It is easiest if you first rotate your coordinate axes so that $\hat{z}$ is parallel to ***d***.

**3. Berry curvature for gapped graphene or $MoS_2$.**

We saw in a previous problem set that the effective Hamiltonian for the electronic states with a given ***k*** vector in graphene can be expanded in the following form near the K and K' points where the band gap goes to zero:

$$H_{\text{eff}}(\boldsymbol{k}) = \begin{pmatrix} 0 & U(k_x - \pm ik_y) \\ U(k_x \pm ik_y) & 0 \end{pmatrix}.$$

Here the two basis states are the tight-binding Bloch states formed from (a) the A orbitals and (b) the B orbitals. $U$ is a constant (assume it is real for simplicity) and the wave vectors are measured relative to the K or K' point (not from zero). The sign in the off-diagonal terms is different for the K and K' points.

If the A-B lattice symmetry in graphene is broken (e.g., as it is in hexagonal boron nitride or transition-metal dichalcogenides like $MoS_2$), there will also be a splitting of the diagonal matrix elements:

$$H_{\text{eff}}(\boldsymbol{k}) = \begin{pmatrix} m & U(k_x - \pm ik_y) \\ U(k_x \pm ik_y) & -m \end{pmatrix}. \tag{4}$$

**(a)** Show that the Berry curvature for the upper band derived from Eq. (4) has the form

$$\Omega_+ \equiv \nabla_{\boldsymbol{k}} \times \boldsymbol{A}_+ = -\frac{1}{2}\frac{(\pm) mU^2\ \hat{z}}{\left(m^2 + U^2\left(k_x^2 + k_y^2\right)\right)^{3/2}}$$

where the $\pm$ corresponds to the K or K' point. You can do this by thinking how the result in problem 2 is changed when all derivatives with respect to $k_z$ are zero (since $k_z$ does not appear in the problem), or by using Eq. (3) with the form of the eigenstates in Eq. (1) and $\theta = \cos^{-1}(m/d)$, $d = \sqrt{m^2 + U^2(k_x^2 + k_y^2)}$, and $\phi = \tan^{-1}(\pm k_y / k_x)$. (Note that you are no longer free to rotate your coordinate axes in this problem without transforming the Hamiltonian because the Hamiltonian is not spherically symmetric in k-space.)

**(b)** By doing a surface integral of the Berry curvature, determine the Berry phase in the upper band for a closed contour about either the K or the K' point as shown in the sketch on the right, in which the wavevector ***k*** traces out a path of constant energy $E = \sqrt{m^2 + U^2 k_0^2}$, where $k_r$ is a constant radius in k-space from the bandgap minimum.

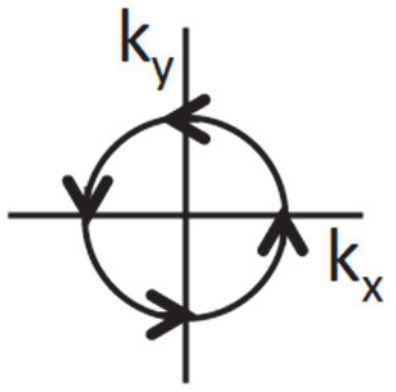

**(c)** Discuss the limits of your answers in (b) for $m/(Uk_r) >> 1$ and $m/(Uk_r) << 1$.

Also, sketch the magnitude of the Berry curvature as a function of distance $k = \sqrt{k_x^2 + k_y^2}$ from the K' point to understand what region of k space contributes significantly to the integrals of Berry curvature.

Note: In the above we have assumed implicitly that the diagonal term *m* (known as the "mass" term") has the same sign for both the K and K' points. In a model with broken time reversal symmetry it is possible to imagine a situation in which the mass terms at K and K' have opposite signs, in which case the total integral of the Berry phase within a band can be nonzero, achieving a "topological" state.

**(d)** The Berry curvature associated with the K or K' valleys causes the carriers there to be deflected sideways relative to an applied electric field (even in the absence of any applied magnetic field), with opposite signs of deflection for the two valleys. The strength of this effect, *e.g.*, the intrinsic valley Hall conductivity associated with occupied states in the conduction band near the K point, can be written in terms of the Berry curvature as

$$\sigma_K^{in} = -\frac{e^2}{h}\left(\frac{1}{2\pi}\right)^2 \int_{\substack{\text{occ. states} \\ \text{near K}}} \mathbf{\Omega}_+(k_x,k_y)\cdot\hat{z}\; d^2k .$$

Assuming a circular Fermi circle and taking the lowest-order Taylor expansion for $Uk_0 / m << 1$ (second order), use the result of part (b) to express the intrinsic valley Hall conductivity in terms of the carrier concentration per unit area

$$n = 2\left(\frac{1}{2\pi}\right)^2 \int_{\substack{\text{occ. states} \\ \text{near K}}} d^2k .$$

The valley Hall effect in $MoS_2$ was first demonstrated experimentally by Kin Fai Mak, Paul McEuen, and collaborators in *Science* **344**, 1489 (2014).

**4**. **Su-Shrieffer-Heeger Model: The Peierls' instability and topological edge states:**.
We discussed in class the Su-Shrieffer-Heeger model which shows that quasi-one-dimensional metals are unstable with respect to dimerization. In this problem I will ask you to work through the physics of this. Consider the specific realization of polyacetylene:

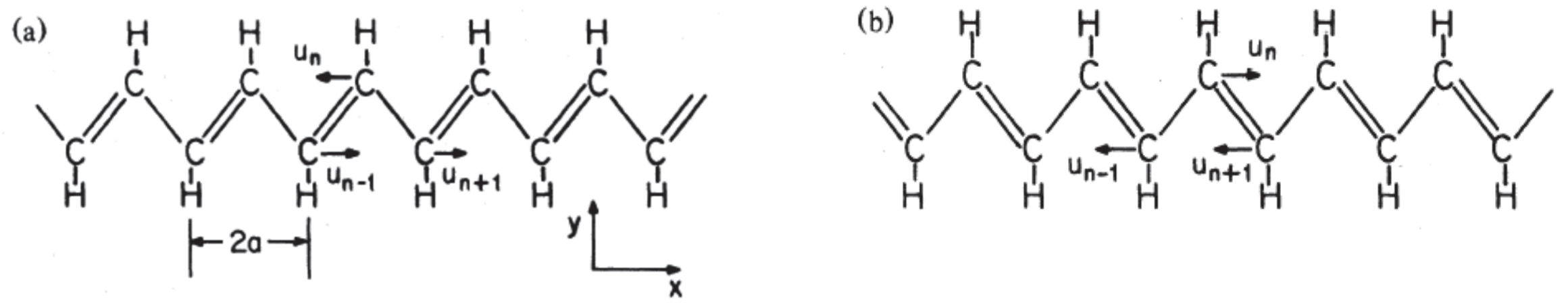

Define the lattice coordinates $u_n$ to denote the horizontal displacement of the n$^{th}$ CH group away from its undisplaced lattice position at *na* (as drawn). (These have positive values for displacements in the +x direction, and negative values for displacements in the -x direction.) For the structure shown in Fig. (a), $u_n < 0$ with $u_{n-1}$ and $u_{n+1} > 0$, leading to a short ("double") bond between the groups n-1 and n, and a long ("single") bond between n and n+1. For the structure shown in Fig. (b), we have the reversed situation. The potential energy of the phonon coordinates can be written

$$V_{lattice} = \frac{1}{2}\sum_n K\left(u_{n+1} - u_n\right)^2 .$$

**(a)** (Easy) What is the potential energy of the lattice for the "dimerized" chain with $u_n = (-1)^n u_0$?

The electronic band structure near the Fermi energy can be calculated approximately using a tight-binding model of $p_z$-type orbitals occupied by one electron per CH group, in a way very similar to carbon nanotubes. (Just as for nanotubes, the s, $p_x$, and $p_y$ orbitals lead to bands well below and well above the Fermi energy.) Let us express the electronic part of the Hamiltonian using the (real-valued) overlap matrix element $t_{n+1,n}$ (in a previous notation these overlap integrals have been called $\gamma$)

$$H_{el} = -\sum_{n,s} t_{n+1,n}\left[\left|n+1,s\right\rangle\left\langle n,s\right| + \left|n,s\right\rangle\left\langle n+1,s\right|\right],$$

where $\left|n,s\right\rangle$ is a state with a $p_z$ electron of spin s=±1/2 on the n$^{th}$ CH group.

**(b)** For the undimerized chain (all the u's = 0), the overlap matrix elements are all the same (set them all equal to $t_0$). Derive the tight-binding band structure (this is easy), and plot it using an unconventional reduced zone scheme for a unit cell having a basis with two CH groups (as I did briefly in class). What levels are filled, if there is one electron in the $p_z$ band per CH group?

**(c)** If the polyacetylene chain happens to become dimerized, then the overlap matrix elements will depend on the length of the bond. To account for this electron-lattice interaction, let us expand this dependence to first order about the undimerized state:

$$t_{n+1,n} = t_0 + \delta t_{n+1,n} = t_0 - \alpha(u_{n+1} - u_n).$$

Given a dimerized displacement amplitude u, such that $u_n = (-1)^n u_0$, derive the tight-binding band structure assuming periodic boundary conditions with an even number of CH groups. You will have a crystal with two atoms per unit cell, so the procedure will be very similar to the nanotube problem. Using a relatively small dimerization, plot the dimerized band structure together on a graph with your result from (b), in a reduced-zone scheme for two CH groups per unit cell. Compare the two different band structures. (The answer should be as expected, given intuition you gained from the nearly-free electron model and what we discussed in class.)

**(d)** For a chain with a number $N$ of CH units (periodic boundary conditions) with temperature T=0, show that the total energy of the system, including both the electronic and the lattice contributions is, after converting a sum to an integral

$$E_0 = \frac{-4Nt_0}{\pi} E\left(1 - z^2\right) + \frac{NKt_0^2 z^2}{2\alpha^2}$$

where $E(1-z^2) = \int_0^{\pi/2} dx \sqrt{1 - (1 - z^2)\sin^2(x)}$ is an elliptic integral and $z = 2\alpha u_0 / t_0$.

**(e)** The important question is: Is it possible to lower the total energy of the system by dimerizing, or does the undimerized configuration give the ground state? Expanding the elliptic integral in small z (or u),

$$E\left(1 - z^2\right) \approx 1 + \frac{1}{2}\left[\ln\left(\frac{4}{|z|}\right) - \frac{1}{2}\right] z^2 + \ldots$$

show that for a half-filled band the undimerized state is always unstable to dimerization at $T$=0. This is the Peierls' instability, and the resulting dimerized state is called a commensurate charge density wave because of the periodic concentration of charge in the "double" bonds, commensurate with some multiple of the underlying lattice of atoms. (There is also such a thing as an incommensurate charge density wave in other materials.) Taking K = 21 eV-Å$^{-2}$, $t_0$ = 2.5 eV $\alpha$ = 4.1 eV/Å, and the lattice constant a=1.22 Å, plot the energy per CH group as a function of the staggered dimerization coordinate u. (We treat the lattice coordinates in the Born-Oppenheimer approximation, fixing them in position on the time scale of electronic motions.) There should be two stable minima at $\pm u_0$ = 0.04 Å, corresponding to the A and B phases shown in the figure. Explain, in terms of the competition between electronic and lattice energies, why a dimerized state is favored at low temperatures.

**(f) Demonstration of edge states**: Now consider a finite 1-d chain with 20 CH groups (so, 10 unit cells), with uncoupled ends instead of periodic boundary conditions. Normalize the energy scale by assuming that the overlap matrix element between unit cells is 1, and let the overlap matrix element within each unit cell be real with a variable strength $\upsilon$ between 0 and 3. The Hamiltonian can therefore be represented as a 20 x 20 matrix with the pattern (here I only give a 10 x 10 version)

$$H = \begin{pmatrix} 0 & \upsilon & 0 & 0 & 0 & 0 & 0 & 0 & 0 & 0 \\ \upsilon & 0 & 1 & 0 & 0 & 0 & 0 & 0 & 0 & 0 \\ 0 & 1 & 0 & \upsilon & 0 & 0 & 0 & 0 & 0 & 0 \\ 0 & 0 & \upsilon & 0 & 1 & 0 & 0 & 0 & 0 & 0 \\ 0 & 0 & 0 & 1 & 0 & \upsilon & 0 & 0 & 0 & 0 \\ 0 & 0 & 0 & 0 & \upsilon & 0 & 1 & 0 & 0 & 0 \\ 0 & 0 & 0 & 0 & 0 & 1 & 0 & \upsilon & 0 & 0 \\ 0 & 0 & 0 & 0 & 0 & 0 & \upsilon & 0 & 1 & 0 \\ 0 & 0 & 0 & 0 & 0 & 0 & 0 & 1 & 0 & \upsilon \\ 0 & 0 & 0 & 0 & 0 & 0 & 0 & 0 & \upsilon & 0 \end{pmatrix}$$

Diagonalize the 20 x 20 version of this matrix numerically and plot the spectrum of energy levels as a function of $\upsilon$ between 0 and 3. The regime of the topological ground state is $\upsilon < 1$. Show that in this regime there are two states with energy very close to zero, but in the non-topological state there are no states near zero energy. For $\upsilon = 0.5$, determine the eigenstates corresponding to these zero-energy levels and verify that they are linear superpositions of states localized near the two edges of the sample.

(As $\upsilon$ approaches 1 from below for a finite-size sample, the two edge states can couple and experience an avoided crossing so that their energies split about E = 0. If you are interested and have time, you can check that for a larger finite chain the energy splittings in this regime are smaller.)